\documentclass[twocolumn]{aastex631}

\usepackage{placeins}
\usepackage{amsmath}
\usepackage{txfonts}
\usepackage{comment}

\newcommand\medMW{$<\dot{M}_{PEW}>$~}

\newcommand\medtnu{$<t_{\nu,0}>$~}
\newcommand\medMO{$<M_0>$~}
\newcommand\medRO{$<R_0>$~}
\newcommand\medMacc{$<\dot{M}_{*}>$~}
\newcommand\MO{$M_0$~}
\newcommand\RO{$R_0$~}
\newcommand\Mw{$\dot{M}_{PEW}$~}
\newcommand\Msunyr{M$_{\odot}$/yr~}
\newcommand\tnu{$t_{\nu,0}$~}
\newcommand\tacc{$t_{acc,0}$~}
\newcommand\alphaDW{$\alpha_{DW}$~}

\newcommand{\BT}[1]{\textcolor{black}{#1}}
\newcommand{\PP}[1]{\textcolor{black}{#1}}
\newcommand{\BTbis}[1]{\textcolor{black}{#1}} %update from August 15 

\newcommand{\revBT}[1]{\textcolor{black}{#1}}

\defcitealias{c}{AGE-PRO I}
\defcitealias{AGEPRO_II_Ophiuchus}{AGE-PRO II}
\defcitealias{AGEPRO_III_Lupus}{AGE-PRO III}
\defcitealias{AGEPRO_IV_UpperSco}{AGE-PRO IV}
\defcitealias{AGEPRO_V_gasmasses}{AGE-PRO V}
\defcitealias{AGEPRO_XI_gas_disk_sizes}{AGE-PRO VI}
\defcitealias{AGE-PRO_externalPE}{AGE-PRO_externalPE}

\shorttitle{AGE-PRO: disk population synthesis}
\shortauthors{Tabone et al.}
\begin{document}

\title{The ALMA Survey of Gas Evolution of PROtoplanetary Disks (AGE-PRO):\\
VII. Testing accretion mechanisms from disk population synthesis}

\correspondingauthor{Benoît Tabone}
\email{benoit.tabone@cnrs.fr}

\author[0000-0002-1103-3225]{Benoit Tabone}
\affiliation{Université Paris-Saclay, CNRS, Institut d’Astrophysique Spatiale, 91405 Orsay, France}

\author[0000-0003-4853-5736]{Giovanni P. Rosotti}
\affiliation{Dipartimento di Fisica, Università degli Studi di Milano, Via Celoria 16, I-20133 Milano, Italy}

\author[0000-0002-8623-9703]{Leon Trapman}
\affiliation{Department of Astronomy, University of Wisconsin-Madison, 
475 N Charter St, Madison, WI 53706, USA}

\author[0000-0001-8764-1780]{Paola Pinilla}
\affiliation{Mullard Space Science Laboratory, University College London, 
Holmbury St Mary, Dorking, Surrey RH5 6NT, UK}

\author[0000-0001-7962-1683]{Ilaria Pascucci}
\affiliation{Lunar and Planetary Laboratory, the University of Arizona, Tucson, AZ 85721, USA}

\author[0000-0003-2090-2928]{Alice Somigliana}
\affiliation{European Southern Observatory, Karl-Schwarzschild-Strasse 2, D-85748 Garching bei München, Germany}

\author[0000-0001-6410-2899]{Richard Alexander}
\affiliation{School of Physics \& Astronomy, University of Leicester, University Road, Leicester, LE1 7RH, United Kingdom}

\author[0000-0002-4147-3846]{Miguel Vioque}
\affiliation{European Southern Observatory, Karl-Schwarzschild-Str. 2, 85748 Garching bei München, Germany}
\affiliation{Joint ALMA Observatory, Alonso de Córdova 3107, Vitacura, Santiago 763-0355, Chile}

\author[0009-0004-8091-5055]{Rossella Anania}
\affiliation{Dipartimento di Fisica, Università degli Studi di Milano, Via Celoria 16, I-20133 Milano, Italy}

\author[0000-0002-6946-6787]{Aleksandra Kuznetsova}
\affiliation{Center for Computational Astrophysics, Flatiron Institute, 162 Fifth Ave., New York, New York, 10025}

\author[0000-0002-0661-7517]{Ke Zhang}
\affiliation{Department of Astronomy, University of Wisconsin-Madison, 475 N Charter St, Madison, WI 53706, USA}

\author[0000-0002-1199-9564]{Laura M. P\'erez}
\affiliation{Departamento de Astronom\'ia, Universidad de Chile, Camino El Observatorio 1515, Las Condes, Santiago, Chile}

\author[0000-0002-2828-1153]{Lucas A. Cieza}
\affiliation{Instituto de Estudios Astrofísicos, Universidad Diego Portales, Av. Ejercito 441, Santiago, Chile}

\author[0000-0003-2251-0602]{John Carpenter}
\affiliation{Joint ALMA Observatory, Avenida Alonso de C\'ordova 3107, Vitacura, Santiago, Chile}

\author[0000-0003-0777-7392]{Dingshan Deng}
\affiliation{Lunar and Planetary Laboratory, the University of Arizona, Tucson, AZ 85721, USA}

\author[0000-0002-7238-2306]{Carolina Agurto-Gangas}
\affiliation{Departamento de Astronom\'ia, Universidad de Chile, Camino El Observatorio 1515, Las Condes, Santiago, Chile}

\author[0000-0003-3573-8163]{Dary A. Ru\'iz-Rodr\'iguez}
\affiliation{National Radio Astronomy Observatory, 520 Edgemont Rd., Charlottesville, VA 22903, USA}

\author[0000-0002-5991-8073]{Anibal Sierra}
\affiliation{Departamento de Astronom\'ia, Universidad de Chile, Camino El Observatorio 1515, Las Condes, Santiago, Chile}
\affiliation{Mullard Space Science Laboratory, University College London, 
Holmbury St Mary, Dorking, Surrey RH5 6NT, UK}

\author[0000-0002-2358-4796]{Nicol\'as T. Kurtovic}
\affiliation{Max Planck Institute for Extraterrestrial Physics, Giessenbachstrasse 1, D-85748 Garching, Germany}
\affiliation{Max-Planck-Institut fur Astronomie (MPIA), Konigstuhl 17, 69117 Heidelberg, Germany}

\author[0000-0002-1575-680X]{James Miley}
\affiliation{Departamento de Física, Universidad de Santiago de Chile, Av. Victor Jara 3659, Santiago, Chile}
\affiliation{Millennium Nucleus on Young Exoplanets and their Moons (YEMS), Chile}
\affiliation{Center for Interdisciplinary Research in Astrophysics and Space Exploration (CIRAS), Universidad de Santiago de Chile, Chile}

\author[0000-0003-4907-189X]{Camilo Gonz\'alez-Ruilova}
\affiliation{Instituto de Estudios Astrof\'isicos, Universidad Diego Portales, Av. Ejercito 441, Santiago, Chile}
\affiliation{Millennium Nucleus on Young Exoplanets and their Moons (YEMS), Chile}
\affiliation{Center for Interdisciplinary Research in Astrophysics and Space Exploration (CIRAS), Universidad de Santiago de Chile, Chile}

% \author{Mat\'ias G\'arate}
% \affiliation{Max-Planck-Institut fur Astronomie (MPIA), Konigstuhl 17, 69117 Heidelberg, Germany}

\author[0000-0001-9961-8203]{Estephani TorresVillanueva}
\affiliation{Department of Astronomy, University of Wisconsin-Madison, 
475 N Charter St, Madison, WI 53706, USA}

\author[0000-0001-5217-537X]{Michiel R. Hogerheijde}
\affiliation{Leiden Observatory, Leiden University, PO Box 9513, 2300 RA Leiden, the Netherlands}
\affiliation{Anton Pannekoek Institute for Astronomy, University of Amsterdam, the Netherlands}

\author[0000-0002-6429-9457]{Kamber Schwarz}
\affiliation{Max-Planck-Institut fur Astronomie (MPIA), Konigstuhl 17, 69117 Heidelberg, Germany}

\author[0000-0003-2090-2928]{Claudia Toci}
\affiliation{European Southern Observatory, Karl-Schwarzschild-Strasse 2, D-85748 Garching bei München, Germany}
\affiliation{INAF, Osservatorio Astrofísico di Arcetri, 50125, Firenze, Italy}

\author{Leonardo Testi}
\affiliation{Dipartimento di Fisica e Astronomia, Universita’ di Bologna, Via Gobetti 93/2, I-40122 Bologna, Italy}

\author[0000-0002-2357-7692]{Giuseppe Lodato}
\affiliation{Dipartimento di Fisica, Università degli Studi di Milano, Via Celoria 16, I-20133 Milano, Italy}

\begin{abstract}
The architecture of planetary systems depends on the evolution of the disks in which they form. In this work, we develop a population synthesis approach to interpret the AGE-PRO measurements of disk gas mass and size considering two scenarios: turbulence-driven evolution with photoevaporative winds and MHD disk-wind-driven evolution. A systematic method is proposed to constrain the distribution of disk parameters from the disk fractions, accretion rates, disk gas masses, and CO gas sizes. We find that turbulence-driven accretion with initially compact disks ($R_0 \simeq 5-20~$au), low mass-loss rates, and relatively long viscous timescales ($t_{\nu,0} \simeq 0.4-3~$Myr or $\alpha_{SS} \simeq 2-4 \times 10^{-4}$) can reproduce the disk fraction and gas sizes. However, the distribution of apparent disk lifetime defined as the $M_D/\dot{M}_*$ ratio is severely overestimated by turbulence-driven models. On the other hand, MHD wind-driven accretion can reproduce the bulk properties of the disk populations from Ophiuchus to Upper Sco assuming compact disks with an initial magnetization of about $\beta \simeq 10^5$ ($\alpha_{DW} \simeq 0.5-1 \times 10^{-3}$) and a magnetic field that declines with time. More studies are needed to confirm the low masses found by AGE-PRO, notably for compact disks that question turbulence-driven accretion. The constrained synthetic disk populations can now be used for realistic planet population models to interpret the properties of planetary systems on a statistical basis. 
\end{abstract}
%\keywords{accretion discs - MHD disc winds – protoplanetary discs – submillimetre: planetary systems}

\section{Introduction}

With more than 5000 exoplanets detected so far\footnote{On April 2024, from \url{https://exoplanets.nasa.gov/}.}, planet formation appears to be a widespread phenomenon in disks orbiting young stars \citep{2019ApJ...874...81F}. The more striking results obtained by the exoplanet community are the large diversity in the properties of exoplanetary systems and the occurrence of specific kinds of systems. The next challenge is to understand how different evolutionary pathways of planet-forming disks lead to the diversity and occurrence of planetary systems \citep{2016JGRE..121.1962M}. 

Recent models have already attempted to include many steps of planet formation in disk evolution models to predict the properties of exoplanets at a population level 
\citep{2023ASPC..534..717D}. Yet the evolution of gaseous disks remains vastly unknown despite its paramount importance in every step of planet formation \citep{2023ASPC..534..501M}. One of the main limiting factors is the need for unbiased surveys of young stars of different ages since an individual disk cannot be followed up across its lifetime. Pioneering surveys have already unveiled some of the basic properties of disks. Infrared excess and accretion signatures in clusters of different ages indicate that disks disperse quickly after a typical timescale of 2-5 Myr \citep{2010A&A...510A..72F,2014A&A...561A..54R}, depending on the mass of the host star and the vicinity of massive stars \citep{2015A&A...576A..52R,2022EPJP..137.1132W}. Surveys from the UV to the near-IR have also shown that almost all disks exhibit signs of accretion with accretion rates scaling with stellar mass and declining with time \citep{2016ARA&A..54..135H,2017A&A...600A..20A,2017A&A...604A.127M,2023ApJ...945..112F}.

Over the past decades, disk dispersal and disk accretion have been explained by two distinct processes. In the classical picture of viscous disks, popularized by \citet{1973A&A....24..337S}, disks accrete via the radial transport of angular momentum mediated by turbulence. The origin of turbulence can be the magnetorotational instability (MRI) or hydrodynamical instabilities such as the vertical shear instability or the gravitational instability \citep[][]{2023ASPC..534..465L,2016ARA&A..54..271K}. On the other hand, disk dispersal is explained by photoevaporative winds launched from the disk atmosphere heated by energetic radiation from the star \citep{1994ApJ...428..654H, 2014prpl.conf..475A}. The relevant energy of the photons remains debated. Hydrogen ionizing UV photons (EUV) would have a limited impact due to efficient screening of the disk layers but either X-ray or FUV photons can penetrate the disk's upper layers and deposit enough energy to drive a thermal wind outward of 1-10~au \citep{2010MNRAS.401.1415O,2009ApJ...690.1539G,2018ApJ...857...57N}. Early models combining turbulence-driven accretion and photoevaporation were able to account for the observed accretion rates and disk dispersal time \citep[e.g.,][]{2006MNRAS.369..229A,2012MNRAS.422.1880O}.

However, theoretical works including the weak coupling between the magnetic field and the gas show that MRI cannot be triggered in large portions of disks called dead-zones \citep{1996ApJ...457..355G,2011ApJ...736..144B, delage2022}. Alternatively, MHD disk-winds can efficiently drive accretion across both the dead zones and MRI active regions by extracting angular momentum vertically \citep{1982MNRAS.199..883B,1997A&A...319..340F,2013ApJ...769...76B}. \revBT{MHD disk winds would naturally account for the observations of winds revealed by a wide range of tracers from the millimeter to the optical \citep[e.g.,][]{2017A&A...607L...6T,2018A&A...618A.120L,2025NatAs...9...81P}}. The only requirement is that the disk is embedded in a magnetic field with a non-vanishing magnetic flux. It should be noted that while early MHD disk-wind models used strongly magnetized disks, resulting in very low disk masses, recent theoretical work demonstrated the ability of MHD disk-winds to operate even with a mid-plane magnetization as low as $\beta = 10^5$ where the $\beta$ parameter is the ratio between thermal and magnetic pressure \citep{2017A&A...600A..75B}. Disk dispersal in MHD wind-driven accretion remains an open question due to the lack of numerical simulations running on the long-time scale. Still, simple disk evolution models show that if the magnetic field strength does not decline too fast, disks can be dispersed in a short dispersal phase \citep{2013ApJ...778L..14A,2016ApJ...821...80B,2022MNRAS.512.2290T}.

\revBT{Today, our lack of knowledge on which mechanism drives disk accretion and dispersal constitutes a fundamental hurdle in our understanding of planet formation and migration.  If accretion is driven by MHD disk-winds, every step of planet formation will be impacted. To name a few examples, MHD disk-winds can create pressure bumps where dust grows \citep{2018MNRAS.477.1239S,2020A&A...639A..95R}, pebble accretion is expected to be amplified in low-turbulence disks \citep{2024Icar..41716085Y}, and planet migration pattern can be profondly impacted by the different local gas flow \citep{2022A&A...658A..32L,2023A&A...677A..70W} with possibly outward type-II planet migration for highly magnetized disks \citep{2020A&A...633A...4K}. 1D disk evolution models, under specific prescriptions, also predict the formation of a gas depletion in the inner disk, altering type-I migration \citep{2015A&A...584L...1O,2018A&A...615A..63O}, a proposal that calls for global numerical simulation running on secular timescale. In other words, if accretion is effectively driven by MHD winds, planet-formation scenarios and planet population synthesis models need to be revised.}

However, infrared excess or accretion signatures are indirect probes of the evolution of the bulk disk. It is only with ALMA that fundamental properties like disk masses and sizes can now be probed on statistically representative samples. The first surveys focused on the millimeter (mm) dust emission giving access to the continuum flux and continuum size in the close-by star-forming regions \citep[e.g.,][]{andrews2005, andrews2013, ricci2010, 2016ApJ...827..142B,2017AJ....153..240A,2019A&A...626A..11C,2022A&A...661A..53V}. 
%Using estimates of dust opacity and assuming optically thin emission, disk dust masses have been inferred. This revealed a decline of dust mass from very young disks (flat spectrum) down to older regions like Upper Scorpius but also a large scatter is disk masses from one region to the other. 
It is unclear though to what extent dust mm flux can inform us about the gas. Because of dust radial drift, one would naturally expect the continuum disk size and the continuum luminosity to be poor proxies for the disk gas size and mass \citep{2014prpl.conf..339T,2020MNRAS.498.2845S,2021MNRAS.507..818T,2022MNRAS.514.1088Z}. 
%Yet, the relatively constant ratio between dust and gas size and the dust-independent gas mass estimates suggest that dust continuum emission provide first constraints on the properties of the bulk gaseous disk. 
Yet, using mm continuum luminosity, \citet{2016A&A...591L...3M} found a correlation between disk mass (assumed to be 100 times the dust disk mass) and accretion rate with a typical ratio of $M_D/\dot{M}_{*} \simeq 1-3~$Myr.

Surveys at the population level offer a unique opportunity to challenge disk evolution scenarios and the recent ALMA observations have triggered a renewed interest in disk evolution models. \citet{2017MNRAS.472.4700L} and \citet{2017ApJ...847...31M} showed that the $M_D - \dot{M}_{*}$ correlation
%found in young star-forming regions ($\simeq 2-3$~Myr) 
can be naturally reproduced by viscous models with a significant spread in the viscous timescale and disk's age. Yet, these models did not consider disk dispersal, which can affect the $M_D - \dot{M}_{*}$ correlation \citep{2020MNRAS.492.1120S}. When photoevaporation is included, viscous models can still reproduce the overall $M_D - \dot{M}_{*}$ but remain dependent on dust evolution models \citep{2019A&A...628A..95M,2020MNRAS.498.2845S,2023A&A...673A..78E}. The development of 1D disk evolution models including MHD disk-winds \citep{2016A&A...596A..74S,2016ApJ...821...80B,2022MNRAS.512.2290T} now permits confronting the predictions of MHD wind-driven accretion with disk demographics. In this context, \citet{2022MNRAS.512L..74T} showed that simple MHD wind-driven evolution models can naturally reproduce the $M_D - \dot{M}_{*}$ correlation and the rapidity of disk dispersal.

Today, the ALMA large program AGE-PRO \citep{AGEPRO_I_overview}, gives access to the gas evolution thanks to deep observations of molecular lines over a sample of 30 disks located in three star-forming regions of different ages: Ophiuchus \citep{AGEPRO_II_Ophiuchus}, Lupus \citep{AGEPRO_III_Lupus}, and Upper Scorpius \citep{AGEPRO_IV_UpperSco}. The outcome of AGE-PRO is twofold. First, the inclusion of N$_2$H$^+$ and CO isotopologues in the spectral setup permits estimates of disk gas masses \citep{2019ApJ...881..127A,2022ApJ...926L...2T} using extensive grids of thermochemical models \citep{AGEPRO_V_gasmasses}. The disk mass estimates are therefore not hampered by assumptions on dust optical properties and gas-to-dust ratio.
%, though significant systematic uncertainties are induced by the use of thermochemical models. 
Secondly, the CO disk sizes, a key diagnostic for disk physical size \citep{2023MNRAS.518L..69T,2023ApJ...954...41T},  are now measured taking into account beam convolution effects \citep{AGEPRO_XI_gas_disk_sizes}.
%which can be converted into physical disk size as defined as the radius encompassing the bulk mass of the disk \cite{2023ApJ...954...41T}. 
%Disk gas size has often been suggested as a discriminant test between radial and vertical transport of angular momentum as the former requires disk spreading.

%In the classical picture of viscous disks, popularized by Sakura and Sunayev and supported by the re-discovery of the magneto-rotational instability, disk can accrete via a raidal tansport of angular momentum mediated by turbulence 

%Recent surveys of star-forming regions of different ages open the possibility to constrain the disk evolution pathways. However, models at disk population level are required to interpret these data and quantify the processes that drive disk evolution and dispersal. 

In this paper, we develop a disk population synthesis approach to interpret the results of the ALMA AGE-PRO large program. A systematic approach is proposed to constrain the distribution of disk parameters from not only the disk mass and size inferred by AGE-PRO but also accretion properties and disk fraction. In Sec. \ref{sec:model}, the disk evolution models are presented along with the observational constraints and the population synthesis approach. The fitting procedure is then applied to each scenario in Sec. \ref{sec:results}. The implications of the present work and its limitations are discussed in Sec. \ref{sec:discusion} and the findings are summarized in Sec. \ref{sec:conclusion}.

\begin{comment}
    
Review of disk pop models, \\
- Disk evolution, angular tr of AM and wind contols the final architechture of pl. system. Yet which process dominate and when remains an open question.
- Many direct observations
- Otherway to adress the question: directly probing the long term evolution of disks.
- Viscous evolaution: from Lodato+2017 and Mulders+2017 with comparision with observations but no PEW. In fact Somigliana+2019 = PEW affect evolution of median disk mass by removing light disks. Also Emsenhuber+2023 with neural network to explore the parameter space but focused on MD and Macc, not disk mass. In our work: propose a simpler fitting procedure. \\
- MHD DW: Tabone+2022a,. Also Weider+2023 but conmplicated.
- recall the assumption: Lupus is reresentative of a disk population of an age of about 2-3Myr.
\end{comment}

\section{Model}
\label{sec:model}

\subsection{Individual disks}
\label{subsec:indiv-disk}

The evolution of an individual disk is computed using a 1D formalism, where the MHD equations are integrated in the vertical direction. The radial transport of angular momentum is described following the \citet{1973A&A....24..337S} parametrization with the generalization proposed by \citet{2022MNRAS.512.2290T}. Our choice in the parameterization of disk evolution has the major advantage of facilitating the comparison between turbulence-driven and MHD wind-driven evolution. The values of the parameters can also be easily compared with the results of numerical simulations. The parameters of the model are summarized in Table \ref{table:model}. In both scenarios, the evolution of an individual disk is set by 4 parameters: initial disk mass $M_0$ and radius $R_0$, the generalized $\alpha$ parameter (or equivalently the accretion timescale $t_{acc, 0}$ or the viscous timescale $t_{\nu, 0}$), and a parameter quantifying the wind mass-loss rate. For the MHD disk-wind model, an additional parameter, $\omega$, controls the secular evolution of the magnetic field. We only consider isolated disks that are not fed by any envelope or streamer. The time $t=0$ can therefore be considered as the end of the Class I phase, and \PP{as several of the AGE-PRO Oph targets are Class I, we assume that the Oph disks represent the time zero in our models}. Neglecting the impact of early accretion is further justified by the mass of the residual envelope around the Ophiuchus sources being smaller than the disk masses \citep{AGEPRO_II_Ophiuchus}. \revBT{We further neglect the impact of mild external UV irradiation, which is futher studied in the case of turbulence-driven accretion by \citet{AGE-PRO_externalPE} for the Upper Sco AGE-PRO sample.}

%represented in the AGE-PRO sample by disks in Ophiuchus ("flat spectrum")

\begin{table}
\centering 
\caption{Summary of the parameters of the population models with the free parameters outlined in bold font.}             % title of Table
\begin{tabular}{c c c c}        % centered columns (4 columns)
\hline\hline           % inserts double horizontal lines
 Parameter & Symbol & Range of median value & Adopted \\
 & & & \PP{spread}$^{(a)}$ \\
\hline      % inserts single horizontal line
 && \textbf{Initial conditions}&\\ 
 \textbf{Initial disk size} &  $R_0$  & 5-20~au & 0.3 \\  
 \textbf{Initial disk mass} &  $M_0$  & $10^{-4}$ - $10^{-1}$~$M_{\odot}$  & 0.6  \\  
 \hline
 && \textbf{MHD wind model}&\\
  \textbf{$\alpha$ parameter} &  $\alpha_{DW}$ & $10^{-4}-10^{-2}$ & 0.2 \\
 Accretion timescale & $t_{acc,0}$ & $0.1-2\,$Myr & ** \\
  \textbf{$\omega$ parameter} & $\omega$ & $0.25-0.5$ & 0 \\
 \textbf{Magnetic lever arm} & $\lambda$ & 2-20 & 0 \\
  Ejection-to-accretion & $f_M$ & 0.1-10 & ** \\
  \hline
  && \textbf{Turbulent model} &\\
    \textbf{$\alpha$ parameter} &  $\alpha_{SS}$ & $10^{-4}-10^{-2}$ & 0.2 \\
 Viscous timescale & $t_{\nu,0}$ & $0.1-5\,$Myr & ** \\
  \textbf{Wind mass-loss rate} & $\dot{M}_{PEW}$ & $10^{-11}$-$10^{-6}$ \Msunyr & 0.3 \\
\hline                  
\hline 
% inserts single 
%inserts single line
\end{tabular}\\
{$^{(a)}$ This corresponds to the spread of the lognormal distributions used for synthetic populations and measured in dex. Zero range corresponds to parameters constant across a disk population. ** correspond to parameters which are derived from other parameters.}  
\label{table:model}
\end{table}

\subsubsection{Turbulence-driven evolution}

\begin{figure*}[ht!]
\includegraphics[width=2\columnwidth]{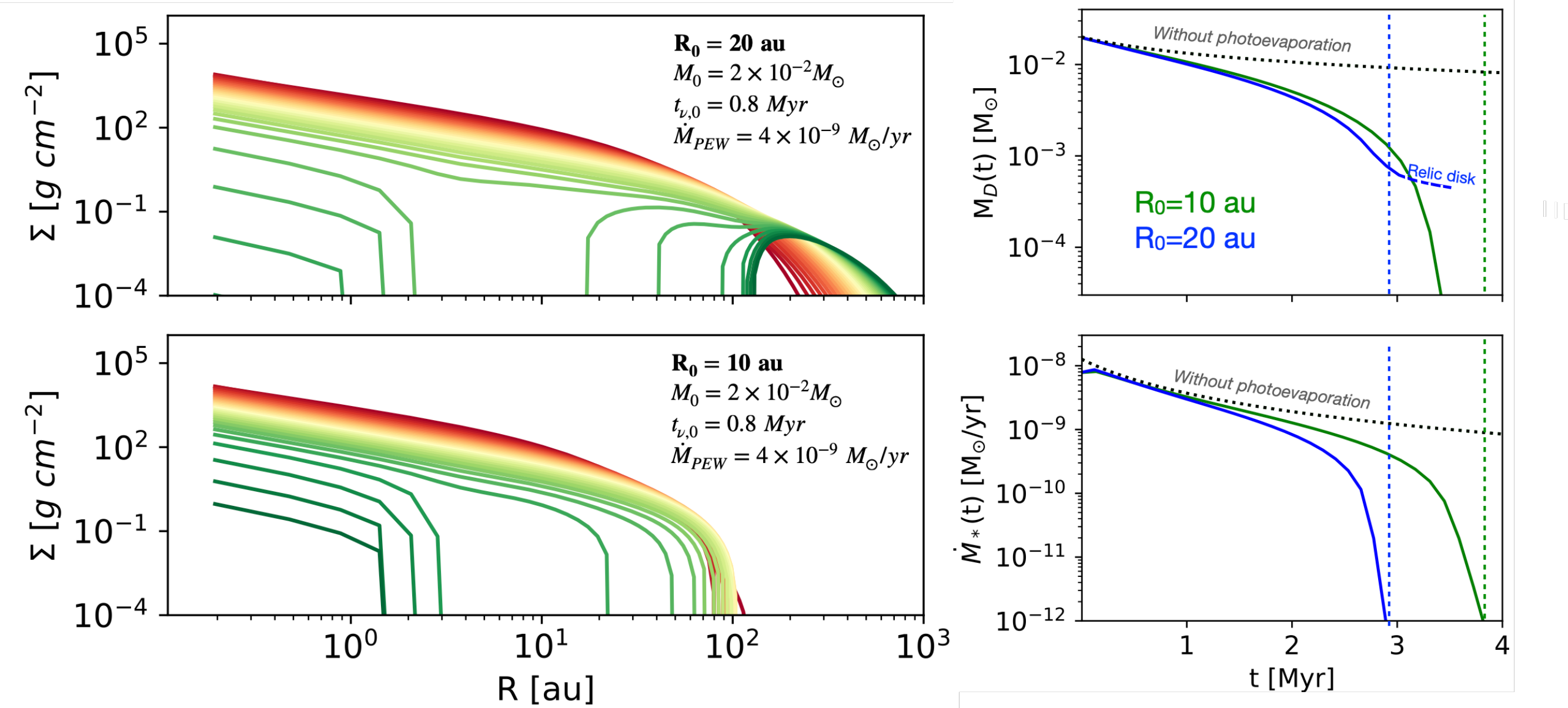}
\centering
\caption{Example of turbulence-driven evolution for a set of parameters illustrating two different pathways of dispersal. \textit{Left:} surface density profiles from initial time (blue) to dispersal (red). \textit{Right:} evolution of the disk mass and accretion rate. The first solution is an example of the popular inside-out dispersal pathway with gap opening occurring for large disks and leading to relic disks. The second solution exhibits an outside-in dispersal route due to the wind extracting the mass residing in the outer disk. The two solutions share the same viscous timescale, initial mass, and wind mass-loss rate but differ in their initial disk size as indicated in the left panels. The evolution of the disk mass and accretion rate in the absence of photoevaporation is depicted in dotted lines.}
\label{fig:example_turbulence}
\end{figure*}   

In the turbulence-driven scenario, the evolution of the surface density profile $\Sigma(r,t)$ is governed by
\begin{equation}
\begin{split}
       \frac{\partial \Sigma}{\partial t} = \frac{3}{R}\frac{\partial}{\partial R} \left\{ \frac{1}{R \Omega} \frac{\partial}{\partial R} \left(R^2\alpha_{SS} \Sigma c_s^2 \right) \right\}
       - \dot{\Sigma}_{PEW}(R),
\end{split}
\label{eq:master-eq-viscous}
\end{equation}
where $\alpha_{SS}$ is the Shakura-Sunyaev parameter and $\dot{\Sigma}_{PEW}(R)$ is the photoevporative wind mass-loss rate per unit surface area. In this work, we adopt the radial profile of $\dot{\Sigma}_{PEW}(R)$ computed by \citet{2021MNRAS.508.3611P} for a stellar mass of $M_*= 0.5~$M$_{\odot}$. The latter work utilizes hydrodynamical simulations including heating by stellar X-rays and neglecting molecule formation or shielding by a putative inner disk-wind ($\lesssim 1~$au). Instead of relying on the values of mass-loss rates computed by numerical simulations, we take the total mass-loss rate as a free parameter to scale the mass-loss rate profile. This allows us to mitigate the discrepancies in terms of mass-loss rates found in different numerical simulations \citep[see review by][]{2023ASPC..534..567P}. The importance of the radial profile of $\dot{\Sigma}_{PEW}(R)$ and the comparison with the total mass-loss rate provided in the literature is discussed in Sec. \ref{sec:discusion}. We further assume that the mass-loss rate is constant over time.
This implies that we neglect the dependency of the photoevaporation rate on the disk structure, in line with the findings of \citet{2012MNRAS.422.1880O} and the effect of the attenuation of the stellar radiation field by an evolving inner wind, as well as an evolving stellar radiation field and flares. \BT{We stress that the photoevaporative mass-loss rates $\dot{M}_{PEW}$ reported in this study are effective mass-loss rates formally calculated as $\dot{M}_{PEW} = \int_0^{+\infty} \dot{\Sigma}_{PEW}(R) 2 \pi R dR$ and used to normalize the $\dot{\Sigma}_{PEW}(R)$ profile. When the local surface density drops to zero, no mass-loss rate is assumed. Therefore, for compact disks for which the surface density rapidly drops to zero in the outer region where $\dot{\Sigma}_{PEW}(R)$ is non-vanishing, or for disks with gaps, the true mass-loss rate is lower than the reported $\dot{M}_{PEW}$ value.} 

\begin{comment}
with a total mass-loss rate proportional to $L_X$
\begin{equation}
    \dot{M}_{PEW} = 6~10^{-9} \left( \frac{L_X}{10^{30} erg/s }\right) M_{\odot}/yr.
\end{equation}
\end{comment}

Numerical simulations of MRI turbulence predict a weak correlation between the value of $\alpha_{SS}$ and the disk magnetisation as quantified by $\beta$, the ratio between the gas and magnetic pressure \citep{2023ASPC..534..465L}. As a consequence, $\alpha_{SS}$ could vary in time as the disk evolves but this also depends on the poorly known evolution of the magnetic field strength. For simplicity, we assume $\alpha_{SS}$ to be constant in time.

The radial profile of $\alpha_{SS}$ is taken to be constant, as often assumed in disk evolution models \citep[e.g.,][]{2017ApJ...847...31M, 2017MNRAS.472.4700L}. Theoretical works predict steep variations in the $\alpha_{SS}$ profile depending on the ability of the MRI to be triggered. Notably within $~0.1-0.5~$au, the ionisation of alkali metals ensures efficient MRI, possibly coupled to MHD disk-winds. The bulk part of the disk and the stellar accretion rate are not expected to be affected by such a high $\alpha_{SS}$ region as the gas slowly advected from the outer to the inner regions will rapidly flow across the inner highly turbulent region down to the magnetosphere of the star. Our value of $\alpha_{SS}$ is therefore representative of the bulk part of the disk and we adopt an inner disk radius of 0.1~au.

Initially, the surface density profile follows the self-similar solution of \citet{1974MNRAS.168..603L} with
\begin{equation}
    \Sigma(R,t) = \frac{M_0}{2\pi R_0^2} \left( \frac{R}{R_0}\right)^{-1} e^{-R/R_0},
\end{equation}
where $M_0$ is the initial disk mass and $R_0$ is the initial disk characteristic size. We further define the initial viscous timescale as
\begin{equation}
    t_{\nu,0} \equiv \frac{R_0}{3 \epsilon_{0} c_{S,0} \alpha_{SS}} % = 0.5 Myr \left( \frac{\alpha_{SS}}{10^{-3}}  \right)^{-1} \left( \frac{R_0}{10 au}  \right),
    \label{eq:def-tnu}
\end{equation}
\BT{where $c_{s,0}$ and $\epsilon_0$ are the sound speed and the disk aspect ratio at $R_0$. In this work, we assume that $T \propto R^{-1/2}$ and $\epsilon(R=1~\text{au}) = 3.33 \times 10^{-2}~(M_*/1M_{\odot})^{-1/2}$. This leads to the following relation}
\begin{equation}
    t_{\nu,0} = 0.34~\text{Myr}~\left( \frac{\alpha_{SS}}{10^{-3}}  \right)^{-1} \left( \frac{R_0}{10~au}  \right) \left( \frac{M_*}{0.5~M_{\odot}}  \right)^{1/2}.
\end{equation}
%\textbf{[To the diskpop team: dependency with stellar mass to be clarified in diskpop for Monte Carlo simulations of viscous disks with PEW (factor 1.4 in tnu0 for $M_*=0.5~M_{\odot}$) .]}

The different evolution pathways under the effect of turbulence and photoevaporative wind are illustrated in Fig. \ref{fig:example_turbulence} with two solutions. The solutions share the same viscous timescale, initial disk mass, and total mass-loss rate but differ in their initial disk size. Both lead to a disk dispersal time, defined as the time when accretion stops, of about 3-4~Myr. The first example (top left panel) has already been extensively described in the literature and corresponds to a relatively large disk of $R_0=20~$au. As the disk evolves, it spreads radially until photoevaporation opens a gap around 2-5~au where the mass-loss rate profile $\dot{\Sigma}_{PEW}$ peaks \citep[see Fig. 7 in][]{2021MNRAS.508.3611P}. Following the gap opening, the inner disk is rapidly drained and a relic disk remains. This behavior is reflected in the evolution of the accretion rate and disk mass (right panels, blue lines): after a decline of a typical timescale of $t_{\nu,0}$, the accretion rate quickly vanishes whereas the disk mass remains relatively constant, even after the accretion stops. We however stress that this so-called "relic disk problem" is only the result of our simple prescription of photoevaporation mass-loss rate profile which does not account for the direct irradiation of the cavity following gap opening. \revBT{In the following, a disk is considered to be dispersed when the stellar accretion rate drops below a threshold of $10^{-12}~M_{\odot}/yr$}.

%could be solved by "thermal sweeping" \citep{2013MNRAS.436.1430O}, a process not described in our model since dispersal is defined as the time when accretion stops.}

In the second solution, the initially more compact disk ($R_0 = 10~$au) does not spread despite radial transport of angular momentum. This surprising feature is because the bulk of the disk and its outermost region reside where the wind-mass-loss rate is significant. The consequence is that the wind does not open a gap but truncates the disk. This results in a drastically different evolution: the outer disk is eroded by the wind and the the inner disk is drained due to turbulent accretion. Therefore, the disk mass, accretion rate, and size jointly decrease before full disk dispersal. This evolution is somewhat similar to that of a large disk exposed to external UV since external photoevaporation removes mass from the outer disk \citep[e.g.,][]{clarke2007}.

\subsubsection{MHD wind-driven evolution}

\begin{figure*}[ht!]
\includegraphics[width=2\columnwidth]{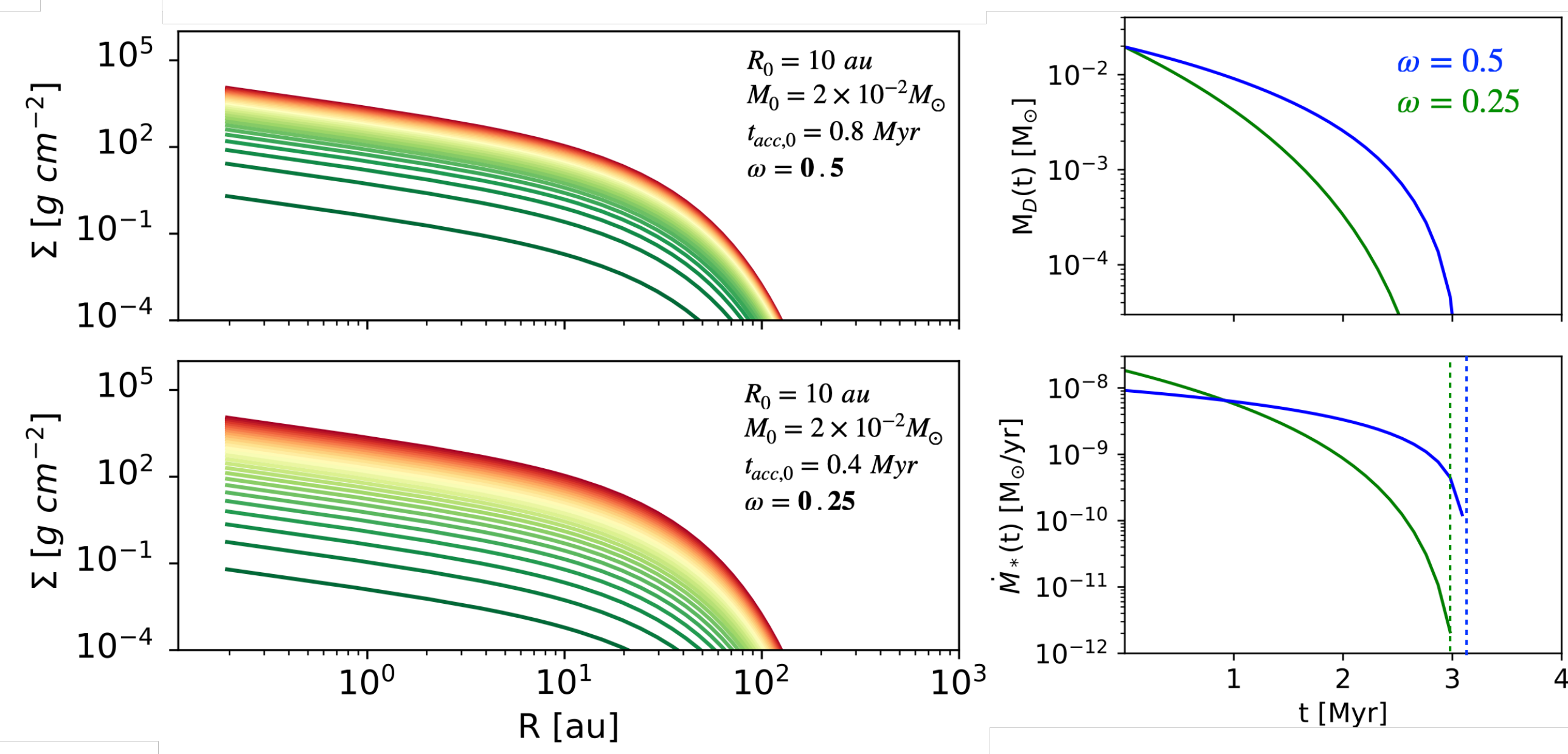}
\centering
\caption{Example of MHD wind-driven evolution for a set of parameters illustrating the impact of time evolution of the disk magnetization as parameterized by $\omega$. The left panels show the surface density profile from initial time (blue) to dispersal (red). The right panels show the evolution of the disk mass and accretion rate. The two solutions share the same value of $\lambda$ and initial mass and size but differ by the value of $\omega$. The accretion timescale is chosen to get the same dispersal time according to Eq. (\ref{eq:tdisp_MHD-DW}). In practice, the disk is dispersed slightly earlier because either the accretion rate or the disk mass dropped to their threshold disk dispersal values.}
\label{fig:example_MHD-wind}
\end{figure*}

Following the formalism of \citet{2022MNRAS.512.2290T} the evolution of the surface density under the effect of an MHD disk-wind is described as

\begin{equation}
       \frac{\partial \Sigma}{\partial t} = 
       \frac{3}{2 R}  \frac{\partial}{\partial R} \left[ \frac{\alpha_{DW} \Sigma c_s^2}{\Omega} \right] + \frac{3}{R}\frac{\partial}{\partial R} \left\{ \frac{1}{r\Omega} \frac{\partial}{\partial R} \left(r^2\alpha_{SS} \Sigma c_s^2 \right) \right\} - \dot{\Sigma}_{MHD-DW},
\label{eq:master-eq-MHD-DW}
\end{equation}
where $\alpha_{DW}$ parameterizes the wind torque and  $\dot{\Sigma}_{MHD-DW}$ the mass-loss rate of the wind. In contrast with photoevaporative winds, the mass-loss rate of the MHD disk-wind depends on the extraction of angular momentum as
\begin{equation}
\dot{\Sigma}_{MHD-DW} = \frac{3 \alpha_{DW} \Sigma c_s^2}{ \textcolor{black}{4} (\lambda-1) R^2 \Omega},
\label{eq:sigma_dot_MHD-DW}
\end{equation}
where $\lambda$, the magnetic lever arm parameter \citep{1982MNRAS.199..883B}, quantifies the efficiency of the wind to carry out angular momentum. The lower the $\lambda$, the more mass is required to be launched to drive disk accretion. A considerable advantage of this parameter is that $\lambda$ can be estimated from wind observations \citep{2006A&A...453..785F,2020A&A...640A..82T}. Observations typically find $\lambda \simeq 2-6$ \citep{2017A&A...607L...6T,2020A&A...634L..12D,2021ApJS..257...16B,2024A&A...686A.201N} and in this work we explore a wide range of $\lambda$ values (Table \ref{table:model}). In Eq. (\ref{eq:master-eq-MHD-DW}), $\alpha_{SS}$ quantifies radial transport by either turbulence or by a laminar torque emerging from the radial component of the magnetic field which is usually negligible compared to vertical torque.%s In the following, we neglect the radial transport of angular momentum.

In this work, we adopt the analytical solutions of \citet{2022MNRAS.512.2290T} in the absence of radial transport of angular momentum ($\alpha_{SS} = 0 $). The analytical solutions further assume a constant $\lambda$ value in space and time and a constant \alphaDW across the disk. The surface density profile is given by
\begin{equation}
       \Sigma(R,t) = \Sigma_c(t) (R/R_0)^{-1+\xi} e^{-R/R_0},
\label{eq:sigma_ansatz}
\end{equation}
where $\Sigma_c(t)$ is the characteristic surface density, $\xi = 1/[2(\lambda-1)]$ is the mass ejection index, and $R_0$ is the characteristic disk radius.  A considerable source of uncertainty in MHD wind-driven disk evolution is the secular evolution of the magnetic field strength since $\alpha_{DW} \propto B_z B_{\phi}/\Sigma$. Here, we use the "$\Sigma_c$ dependent wind torque" solutions of \citet{2022MNRAS.512.2290T} where the evolution of the wind torque is parameterized as $\alpha_{DW} = \alpha_{DW}(0) (\Sigma_c(t)/\Sigma_c(0))^{-\omega} $, where $\alpha_{DW}(0)$ is the initial $\alpha_{DW}$ and $\omega$ is a free parameter. The case $\omega=1$ describes a constant magnetic field strength (assuming a constant $B_z/B_{\phi}$ ratio). This leads to the following evolution of the disk mass and accretion rate:
\begin{equation}
\begin{split}
    M_{D}(t)& = M_0 \left( 1-\frac{\omega}{2 t_{acc,0}} t \right)^{1/\omega}, \\
    \dot{M}_{*}(t)& = \frac{M_0}{2 t_{acc,0} (1+f_{M})} \left( 1-\frac{\omega}{2 t_{acc,0}}t \right)^{-1+1/\omega},
\end{split}
\label{eq:MHD-DW_anal-solution}
\end{equation}

where 
\begin{equation}
    f_M  \equiv \dot{M}_W/\dot{M}_*= \left(R_0/R_{in} \right)^{\xi}-1
\end{equation}
is the mass ejection-to-accretion ratio. $t_{acc,0}$ is a generalization of the viscous timescale (see Eq. (\ref{eq:def-tnu})). Following the same assumption about the disk thermal structure as for the turbulent case, the accretion timescale is:
\begin{equation}
    t_{acc,0} = 0.34~\text{Myr}~\left( \frac{\alpha_{DW}}{10^{-3}}  \right)^{-1} \left( \frac{R_0}{10~\text{au}}  \right) \left( \frac{M_*}{0.5~\text{M}_{\odot}}  \right)^{1/2}.
\end{equation}

\BT{The essential features of the solution is illustrated in Fig. \ref{fig:example_MHD-wind} with two values of $\omega$. In both cases, the surface density profiles preserve their shape while decreasing in absolute value. The angular momentum is indeed transported vertically without the need for radial expansion. We however stress that this is a property of the solution related to its initial condition. For sharp cuts in the outer disk, the disk can shrink since wind-driven accretion acts as an advection of gas. The latter situation is however not realistic since any turbulent or laminar radial stress in the early phase leads to a smoothly declining outer surface density profile. \revBT{We note that the alternative model of \citet{2016A&A...596A..74S} predicts the formation of a cavity, reminiscent of transition disks. Whereas cavities sustained by MHD disk-winds are a stable configuration \citep{2022A&A...667A..17M}, gap opening has never been characterized in global numerical simulations. Pending simulations on secular timescales, we adopt here disk evolution solutions that do not introduce artificial features in the surface density profile.} The second property of MHD wind-driven evolution is the dispersal of the disk after a finite time due to rapid disk draining. Therefore, our MHD wind-driven solution accounts for both, accretion and dispersal. More specifically, the disk dispersal time depends on $t_{acc,0}$ and $\omega$ as
\begin{equation}
   t_{disp} = 2 t_{acc,0}/ \omega.
\end{equation}
In both solutions, we choose $t_{acc,0}$ so that the two cases share the same dispersal time (see Fig. \ref{fig:example_MHD-wind}, right panels). This highlights the essential effect of the $\omega$ parameter, which describes the evolution of the magnetic field strength. Higher values of $\omega$ lead to a rather constant disk mass and a high stellar accretion rate before dispersal. Therefore, for a given dispersal time, a disk with a higher $\omega$ value spends more time in a high mass and high accretion rate phase. This property is essential for the evolution of synthetic populations. Finally, we note that from Eq. \ref{eq:MHD-DW_anal-solution} the value of $\lambda$ does not affect the disk mass but the accretion rate by a factor $f_M +1 = (R_0/R_{in})^{1/2[\lambda-1]}$.}

\subsection{Population synthesis with \texttt{Diskpop}}

The observations provide us with the properties of disk populations of different ages and not with the evolution of individual disks. To simulate the evolution of disk populations, we use the \texttt{Diskpop} python package \citep{2024arXiv240721101S}. As a first step, the code randomly picks the values of the parameters for a sample of disks following a user-defined probability distribution and possible correlations between these parameters. The gas evolution of each disk is then computed by solving numerically Eq. (\ref{eq:master-eq-viscous}) or (\ref{eq:master-eq-MHD-DW}), or using analytical solutions with the option of including dust evolution. The output of \texttt{Diskpop} is then analyzed using the \texttt{Popcorn} package \citep{2024arXiv240721101S}.

Throughout this paper, the synthetic populations are made of 400 disks by default and we consider only gas evolution.  
When building a synthetic disk population, the 4 free disk parameters outlined in bold font in Table \ref{table:model} and detailed above translate into 4 probability distributions. We assume that the distributions of the 4 disk parameters are not correlated and follow a log-normal distribution. 
Therefore, our population model is controlled by 8 free population parameters: a median and a spread for each disk parameter. The MHD disk-wind model has a fifth free parameter, $\omega$, for which only two values are explored. Following \citet{2022MNRAS.512L..74T}, a too-low value of $\omega$ produces disks with no detectable accretion signatures but significant mass which would lead to a strong difference between the dispersal time inferred from accretion signatures and IR excess. $\omega >1$ are also unrealistic since they would correspond to disks with increasing magnetic field strength \citep{2022MNRAS.512.2290T}. 

Fitting the properties of disk populations using a population synthesis approach can appear as an under-constrained problem. Here, we propose a rationalized approach to map the parameter space and alleviate the degeneracies in a step-by-step process. 

\subsubsection{Observational data}

\begin{figure}[ht!]
\includegraphics[width=\columnwidth]{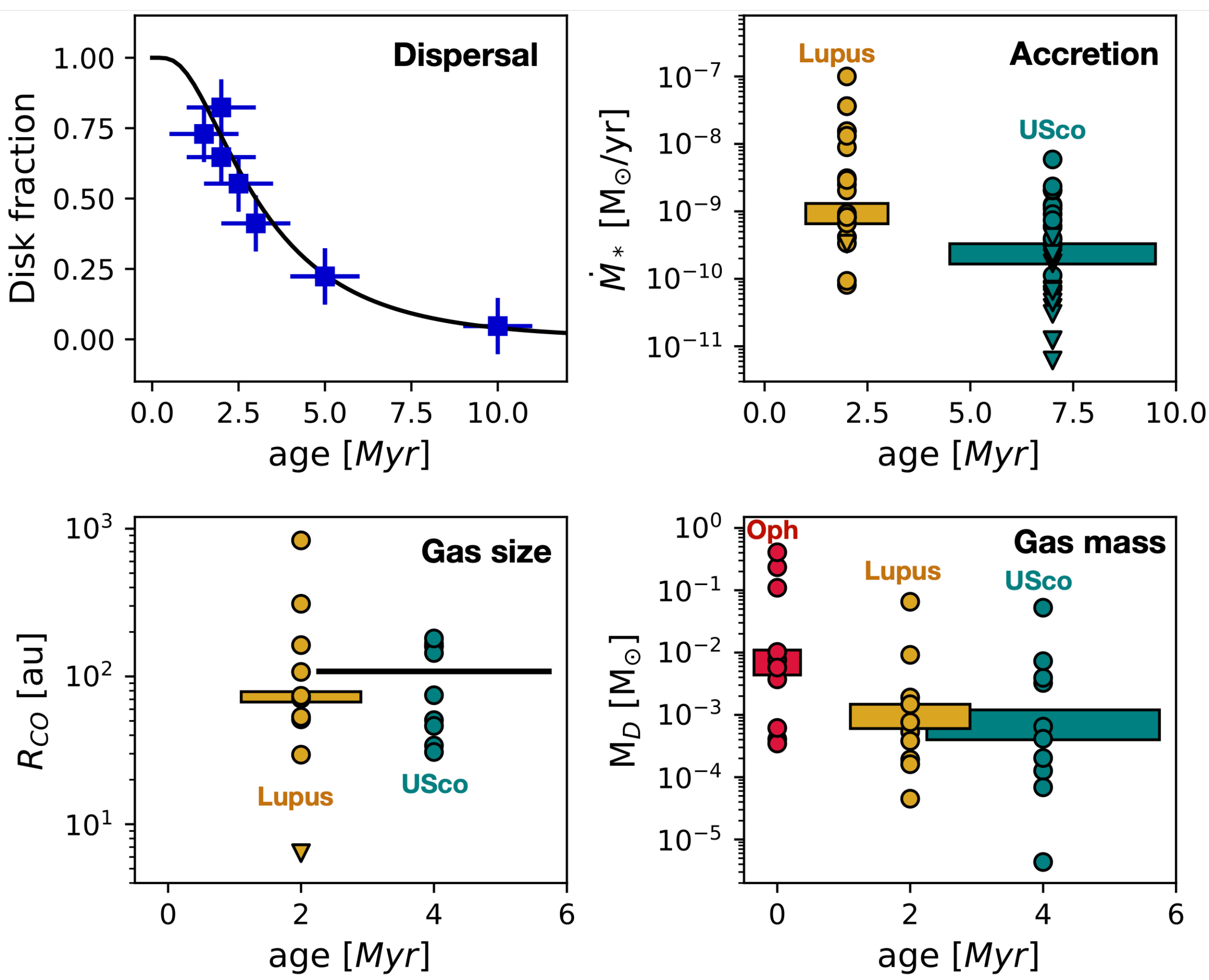}
\centering
\caption{Summary of the observational data to be reproduced by our population synthesis model. \textit{Top left:} disk fraction probed by the mid- and near-infrared excess of several star-forming regions. The disk fractions compiled by \citet{2010A&A...510A..72F} are rescaled by a factor of 1.2 to account for the short-lived disks around binaries. \textit{Top right:} stellar accretion rates from the nearly complete surveys of the Lupus and Upper Sco star-forming regions \citep{2017A&A...600A..20A,2023ApJ...945..112F}. The boxes represent the median accretion rates and the dots and triangles are the measurement and upper limits of the individual sources. \textit{Bottom:} AGE-PRO estimates of the median CO gas size \citep{AGEPRO_XI_gas_disk_sizes} and gas mass \citep{AGEPRO_V_gasmasses}, along with the measurements for individual sources. \revBT{We note that the AGE-PRO Upper Sco sample is somewhat younger than the bulk region over which the accretion rates are measured.}}
\label{fig:observables}
\end{figure}    

The main observational data to be confronted with population synthesis are shown in Figure \ref{fig:observables}. Our synthetic populations are designed to describe the evolution of isolated disks. Therefore, we consider the t=0 age as the end of the Class I phase and take the Ophiuchus AGE-PRO sample as representative of this initial condition. \BTbis{As discussed in \citet{AGE-PRO_I-overview}, there is a significant spread in the isochronal ages of each source for the Lupus and Upper Sco AGE-PRO samples with overlap between the two regions.} Considering the considerable systematic uncertainties in the age determination, we adopt a median age of 2~Myr for the Lupus and 4~Myr for the Upper Scorpius AGE-PRO sample with a typical range of 2 and 4~Myr, respectively. We note that the sources selected by AGE-PRO in Upper Sco are part of the youngest sources of this star-forming region. Therefore, our adopted median age of the AGE-PRO Upper Sco sample is \revBT{shorter than the median age of the entire cluster which is taken to be $7~Myr$. This results in an age difference between the median disk mass and size, estimated from the AGE-PRO sub-sample (Fig. \ref{fig:observables}, bottom panels), and the median accretion rate evaluated from a less biased sample (top right panel).}

The disk fraction (top left panel) stems from the compilation of \citet{2010A&A...510A..72F} of infrared excess which is consistent with other works \citep{2014A&A...561A..54R} and representative of the stellar mass targeted by AGE-PRO \citep{2015A&A...576A..52R}. \BTbis{In our models, we neglect the effect of multiplicity which significantly reduces the disk lifetime for close-in binaries \citep[$\lesssim 40~$au,][]{kraus2012}. We corrected the disk fraction of \citet{2010A&A...510A..72F} by applying a multiplicative factor of 1.2 to match the disk fraction around single or wide binaries determined by \citet{kraus2012}.} The disk fraction is interpreted as the cumulative probability distribution of the disk lifetime. By assuming that disk lifetimes follow a lognormal distribution, we fit the disk fraction by an error function (see black line Fig. \ref{fig:observables}, top left). We estimate a spread of about 0.3 dex and a median value of the disk lifetime of $3\,$Myr.

The accretion rates shown in Figure \ref{fig:observables} stem from the compilation of \citet{alcala2014, alcala2017, 2023ASPC..534..539M} for Lupus and \citet{2023ApJ...945..112F} for the Upper Sco star-forming regions. We selected the stars that are within the bin of AGE-PRO stellar masses ($M_*=0.3-0.7M_{\odot}$). We stress that the sample of \citet{2023ApJ...945..112F} reports upper limits for $50\%$ of the sources. Therefore, the median value of the accretion rate should be considered as an upper limit. Compared to the pioneering survey of \citet{2020A&A...639A..58M} the sample of \citet{2023ApJ...945..112F} is larger (12 versus 29 sources within the AGE-PRO stellar mass range) thought using only H$\alpha$ line flux as a proxy of the accretion rate. Still, the two studies lead to similar median accretion rates for Upper Sco.
%\revBT{We adopt a median age of 7~Myr for the Upper Sco sample used to measure the median accretion rate, recalling that the AGE-PRO sub-sample used to estimate disk mass and size is younger.} 
Overall, we find that the median accretion rate decreases by a factor of at least 3 from Lupus to Upper Sco with a large spread of $0.8$~dex for each region.

The distribution of disk size and mass obtained by the AGE-PRO program is also represented in Figure \ref{fig:observables} as derived in \citet{AGEPRO_XI_gas_disk_sizes} and \citet{AGEPRO_V_gasmasses}, respectively. The observed CO gas sizes, denoted as $R_{CO,90\%}$ or simply $R_{CO}$,  correspond to the radius that encompasses 90$\%$ of the total CO flux and are evaluated by correcting for the beam of the observations. A major caveat when analyzing disk size is the conversion between the actual disk size, i.e. the radius encompassing the majority of disk mass, and the observed CO disk size \citep{2023ApJ...954...41T}. A possible strategy to extract the predicted value of $R_{CO,90\%}$ in the population model is to run radiative transfer calculations using the surface density calculated by the 1D disk evolution model and assuming either standard CO abundances or CO abundances computed with thermochemical models. \revBT{This method is however computationally prohibitive and in this work, we use a simple analytical formula derived from \citet{2023ApJ...954...41T} who show that the $R_{CO,90\%}$ size is given by the radius where the total gas column density drops below a given critical column density of hydrogen denoted as $N_{gas}(R_{CO,90\%})$. However, \citet{2023ApJ...954...41T} does not consider the depletion of gas-phase carbon, which can be substantial in Class II disks \citep{2013ApJ...776L..38F}. Following \citet{2023MNRAS.518L..69T}, we increased the threshold column density by the carbon depletion fraction, as:} 
\begin{equation}
       N_{gas}(R_{CO,90\%}) = \PP{3.7} \times 10^{21} \delta_C^{-1} \left( M_{D} / M_{\odot}\right)^{0.34} \text{cm}^{-2}.
\label{eq:trapman_2023}
\end{equation}
\revBT{Throughout this work we adopt $\delta_C = 0.2$, in line with the under-abundance of CO found in \citet{AGEPRO_V_gasmasses} for the Lupus and Upper Sco AGE-PRO sample.} Therefore, by finding the radius where the column density reaches $N_{gas}(R_{CO,90\%}) $, we determine for each simulated disk the $R_{CO,90\%}$ radius. 

The disk masses in Lupus and Upper Sco are calculated by matching the N$_2$H$^+$ and CO line fluxes, and CO gas size \citep{AGEPRO_V_gasmasses}. For Ophiuchus, the younger disk population, only CO isotopologue lines are used. Overall, we find a drop in disk mass from Ophiuchus to Lupus by about an order of magnitude. From Lupus to Upper Sco, the median CO size and mass do not change significantly. 

\subsubsection{Fitting procedure}

\begin{figure}[ht!]
\includegraphics[width=0.75\columnwidth]{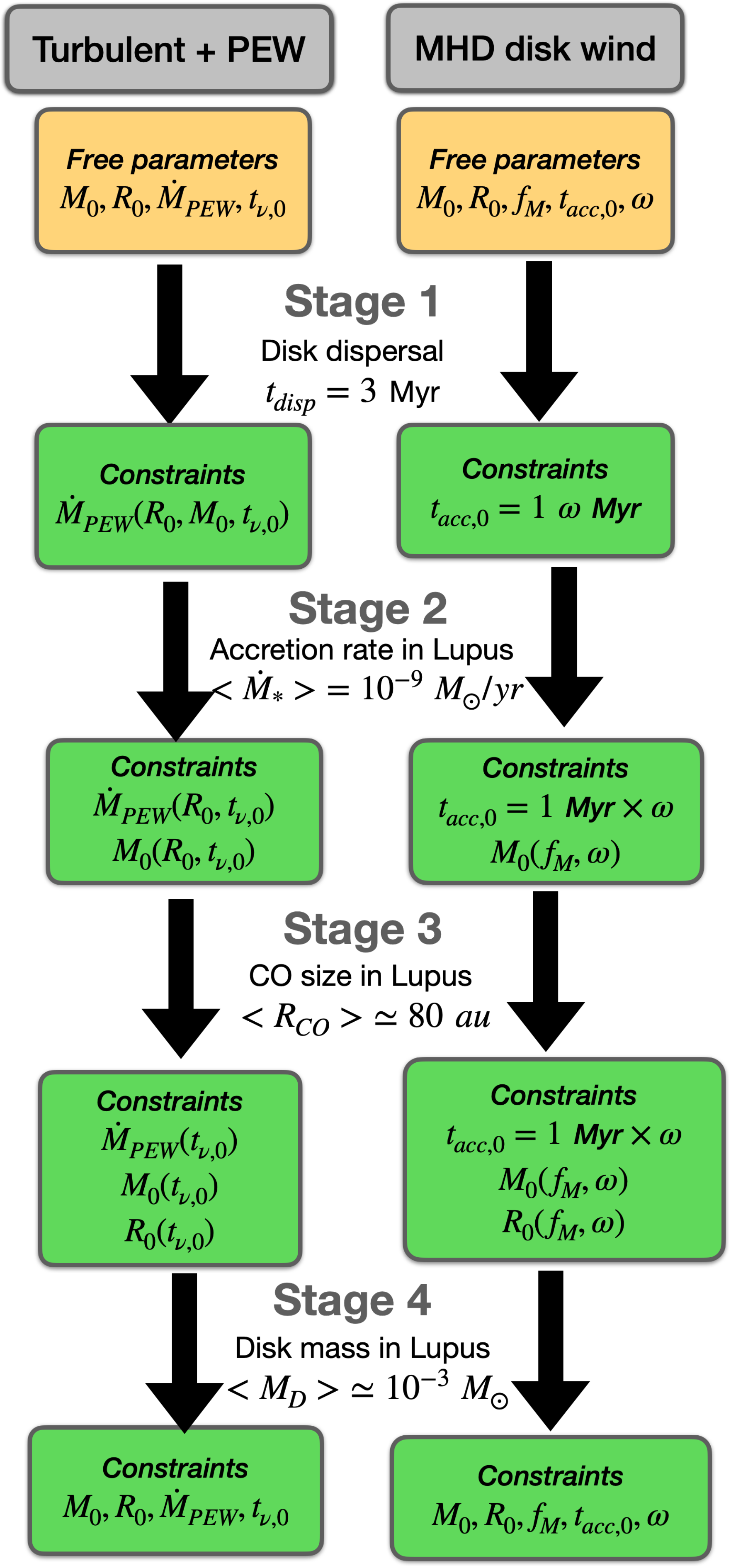}
\centering
\caption{Schematic view of the fitting stages for the turbulent (left) and the MHD disk-wind (right) models. This step-by-step approach allows us to build synthetic populations consistent with the observations in a rationalized approach.
\label{fig:scheme_fitting}}
\end{figure}

Our goal is to reproduce the disk fraction and the median values of the accretion rate, disk mass, and disk size. The spreads of the parameters are therefore of less importance and we keep the 4 spreads fixed to meaningful values provided in Table \ref{table:model}. As explained below (see Sec.~\ref{sec:results} and Fig. \ref{fig:CDF_best_fit}), these values adequately fit the dropping rate of the disk fraction, the spread in observed disk mass, and disk sizes. The main challenge is then to constrain 4 free population parameters. These 4 free parameters are constrained by sequentially reproducing (1) the disk fraction (through the disk dispersal timescale), (2) the median accretion rate in Lupus, (3) the median CO gas size inferred in Lupus, and (4) the median gas mass estimated in Lupus. The workflow is summarized in Fig. \ref{fig:scheme_fitting} for the two scenarios. \BT{The middle-aged Lupus population is used to constrain the parameters before a detailed comparison with the old Upper Sco and the young Ophiuchus populations. The choice of the Lupus population as a reference point is justified by the systematic measurement of accretion rates for all the Lupus AGE-PRO sources, the relatively narrow range in stellar age compared with Upper Sco, and the measurement of the CO disk size, which is challenging in Ophiuchus due to the presence of envelopes \citep{AGEPRO_II_Ophiuchus}.}

Because disk dispersal is not controlled by the same parameters (see Sec. \ref{subsec:indiv-disk}), the parameter adjusted for each stage differs in the two scenarios. In the first stage (fit of the disk fraction), it is the median mass-loss rate $<\dot{M}_{PEW}>$ that is adjusted for the turbulence-driven scenario, while for the MHD disk-wind scenario it is the accretion timescale $<t_{acc,0}>$.  
For the second stage (fit of the median accretion rate in Lupus), the initial disk mass is adjusted in both scenarios, since accretion rates are proportional to $M_0$. The third stage (fit of the median CO gas size in Lupus) constrains the initial disk size $R_0$, and the last stage constrains the remaining parameters, namely the viscous timescale for the turbulence-driven case and the wind-to-accretion mass ratio in the MHD disk-wind case. These four stages allow us to confront the predictions of the population synthesis models to the full AGE-PRO data with very limited degeneracy in the parameters.

\section{Results}

\label{sec:results}

\subsection{MHD wind driven evolution}

\subsubsection{Disk dispersal}

In the analytical solution of \citet{2022MNRAS.512.2290T}, the theoretical disk dispersal time, as defined by the time when disk mass drops to zero, depends only on the accretion timescale $t_{acc,0}$ and the $\omega$ parameter, which quantifies the evolution of the magnetic field strength. For $\omega > 0$ and a stellar mass of $M_* = 0.5~M_{\odot}$, an individual disk disperses at a time
\begin{equation}
   t_{disp} = 2 t_{acc,0}/ \omega = 0.68~\text{Myr} \left( \frac{\alpha_{DW}}{10^{-3}}  \right)^{-1} \left( \frac{R_0}{10~\text{au}}  \right) \omega^{-1}.
 \end{equation}
In this work, our definition of disk dispersal time as the time when the stellar accretion rate drops below $10^{-12}~M_{\odot}/yr$, leads to very similar disk dispersal time for $\omega \gtrsim 0.2$.
Therefore, one can associate for each individual disk, a dispersal time $t_{disp}$. Because $\alpha_{DW}$ and $R_0$ follow log-normal probability distributions, the disk dispersal time will also follow a log-normal distribution with a median value and a spread of
\begin{equation}
\begin{split}
<t_{disp}> &= 0.68~\text{Myr} \left( \frac{<\alpha_{DW}>}{10^{-3}}  \right)^{-1} \left( \frac{<R_0>}{10~\text{au}}  \right) \omega^{-1} \\
\sigma_{t_{disp}}^2 &= \sigma_{\alpha_{DW}}^2  + \sigma_{R_0}^2,
\label{eq:tdisp_MHD-DW}
\end{split}
\end{equation}
where $\sigma_X$ is the spread of the quantity $X$ measured in dex.

In the first stage of the fitting process, we consider $R_0$ as a free parameter which will be constrained from the observed CO size in stage 3 (see Fig. \ref{fig:scheme_fitting}). Therefore, to fit a median disk dispersal time of 3~Myr, one needs to set the median value of $\alpha_{DW}$ for each value of $R_0$ to
\begin{equation}
    <\alpha_{DW}> = 2.3 \times 10^{-4}  \left( \frac{<R_0>}{10~\text{au}}  \right) \omega^{-1}. 
    \label{eq:alphaDW_for-dispersal}
\end{equation}
Equation (\ref{eq:tdisp_MHD-DW}) also justifies the adopted spread in $R_0$ and \alphaDW (see values in Table \ref{table:model}). We indeed choose the spreads such that $\sqrt{\sigma_{\alpha_{DW}}^2  + \sigma_{t_{R_0}}^2} = 0.36$, a value close to the spread in the disk lifetime inferred from the decline of the disk fraction.

\subsubsection{Accretion rates}

\label{subsubsec:MHD-DW_stage2}

\begin{figure}[ht!]
\includegraphics[width=\columnwidth]{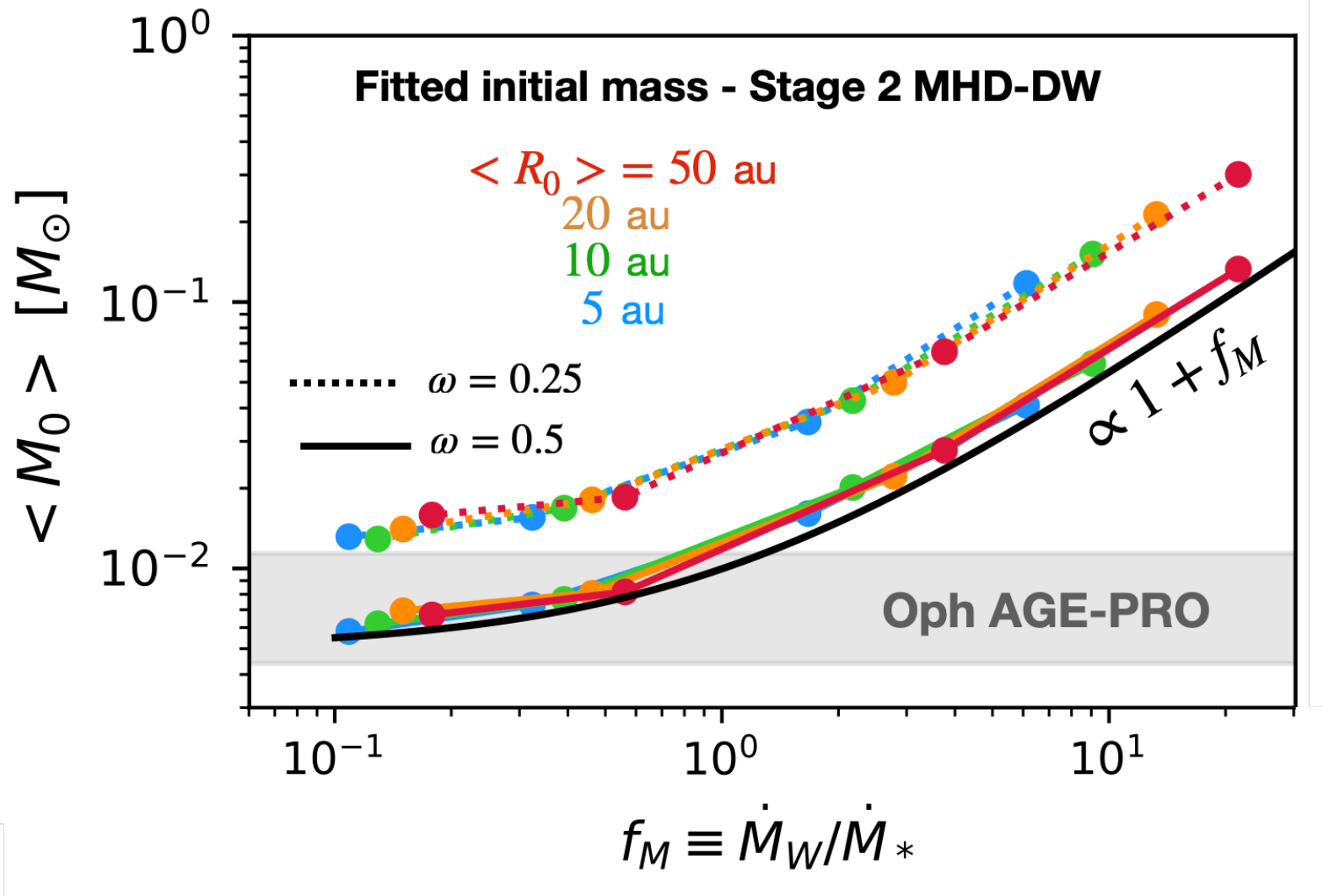}
\caption{Constraints on the initial median disk mass $<M_0>$ obtained by reproducing the median accretion rate of Lupus ($\dot{M}_* \simeq 10^{-9} M_{\odot}/yr$). The grey area indicates the median disk mass derived by AGE-PRO in Ophiuchus. Each dot corresponds to a synthetic population with a medial \alphaDW given by Eq. (\ref{eq:alphaDW_for-dispersal}) matching the observed disk dispersal.}
\label{fig:stage2_MHD-DW}
\end{figure} 

As discussed in \citet{2022MNRAS.512L..74T}, once the distribution of \tacc is set, some essential properties of accretion are set since \tacc also controls the ratio between the accretion rate and the disk mass. In particular, Eq. (\ref{eq:MHD-DW_anal-solution}) shows that the absolute value of the accretion rate scales with the initial disk mass. Therefore, in the second stage of our fitting procedure, we adjust the median disk mass $<M_0>$ for each value of $\omega$, $f_M$, and $<R_0>$ to match a median accretion rate of $\dot{M}_* = 10^{-9}$ M$_{\odot}$/yr measured in Lupus.
%with $<\alpha_{DW>}$ set by Eq. \ref{eq:alphaDW_for-dispersal}. 
Practically, we start with a first guess of the initial disk mass to compute the evolution of a population and measure the median accretion rate at an age of $2\,$Myr. The initial median disk mass is subsequently increased (resp. decreased) if the median accretion is higher (resp. lower) than $10^{-9}$ M$_{\odot}$/yr. A tolerance of $10 \%$ on the median accretion rate is adopted.

The resulting values of $<M_0>$ as a function of the remaining free parameters ($<R_0>$, $\omega$, $<f_M>$) are shown in Fig. \ref{fig:stage2_MHD-DW}. When plotted as a function of the ejection-to-accretion mass ratio $f_M = (R_0/R_{in})^{1/[2(\lambda-1)]}-1$, the results do not depend on the assumed initial disk sizes. This is because the time evolution of the accretion rate is primarily driven by the accretion timescale \tacc and not directly the disk size. 

From Eq. \ref{eq:MHD-DW_anal-solution}, increasing the ejection-to-accretion mass ratio $f_M$ reduces all the accretion rates by a factor of $1/(1+f_M)$. Therefore, the value of  $<M_0>$ required to reproduce the accretion rate in Lupus increases with $f_M$ as $1+f_M$ (see black curve in Fig. \ref{fig:stage2_MHD-DW}), and the impact of mass-loss rate is significant only for $f_M \gtrsim 1$.  Finally, lower values of $\omega$ require higher disk masses to fit the accretion rates. The effect of $\omega$ is more subtle to grasp but can be understood from the inspection of Fig. \ref{fig:example_MHD-wind}. For the same disk dispersal time, a lower value of $\omega$ leads to an accretion rate with a steeper drop before dispersal. This implies that lower values of $\omega$ lead to lower accretion rates at $2\,$Myr. As a consequence, the initial disk mass $<M_0>$ required to fit the observed accretion rate in Lupus needs to be increased when decreasing the $\omega$ value.

Overall, we find that initial median disk masses of at least $<M_0> \simeq 8 \times 10^{-3}~M_{\odot}$ are required to reproduce the median accretion rate measured in Lupus. The median gas mass of $<M_D> \simeq 7^{+4}_{-2} \times 10^{-3} M_{\odot}$ obtained by AGE-PRO in Ophiuchus \citet{AGEPRO_II_Ophiuchus} favors low ejection-to-accretion ratios $f_M \lesssim 1$ and relatively high values of $\omega$.
Interestingly, large mass ejection ratios $f_M$ are excluded since this would require disks that are initially unstable against gravitational instability.

\subsubsection{Constraints from AGE-PRO results on Lupus}
\label{subsubsec:MHD-AGE-PRO_Lupus}

\begin{figure}[ht!]
\includegraphics[width=\columnwidth]{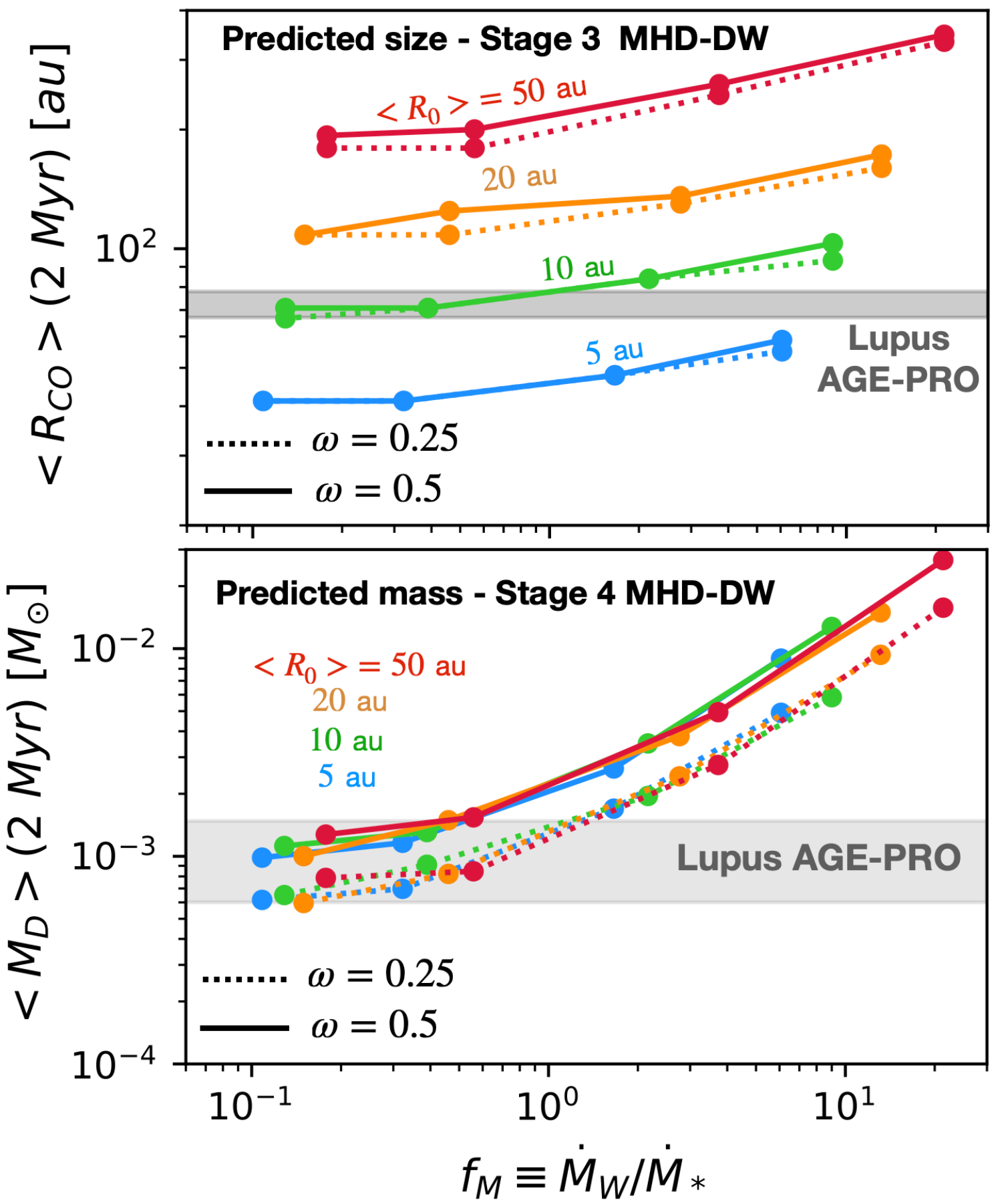
}
\caption{Predicted median CO gas size and disk mass at the age of Lupus for synthetic populations matching disk fraction (see Eq. (\ref{eq:alphaDW_for-dispersal})) and the Lupus median accretion rate (see Fig. \ref{fig:stage2_MHD-DW}). \BT{The grey area shows the median CO disk size and disk mass as measured by AGE-PRO.}}
\label{fig:stage3_MHD-DW}
\end{figure}

At this stage, we are in a position to compare the results of AGE-PRO to synthetic populations which are consistent with disk dispersal and the accretion rate in Lupus (stage 3 and 4 in Fig. \ref{fig:scheme_fitting}). The initial disk mass and accretion timescale are already constrained and only the initial disk size, the $f_M$ parameter, and $\omega$ remain free parameters. In Fig. \ref{fig:stage3_MHD-DW}, we show the predicted median CO disk size and the median disk mass after 2\,Myr, corresponding to the age of Lupus. As expected, a population of initially larger disks has a larger $<R_{CO}>$ size after $2\,$Myr (Fig. \ref{fig:stage3_MHD-DW} top panel).
%Populations with larger $f_M$ have slightly larger $<R_{CO}>$ due to the larger disk mass. 
It is immediately apparent that only initially compact disks can fit the observed CO disk sizes in Lupus with a median disk radius of $<R_0> \simeq 10~$au. We stress that this does not imply that the characteristic size at the age of Lupus is that small since compact disks have shorter accretion timescales and therefore are dispersed first \revBT{(see Fig. \ref{fig:survivors})}.

The median disk mass $<M_D>$ at $2$\,Myr increases with the mass ejection-to-accretion ratio $f_M$. This is simply because higher $f_M$ values correspond to higher initial mass $<M_0>$; a requirement to fit the median accretion rate in Lupus (see Fig. \ref{fig:stage2_MHD-DW} and Sec. \ref{subsubsec:MHD-DW_stage2}). Interestingly, we find that lower values of $\omega$ produce lower disk masses even though the initial disk mass is larger. This is an essential effect of $\omega$: for a single disk, the mass decreases more steeply before disk dispersal for low $\omega$ values (see example Fig. \ref{fig:example_MHD-wind}). Therefore, lower value of $\omega$ result in lower disk mass at the age of Lupus. 

When compared with the median disk mass estimated by AGE-PRO in Lupus (Fig. \ref{fig:stage3_MHD-DW}, bottom panel), the half of the synthetic populations that fit disk dispersal, accretion rate, and disk size are excluded. Most of the predicted median disk masses lie indeed above the median disk mass of Lupus. For $\omega = 0.25$, values of $f_M$ smaller than 1 reproduce well the low disk masses measured in Lupus even though $\omega = 0.5$ and $f_M \lesssim 1$ remains a good fit considering the large uncertainty on the estimated median disk mass.

%For a single disk, the ratio between the mass at any time and its initial mass depends only on $t_{acc,0}$. Therefore two populations of the same distribution of $t_{acc,0}$ have the same ratio. This means that $f_M$

%It is immediately apparent that only initially compact disks can fit the observed CO disk size in Lupus with a median disk radius of $<R_0>5-10~$au. Therefore, disk gas size presented in \citet{AGEPRO_XI_gas_disk_sizes} 

\begin{figure*}[ht!]
\centering
\includegraphics[width=1.6\columnwidth]{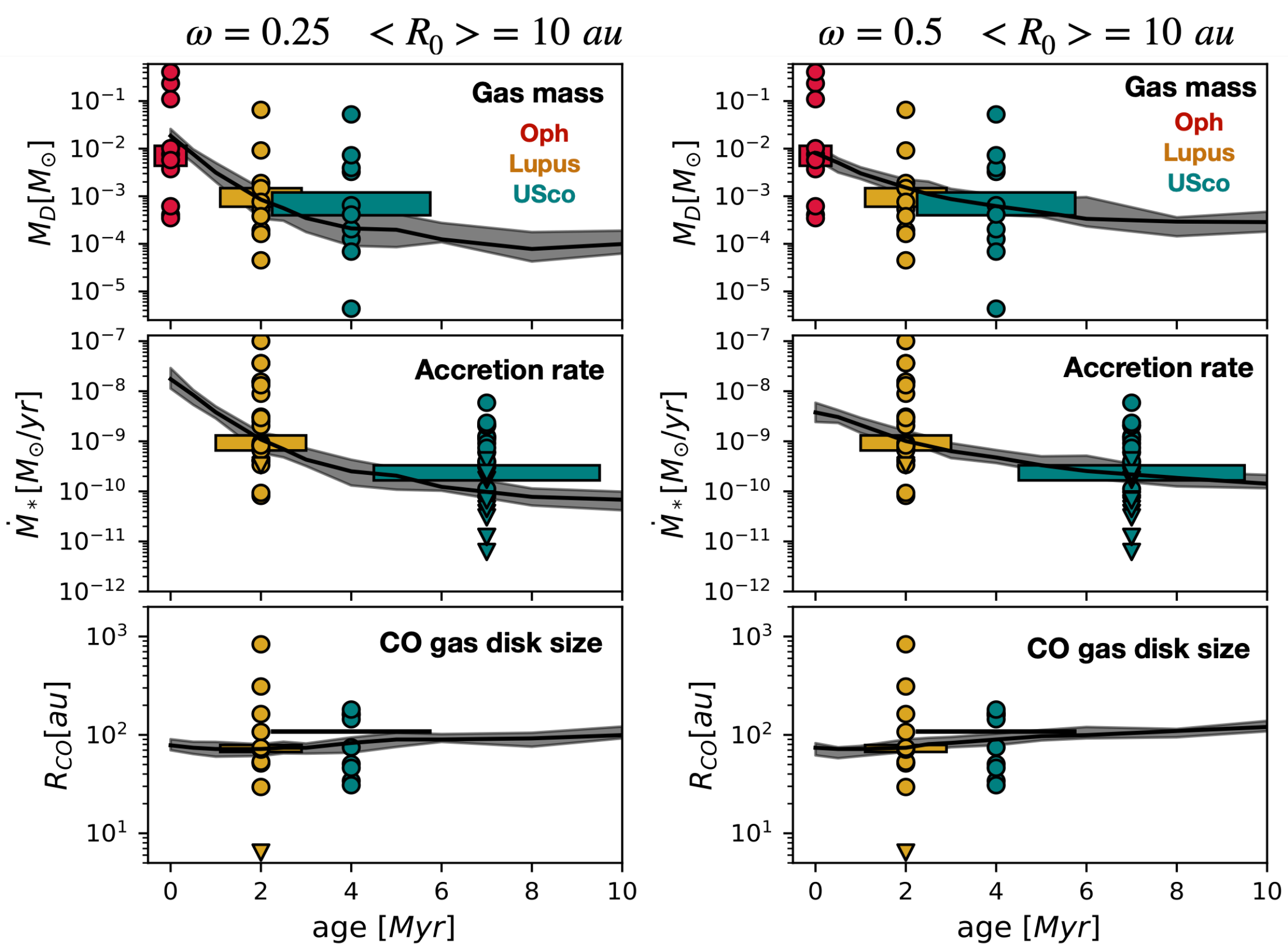}
\caption{Best fit MHD-DW model for $\omega=0.25$ (left) and $\omega=0.5$ (right) compared with the observations. The top panels show the total disk gas mass, the middle panels the accretion rates, and the bottom panels the CO disk size. Individual sources are represented by dots and the corresponding median with uncertainties by colored boxes. AGE-PRO measurements are from \citet{AGEPRO_V_gasmasses} (top panels) and \citet{AGEPRO_XI_gas_disk_sizes} (bottom panels) whereas the accretion rates compiled by \citet{2023ASPC..534..539M} for Lupus and estimated by \citet{2023ApJ...945..112F} in Upper Sco. The grey shaded areas depict the 1st and 3rd quartiles of synthetic sub-populations made of 10 disks. This allows us to depict the uncertainties related to the limited size of the samples. The synthetic populations are made of 2,000 disks. We stress that the medians include the survivor bias since they are calculated over the disk-bearing sources of the synthetic population.}
\label{fig:best-fit-MHD-DW}
\end{figure*} 

\subsubsection{Best fit MHD disk-wind model}

Putting all the constraints stemming from the properties of the Lupus population and disk fraction,  we find best-fit models with an $\omega = 0.25-0.5$, low mass ejection-to-accretion $f_M \lesssim 1$, initially compact disks $<R_0> = 10~$au, and $\alpha_{DW} \simeq 5 \times 10^{-4} - 10^{-3}$. Reproducing the essential features of the Lupus population constitutes the first success of the MHD disk-wind models. This also considerably reduces the parameter space before comparing the time evolution of the median accretion rate, mass, and CO gas size. Here, we retained two best-fit models from stage 4, which corresponds to $\omega=0.25$ (steep decline) and $\omega=0.5$ (shallow decline).

In Fig. \ref{fig:best-fit-MHD-DW}, we present the time evolution of the median disk mass $<M_D>$, accretion rate $<\dot{M}_*>$, CO gas size $<R_{CO}>$ and compare them to the available observational constraints. By construction, the two synthetic populations reproduce the Lupus population at 2\,Myr. We recall that higher $\omega$ values (right panels) correspond to a shallower evolution of the medians. Of the two models, the synthetic population with a higher value of $\omega=0.5$ reproduces better the evolution from Ophiuchus to Upper Sco. Overall, the match between the observations and the best-fit model is excellent. This is particularly clear with the disk mass. AGE-PRO finds a decrease in median disk mass of about 10 from Ophiuchus to Lupus which is well reproduced by $\omega =0.5$ and overestimated by $\omega =0.25$. The importance of having estimates for a young star-forming region was also highlighted in Fig. \ref{fig:stage2_MHD-DW} where we present the initial disk mass required to fit disk fraction and accretion rate; only $\omega=0.5$ and low $f_M$ values led to initial disk masses consistent with Ophiuchus. An $\omega=0.5$ value is also consistent with the relatively constant median disk mass found from Lupus to Upper Sco (Fig. \ref{fig:best-fit-MHD-DW}).

Both models also reproduce the shallow evolution of the CO disk size with a small increase in $<R_{CO}>$ from Lupus to Upper Sco. In the models, the increase in $<R_{CO}>$  is exclusively a survivorship bias. In fact, for an individual disk, $R_{CO}$ is rather slowly decreasing as an effect of the decrease in the disk mass. However, small disks are dispersed first due to their smaller \tacc, driving an increase in median disk size. The result on the median accretion rate is similar to the disk mass; an $\omega=0.5$ value does a slightly better job at reproducing the shallow evolution from Lupus to Upper Sco. %Still, one should keep in mind the Upper Sco sample used to compute the median accretion rate is biased and the true median disk mass is likely smaller.

\subsection{Turbulence-driven evolution with internal photoevaporation}

\subsubsection{Disk dispersal}

%\textbf{Should now fit for MW: script fit_pop_stage1bis_paper.py to be debugged...}

In the case of turbulent disks, the dispersal time for an individual disk depends not only on the accretion timescale (often also referred to as viscous timescale), but also on the 3 other disk parameters: the wind mass-loss rate \Mw, the initial disk mass \MO and, to a lesser extent, the disk size \RO. This means that fitting disk dispersal does not provide a constraint on only one parameter, as in the case of MHD wind-driven accretion, but a joint constraint on the four parameters. In the first stage (see workflow in Fig. \ref{fig:scheme_fitting}), we determine the median photoevaporative wind mass-loss rate \medMW required to fit disk fraction for all possible values of the viscous timescale \medtnu (or, equivalently $<\alpha_{SS}>$), initial disk size \medRO, and mass \medMO.

Quantitatively, we find that for $t_{disp} > t_{\nu,0}$ the dispersal time of an individual disk is roughly a power-law of the disk parameters: % (see Appendix \ref{appendix:disk_dispersal_time}):
\begin{equation}
    t_{disp} \simeq 2.9~\text{Myr} \left( \frac{t_{\nu,0}}{1~\text{Myr}} \right)^{1/3} \left(\frac{M_{0}}{10^{-1}~\text{M}_{\odot}} \right)^{2/3}  \left( \frac{\dot{M}_{W}}{1 \times 10^{-8}~\text{M}_{\odot}/\text{yr}} \right)^{-2/3}.
    \label{eq:disk_dispersal_time}
\end{equation}
This equation is motivated by the idea that a disk disperses when its accretion rate as predicted by the pure viscous evolution drops below the wind mass-loss rate \citep{clarke2001, alexander2008}. This reasoning provides the power-law indexes in Eq. (\ref{eq:disk_dispersal_time}). We also note that this simple equation predicts that to first order, the dispersal time is sensitive to the viscous timescale, which is proportional to the ratio $R_0/\alpha_{SS}$ (Eq. \ref{eq:def-tnu}).

\begin{figure}[ht!]
\includegraphics[width=0.95\columnwidth]{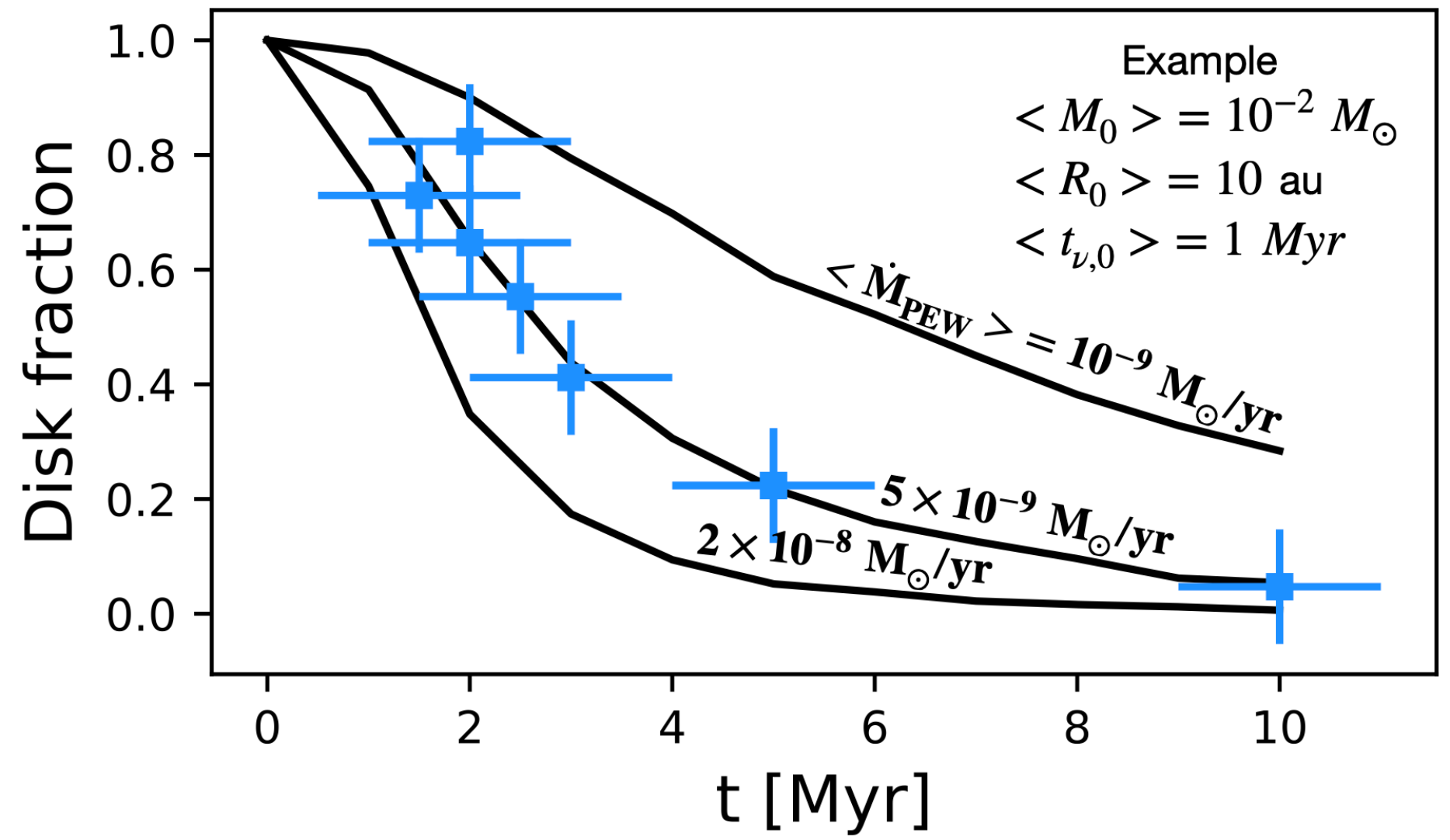}
\caption{Illustration of the method used to constrain the median mass-loss rate \medMW from the disk fraction, assuming a fixed value of $R_0$, \medMO, and \medtnu. The predicted disk fraction as defined by non-accreting disks is shown in solid line for various values of the median mass-loss rate. The other population parameters are indicated in the top left. The fitted values of \medMW for $R_0=10~$au and a wide range of \medMO and \medtnu are provided in Fig. \ref{fig:PEW_stage1}. The data compiled by \citet{2010A&A...510A..72F} are in blue.}
\label{fig:PEW_constraints_disk_fraction_example_tdisp}
\end{figure} 

The idea behind the fitting procedure is illustrated in Fig. \ref{fig:PEW_constraints_disk_fraction_example_tdisp}: for given values of the median \medRO, \medMO,~and \medtnu, the decline of the disk fraction with the age of the population depends on the value of \medMW. We therefore assume a fixed value of \medRO, \medMO,~and \medtnu, and run a synthetic disk population using a guessed value for \medMW. From Eq. (\ref{eq:disk_dispersal_time}), the guessed value to obtain a disk dispersal time of $t_{disp} = 3~$Myr is:
\begin{equation}
\begin{split}
    <\dot{M}_{PEW}> \simeq 1 \times 10^{-8}~\text{M}_{\odot}/\text{yr} \left( \frac{<t_{\nu,0}>}{1~\text{Myr}} \right)^{1/2}   \left( \frac{<M_{0}>}{0.1~\text{M}_{\odot}} \right) \\
    \simeq 5.8 \times 10^{-9}~\text{M}_{\odot}/\text{yr} \left( \frac{\alpha_{SS}}{10^{-3}} \right)^{-1/2} \left( \frac{R_0}{10~\text{au}} \right)^{1/2}   \left(\frac{<M_{0}>}{0.1~\text{M}_{\odot}} \right) 
    \label{eq:disk_dispersal_M0}
\end{split}
\end{equation}
If the disk fraction at the age of $3\,$Myr is larger (resp. smaller) than 50$\%$, the mass-loss rate is reduced (resp. increased) in proportion, and a new population is calculated. Convergence is typically reached after 2-4 iterations with a tolerance on the disk fraction of 6$\%$. 

\begin{figure}[ht!]
\includegraphics[width=\columnwidth]{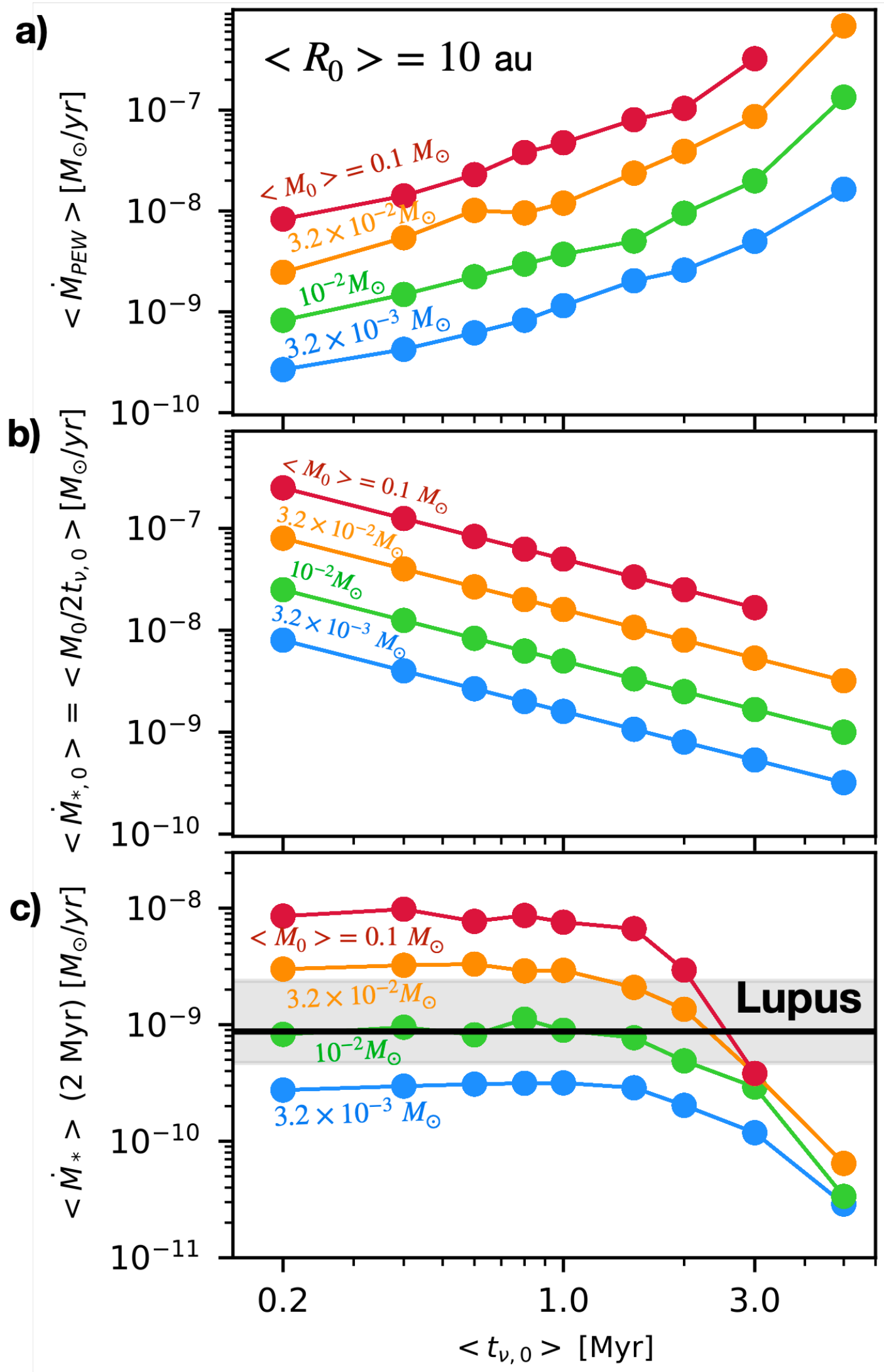}
\caption{Constraints on \medMW obtained by fitting a median disk dispersal time of $t_{disp} = 3~$Myr and the resulting median accretion rate at initial time and after $2~$Myr for a median initial disk size of \medRO$=10~$au. The method used to find the value of \medMW~ that fits disk fraction for each value of \medMO, \medtnu, and \medRO is illustrated in Fig. \ref{fig:PEW_constraints_disk_fraction_example_tdisp}. For \medRO$=10~$au, viscous timescales longer than 3~Myr are not able to reproduce simultaneously disk fraction and median accretion rates at 2~Myr. Results of the fit for other values of the initial median disk size \medRO$=5, 20, 50~$au are provided in Appendix. \revBT{The solution with unrealistically large mass-loss rates of \medMW $\ge 10^{-5}~M_{\odot}/yr$ are not displayed.}}
\label{fig:PEW_stage1}
\end{figure}

The inferred median mass-loss rate \medMW as a function of the assumed \medtnu and \medMO is shown in Fig. \ref{fig:PEW_stage1}-a for $<R_0> =10~$au. The global evolution of \medMW with \medtnu is similar for different values of \medRO as shown in the Appendix. As expected from Eq. (\ref{eq:disk_dispersal_M0}), the inferred mass-loss rate \medMW is larger for higher disk mass \medMO, and longer viscous timescale \medtnu. The mass-loss rate is also relatively insensitive to the initial disk size \medRO solely but on the viscous timescale \medtnu, which is the ratio between the initial characteristic size $R_0$ and $\alpha_{SS}$. Interestingly, \PP{the increase in the mass-loss rate  \medMW with $<t_{\nu,0}>$} is steeper above $<t_{\nu,0}> \simeq 2\,$Myr. This is because, for very low values of $\alpha_{SS}$ (i.e., long $t_{\nu,0}$), the timescale to drain the innermost regions of the disk becomes of the same order of magnitude as the dispersal timescale. Therefore, the outer disk, where the photoevaporative wind is emitted from, needs to be very quickly dispersed, explaining the steep increase in wind mass-loss rate\footnote{We recall here that the parameter $\dot{M}_{PEW}$ is an effective mass-loss rate used to scale the $\dot{\Sigma}_{PEW}$ profile. When a disk is compact, which is the case for long $t_{\nu,0}$ and small $R_0$, the true mass-loss rate is smaller.} with $t_{\nu,0}$. Even longer viscous timescales are excluded since they lead to too long dispersal time due to the long lifetime of the inner regions that are not affected by photoevaporation.

%Interestingly, short viscous timescales \tnu$ \lesssim 1~$Myr with mass-loss rates above $10^{-9}$~\Msunyr are readily excluded since it would require strongly gravitationally unstable disks.

%Disk dispersal depends on MD, MW, tnu (little on R0) $\rightarrow$ plot the initial disk mass as function of tnu to fit disk dispersal.  Plus as an introduction, the disk fraction for various M0 to illustrate the method.

\begin{figure*}[ht!]
\includegraphics[width=2\columnwidth]{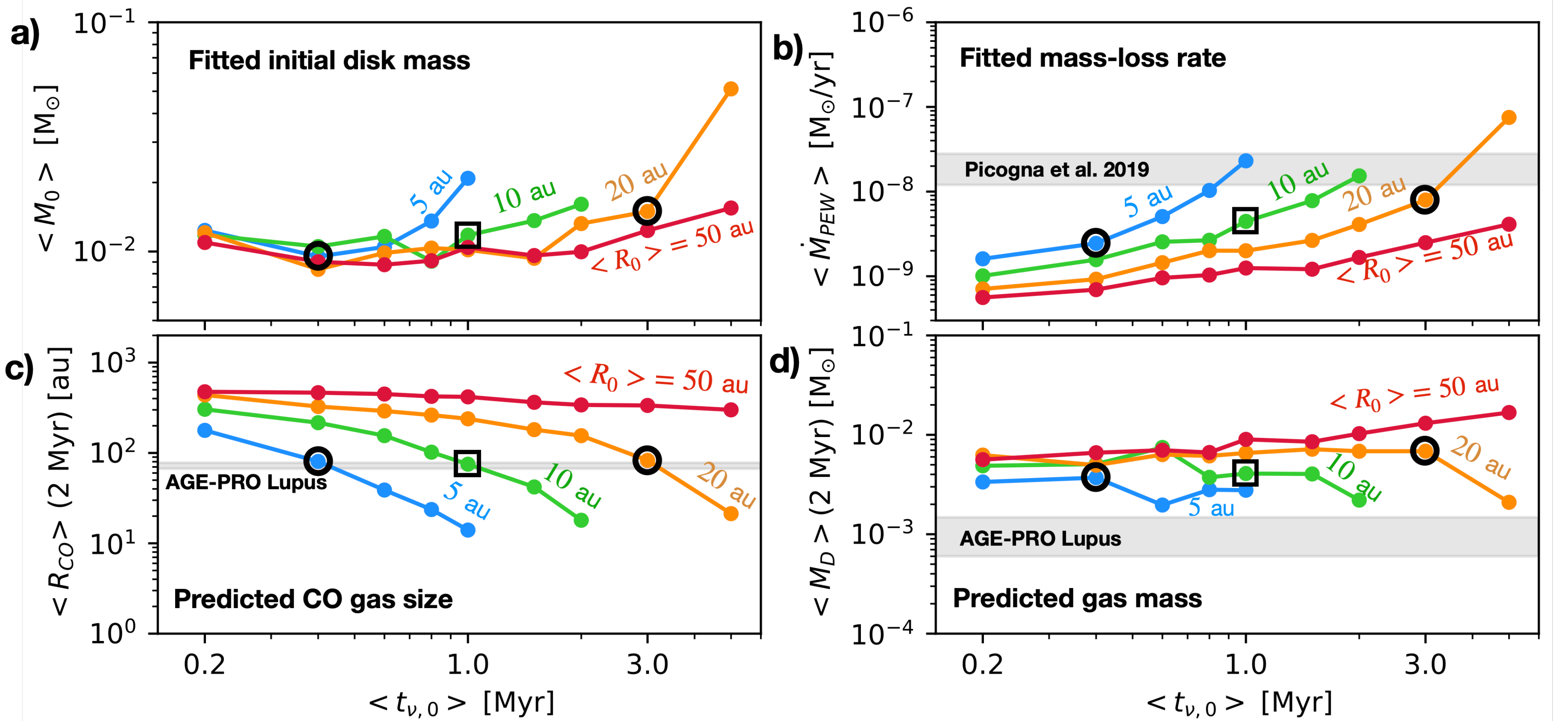}
\centering
\caption{Summary of the constraints on turbulence-driven disk evolution obtained from disk dispersal time and median accretion rate measured in Lupus (namely, after 2).
The fitted values of \medMO (panel a) and \medMW  (panel b) parameters are plotted as a function of \medtnu for various values of \medRO. The calculated median \medRO and $<M_D>$ at the estimated age of Lupus (2~Myr) are shown in panels c) and d). Median CO gas size and disk mass estimated by AGE-PRO are represented by a grey-shaded area. None of the synthetic populations can simultaneously reproduce the low disk mass and the disk size. Focusing on $<R_{CO}>$, we selected three best-fit models outlined by black circles and boxes.}
\label{fig:fig_stage_2-3_PEW}
\end{figure*} 

\subsubsection{Accretion rates}
%We recall that each point in Fig. \ref{fig:PEW_stage1}-a corresponds to a synthetic population that fits the disk fraction.
%To find the median mass-loss rate $<\dot{M}_{PEW}>$ that fits the median disk accretion in Lupus we plot in figure \ref{fig:PEW_stage1}-b the accretion rate at 2~$Myr$ of each synthetic population.

The constraints on the median mass-loss rate obtained from disk fraction are converted into median initial accretion rates in Fig. \ref{fig:PEW_stage1}-b for a reference initial disk size of $<R_0> = 10~$au. 
The initial \medMacc scales as $\propto t_{\nu,0}^{-1  }$ since \medMacc=$<M_0/2 t_{\nu,0}>$ and since the mass loss-rate does not influence the initial accretion rate. In  Fig. \ref{fig:PEW_stage1}-c, we further plot the median accretion rate of each synthetic population after $2\,$Myr. Due to the decline of the accretion rate of each disk, the median accretion rate is shown to drop between $t=0$ and $2\,$Myr. The drop is more pronounced for short viscous timescales due to the faster disk evolution. We also note that photoevaporation tends to disperse the lightest disks, that have higher accretion rates, limiting the drop in the median accretion rates. Overall, this leads to the counterintuitive result that the viscous timescale has little impact on the median accretion rate at 2~Myr, except for very long viscous timescales.

Figure \ref{fig:PEW_stage1}-c illustrates our method to constrain the initial disk mass from the median accretion rate estimated in Lupus using the example of an initial disk size of $<R_0> =10~$au. The median accretion rate of the Lupus disk population is depicted by a grey area. From this figure, it is evident that for $<R_0> = 10~$au and \medtnu$ \simeq 0.2-1.5\,$Myr, an initial median disk mass of about $1 \times 10^{-2}~M_{\odot}$ is required to match the Lupus median accretion rate. For long viscous timescales, the behaviors change. For \medtnu$ = 2\,$Myr, slightly higher disk masses of $\simeq 3 \times 10^{-2}$ are required. Longer viscous timescales are excluded since they produce too low accretion rates at $2~$Myr even for gravitationally unstable disks ($<M_0> =0.1~M_{\odot}$).  As mentioned above, for even longer viscous timescales, the synthetic population cannot match the disk fraction.

To find the value of the median disk mass that matches the median accretion rate for every value of \medtnu and \medRO, we interpolate the value of \medMO and \medMW. For that, we start with the grid of \medRO, \medtnu, and \medMO over which the value of \medMW has been computed in stage 1. We then find, for each value of \medtnu and \medRO the two populations that predict accretion rates at the age of Lupus bracketing the observed rate of $10^{-9}~M_{\odot}/yr$. We finally compute the value of \medMO and \medMW using a linear interpolation. We further check that the populations with the interpolated value of \medMO and \medMW does predict a disk dispersal time of about $3\,$Myr and a median accretion rate of $10^{-9}~M_{\odot}/yr$ at $2\,$Myr.

The result of the method is shown in Fig. \ref{fig:fig_stage_2-3_PEW}, top panels. It highlights that the two observational constraints used so far, namely disk fraction and accretion rate in Lupus, can be framed as constrained mass-loss rates \medMW and initial disk masses \medMO as a function of the two free parameters: the viscous timescale \medtnu and initial disk size \medRO. In Fig. \ref{fig:fig_stage_2-3_PEW}, we recover the result of Fig. \ref{fig:PEW_stage1}-c that the initial median disk mass required to fit disk dispersal and accretion rate in Lupus for $<R_0> =10~$au is constant for short viscous timescale and increases at longer viscous timescales. The corresponding mass-loss rate also increases with the viscous timescale as suggested by Eq. (\ref{eq:disk_dispersal_M0}).  Its value is typically above $10^{-9}~M_{\odot}/yr$ and can reach few times $10^{-8}~M_{\odot}/yr$ for long viscous timescales and compact disks. Interestingly, compared to the value predicted by the hydrodynamical model of \citet{2021MNRAS.508.3611P}, we find that only very specific values of \medtnu and \medRO are in line with their theoretical results.

\subsubsection{Constraints from AGE-PRO results on Lupus}

\begin{figure*} %[ht!]
\centering
\includegraphics[width=2.1\columnwidth]{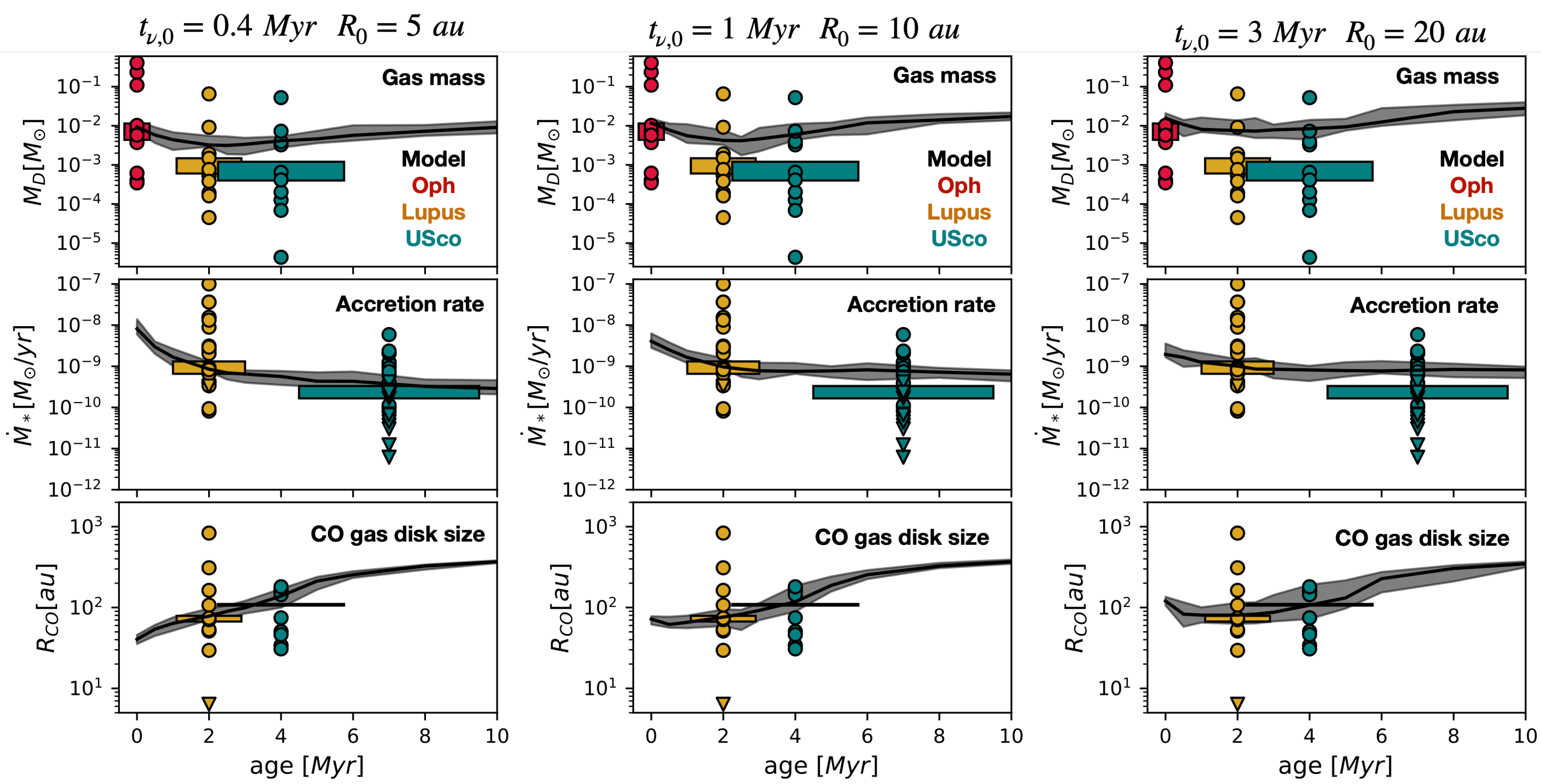}
\caption{Best fit viscous model for \PP{\tnu\revBT{$=0.4~$Myr }(left), \tnu\revBT{$=1.0\,$Myr} (middle) and \tnu\revBT{$=3.0\,$Myr} (left) } compared with the observations. The figure follows the same convention as Fig. \ref{fig:best-fit-MHD-DW}.}
\label{fig:best-fit-viscous}
\end{figure*}

At this stage, we are in a position to compare the results of AGE-PRO to synthetic populations which are consistent with disk dispersal and the median accretion rate in Lupus. The leftover free parameters are \medtnu (or, equivalently, $<\alpha_{SS}>$) and \medRO. In Fig. \ref{fig:fig_stage_2-3_PEW} bottom panels, we show the Lupus median CO size and median gas mass for all the synthetic populations that fit the disk fraction and the median accretion rate in Lupus. As expected, a population of initially larger disks or shorter viscous timescales predicts a larger $<R_{CO}>$ size after $2\,$Myr. 
%For very short viscous timescales, the impact of the initial disk size is less important since disk size at $2\,$Myr is dominated by viscous spreading. 
The median CO gas size is particularly sensitive to \medtnu for initially compact disk populations with a drop in $<R_{CO}>$ with \medtnu for \medRO$\le 20~$au. This is not only due to limited viscous spreading but also to the onset of disk dispersal by outside-in pathways occurring for disks that remain compact over their lifetime.

Compared with the AGE-PRO measurements of the CO gas size in Lupus, it is immediately apparent that only initially compact disks can fit the observations with a median disk radius of $<R_0> \simeq 5-20~$au (see models outlined by black circles and squares in Fig. \ref{fig:fig_stage_2-3_PEW}-c). We also find a degeneracy between initially compact disks with short viscous timescales and larger disks with longer viscous timescales. This results in a relatively well constrained $\alpha_{SS}$ parameter of $\alpha_{SS} = 2-4\times 10^{-4}$. In the next section, we adopt the three best-fit models highlighted by black points in Fig. \ref{fig:fig_stage_2-3_PEW}.

The predicted median disk mass at $2\,$Myr appears to be the main tension with the observations. All the models that reproduce the median CO size consistently produce a median disk mass of about $5 \times 10^{-3}~M_{\odot}$ (see models outlined by black circles in Fig. \ref{fig:fig_stage_2-3_PEW}-d). This is a factor five to ten above the disk mass estimated in Lupus by AGE-PRO (see the grey shaded area in Fig. \ref{fig:fig_stage_2-3_PEW}, bottom right). The only synthetic populations that are closer to the median Lupus disk mass have compact disks and long viscous timescales. In this extreme regime, photoevaporation removes a significant fraction of the disk mass because most of the disk mass resides where the mass loss-rate profile peaks. However, these models strongly underpredict the $<R_{CO}>$ size since the low disk masses are reproduced at the expense of a dramatic shrinking of the disk.

\subsubsection{Best fit viscous model with PEW}

Putting together all the constraints stemming from the properties of the Lupus population and disk fraction, we find that the disks should be initially compact ($R_0 \simeq 5-20~$au) with relatively long viscous timescales of \revBT{$t_{\nu,0} \simeq 0.4-3\,$Myr} even though no models can simultaneously reproduce disk accretion rate, disk mass, and size in Lupus. We show three best-fit models in Fig. \ref{fig:best-fit-viscous} encompassing the parameter space of the best fit and already outlined in Fig.\ref{fig:fig_stage_2-3_PEW} (see black circles and squares). By construction, the model reproduces the median value of $<R_{CO}>$ and $<\dot{M}_*>$ at the age of Lupus ($2\,$Myr).

We recover that the essential tension between the models and the observations is the median disk mass. In both models, the median disk mass is remarkably constant even after $10\,$Myr. This is not only the result of a slow evolution of individual disks since the disk mass is rather flat even after 10 times the viscous timescale but also the result of disk dispersal which preferentially removes light disks. Such a survivorship bias on median disk mass or accretion rates can also be seen in former population models but, to our knowledge, has never been discussed \citep[e.g.,][]{2009ApJ...704..989A}. The drop in median disk mass between Ophiuchus and Lupus cannot be reproduced by the turbulence-driven model.

The median accretion rate is predicted to drop by about a factor of 5-10 within the first $2\,$Myr before experiencing a relatively shallow decline at a longer time. Compared with the evolution of the median accretion rates between Lupus and Upper Sco, the best-fit models tend to overestimate the accretion rate in Upper Sco. 

In the three models, we predict an increase in the CO gas size with time but the Upper Sco AGE-PRO population is not old enough to test this increase. Therefore, the median CO disk size for Upper Sco is consistent with the models. An increase in gas size is expected from turbulence-driven models due to disk spreading \citep{2020A&A...640A...5T,2023MNRAS.518L..69T}. However, disk dispersal also removes disks with small radii (i.e. short viscous timescales), enhancing the increase in disk size. \revBT{We also refer to \citet{AGE-PRO_externalPE} for a detailed discussion of the effect of external photoevaporation in Upper Sco.}

\section{Discussion}

\label{sec:discusion}

\subsection{Constraints on disk evolution mechanisms}

\begin{table}
\centering 
\caption{Summary of the best-fit parameters for the two scenarios.}             % title of Table
\begin{tabular}{c c c }        % centered columns (4 columns)
\hline\hline           % inserts double horizontal lines
 Parameter & Symbol & Best fit value \\
\hline                     
 \hline
 & \textbf{MHD wind model}&\\
  \textbf{Initial disk size} &  $R_0$  & \revBT{10~au}  \\  
 \textbf{Initial disk mass} &  $M_0$  & $7.5 \times 10^{-3}$~$M_{\odot}$   \\ 
  \textbf{$\alpha$ parameter} &  $\alpha_{DW}$ & $4.6 \times 10^{-4}$ \\
 Accretion timescale & $t_{acc,0}$ & $0.75\,$Myr  \\
 \textbf{$\omega$ parameter} & $\omega$ & $0.5$ \\
 \textbf{Magnetic lever arm} & $\lambda$ & $> 8$ $^{(b)}$ \\
  Ejection-to-accretion & $f_M$ & $<0.5$  \\
  \hline
  & \textbf{Turbulent model$^{(a)}$} &\\
   \textbf{Initial disk size} &  $R_0$  & 10~au \\  
 \textbf{Initial disk mass} &  $M_0$  & \revBT{$1.2 \times 10^{-2}~M_{\odot}$}   \\ 
    \textbf{$\alpha$ parameter} &  $\alpha_{SS}$ & $3.4 \times 10^{-4}$  \\
 Viscous timescale & $t_{\nu,0}$ & $1.0 \,$Myr  \\
\textbf{Wind mass-loss rate} & $\dot{M}_{PEW}$ & \revBT{$4.4 \times 10^{-9}$} \Msunyr \\
\hline                  
\hline 
% inserts single 
%inserts single line
\end{tabular}\\
{$^{(a)}$ No set of parameters for turbulence-driven model fits all the observational constraints. The parameters correspond to a model that fits the accretion rate but not the disk mass. There is also a degeneracy between $R_0$ and $t_{\nu,0}$ outlined in Fig. \ref{fig:fig_stage_2-3_PEW}.}\\
{$^{(b)}$ Above this value the accretion-to-ejection ratio is small and $\lambda$ has a negligible impact on the properties of the synthetic population.}  
\label{table:best-fit}
\end{table}

\subsubsection{Summary of the results}

Our study demonstrates that reproducing only the median values of the fundamental population parameters such as accretion rate, disk fraction, disk size, and disk mass is a discriminant test for disk evolution models. We choose to follow a step-by-step approach summarized in Fig. \ref{fig:scheme_fitting}, adopting simple disk evolution models and assuming that  Ophiuchus, Lupus, and Upper Scorpius are representative of the same population at different ages. The best-fit parameters obtained in the previous section are summarized in Table \ref{table:best-fit}.

We find that the adopted MHD wind-driven model can reproduce very well the main properties of the three populations with the gas mass and gas size estimates provided by the AGE-PRO ALMA large program (see Fig. \ref{fig:best-fit-MHD-DW}). Notably, the best-fit accretion timescale of $t_{acc,0} \simeq 0.75~$Myr along with an initial disk size of $<R_0> = 10~$au give an initial value of $\alpha_{DW}=5 \times 10^{-4}$. Following \citet{2022MNRAS.512.2290T}, $\alpha_{DW}$ is related to the disk magnetization often parameterized by the thermal-to-magnetic pressure ratio denoted as $\beta$.  Our constraints translate to $\beta_0 \simeq 10^5$, depending on the unknown $B_{\phi}/B_z$ in the disk atmosphere \citep[see Sec. 4.3 in][for a discussion]{2022MNRAS.512.2290T}.

In contrast, following our fitting procedure, we find that turbulence-driven models with a constant $\alpha_{SS}$ and a photoevaporative mass-loss rate profile of \citet{2021MNRAS.508.3611P} can match successively disk fraction, disk accretion rate, and disk size (Fig. \ref{fig:best-fit-viscous}). This requires relatively long viscous timescales $t_{\nu,0}=0.4-3~$Myr, compact disks of $<R_0> \simeq 5-20$~au, and low wind mass-loss rates of $<\dot{M}_{PEW}> \lesssim 10^{-8}~M_{\odot}/yr$. However, when the three previously mentioned observational constraints are matched, the median disk mass at the age of Lupus and Upper Scorpius is overestimated by a factor of 5-10.

\subsubsection{Origin of the discrepancy with turbulent models}

\begin{figure*}[ht!]
\includegraphics[width=2.1\columnwidth]{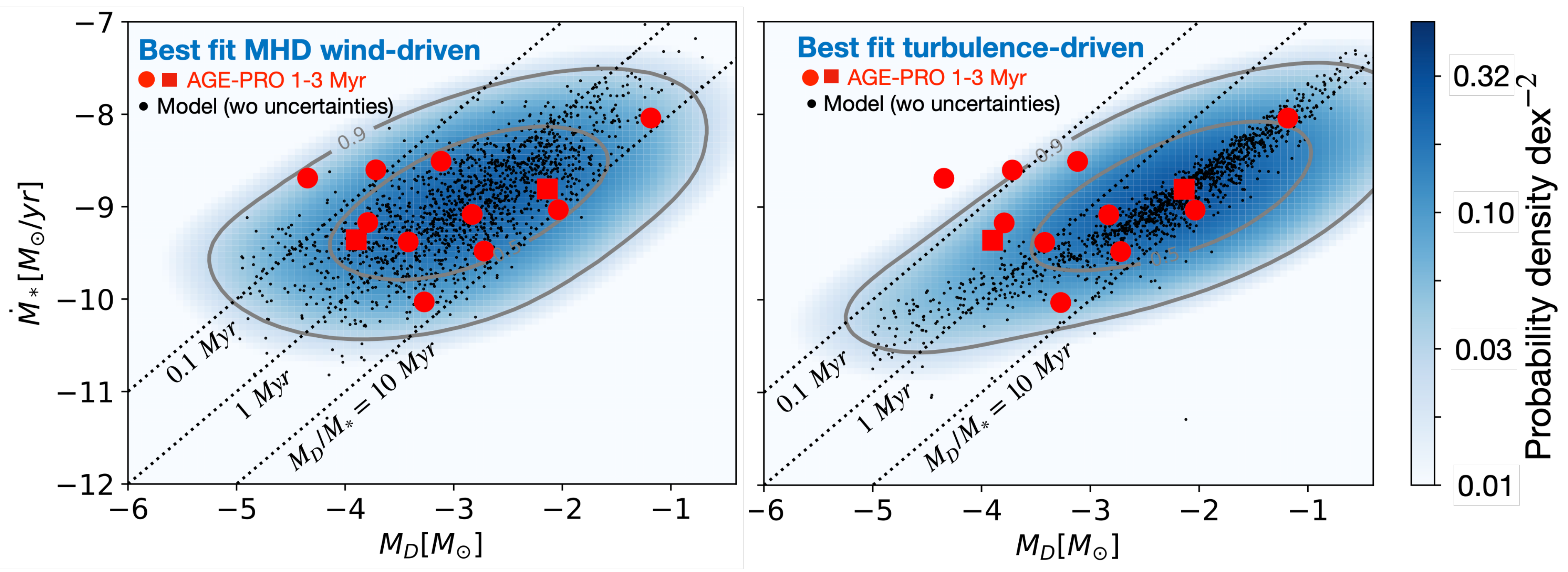}
\caption{Comparison in the disk mass-accretion rate plane between the Lupus sources and our best-fit population models for the turbulence-driven and MHD wind-driven case. The AGE-PRO sources with accretion rates from \citet{2020A&A...639A..58M} are shown in red. The simulated disks are presented in black points \revBT{and correspond to a population of 2,000 disks}. The probability density is plotted in blue shade and the 2D cumulative density distribution is in grey contours (contours encompassing 50\% and 90\% of the disks) To take into account the spreads due to the variability of the accretion rate, and the uncertainties on the measured accretion rate and disk mass, to assume that each modeled disk generates the log-normal probability distribution in the $\dot{M}_* - M_D$. The best fit MHD disk-wind model corresponds to $\omega = 0.5$ and the best fit turbulence-driven model to \revBT{\medtnu$=1.0\,$Myr}, and \medRO$=10~$au. The modeled distributions are computed assuming a detection threshold of $M_D= 10^{-5}~M_{\odot}$.}
\label{fig:MD-Macc}
\end{figure*} 

\begin{figure}[ht!]
\includegraphics[width=1.0\columnwidth]{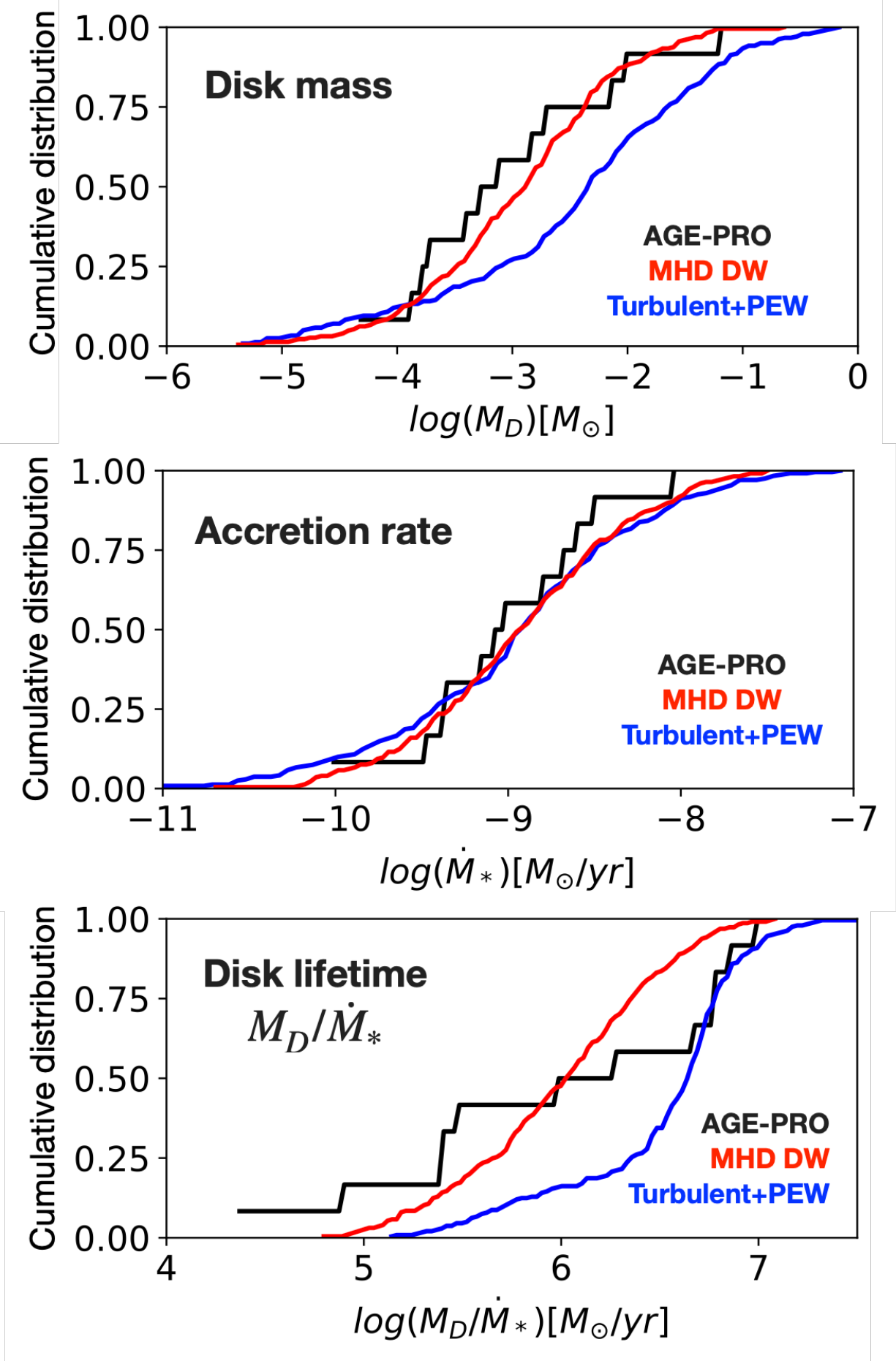}
\caption{Cumulative distribution of the accretion rate, the disk mass, and the disk lifetime for the best fit MHD disk-wind (red) and turbulence-driven model (blue) versus the AGE-PRO sample focusing on 1-3~Myr disks (black). The MHD wind model reproduces well the distributions. The salient discrepancy between the turbulence-driven model and the observations lies in the $M_D/\dot{M}_*$ ratio. The modeled cumulative distributions are computed assuming a detection threshold of $M_D= 10^{-5}~M_{\odot}$.}
\label{fig:CDF_best_fit}
\end{figure} 

Compared with e.g., MCMC fits, our step-by-step fitting method summarized in Fig. \ref{fig:scheme_fitting} has the advantage of being intuitive and computationally efficient at finding best-fit solutions. However, in the absence of a set of parameters that fits all the constraints, the order in which the observations are fitted matters. In the case of turbulent models, one can naively conclude that the disk mass is the \textit{experimentum crucis}. However, the observed Lupus median disk mass can also be matched in stage 2 instead of adjusting for the observed accretion rate. By doing so, we identified the two fundamental origins of the discrepancy between the turbulence-driven scenario and the observations.

First, matching median disk mass derived by AGE-PRO in stage 2, requires low initial disk masses in the range of $1-2 \times 10^{-3}~M_{\odot}$. This is a factor 2-5 smaller than the median disk mass inferred in the young Ophiuchus sample. In other words, the turbulence-driven model fails to reproduce the significant drop in median disk mass between Ophiuchus and Lupus. This is primarily due to the slow evolution of disk mass for viscous disks: ignoring dispersal, the mass of individual disks drops by a factor of $(1+t/t_{\nu,0})^{1/2}$. Even for $t_{\nu,0} = 0.2~$Myr, this only gives a drop of a factor of 3 from Ophiuchus to Lupus. Disk dispersal worsens this trend by preferentially removing light disks and reducing the dropping rate of the median disk mass, in particular at the age of Upper Sco. In contrast, MHD wind-driven accretion can produce a quick drop in disk mass and match the decline of the median disk mass from Ophiuchus to Lupus. However, we acknowledge that the disk masses in Ophiuchus versus Lupus and Upper Sco are estimated following slightly different approaches \citep{AGEPRO_V_gasmasses} such that the dropping rate of disk mass remains a loose constraint.

The major tension between the turbulence-driven model and the observations lies in the ratio between the disk mass and the accretion rate, also called apparent disk lifetime $t_{lt}= M_D/\dot{M}_*$. \revBT{Specifically, the fact that the apparent disk lifetime estimated by AGE-PRO is on average shorter than the age of the disk excludes turbulence-driven accretion with a constant $\alpha_{SS}$ profile.} We show in Figure \ref{fig:MD-Macc} the AGE-PRO sources in the $M_D-\dot{M}_*$ plane, focusing on stellar ages between 1 and 3~Myr. This includes the 10 Lupus and two Upper Sco sources. Figure \ref{fig:MD-Macc} also depicts the predictions for the best-fit MHD disk-wind model and turbulence-driven models as probability density maps. Because the uncertainties on the disk mass and the variability in the accretion rates can strongly affect the correlations in the $M_D-\dot{M}_*$ plane, we take into account a lognormal error on the disk mass of $\sigma=0.7$ dex \citep{AGEPRO_V_gasmasses} and on the accretion rate of $\sigma=0.3$ dex \citep{2014A&A...570A..82V}.

We recover that the turbulence-driven model predicts a correlation between disk mass and accretion rate \citep{2017MNRAS.468.1631R}. 
This is reminiscent of the population synthesis models of \citet{2017MNRAS.472.4700L} and \citet{2017ApJ...847...31M} with the addition that our population model also includes disk dispersal by photoevaporation, which alters the simple linear relationship. In particular, the predicted correlation flattens toward low-mass disks producing a subpopulation of low-mass disks ($M_D \lesssim 10^{-4}~M_{\odot}$) with a short apparent disk lifetime of $M_D/\dot{M}_* \simeq 0.1-1~$Myr. 
\revBT{This behavior contrasts with the results of \citet{2020MNRAS.492.1120S} who find that photoevaporation tends to produce lower accretors in the low-mass part of the $M_D-\dot{M}_*$ plane (bottom left region in Fig. \ref{fig:MD-Macc}). This difference is due to the initial disk sizes, which are now constrained by the AGE-PRO data. In our best-fit model, the disks, which are on average compact, tend to experience an outside-in dispersal pathway outlined in Fig. \ref{fig:example_turbulence}, leading to short $M_D/\dot{M}_*$ before dispersal.}

Our AGE-PRO sample does not recover the correlation between the disk gas masses and the accretion rates found by \citet{2016A&A...591L...3M} on larger samples and using dust emission. This is likely due to too low statistics and large uncertainties in the retrieved gas mass. As detailed in Sec. \ref{sec:results} and further highlighted in Fig. \ref{fig:CDF_best_fit} (top and middle panels) the turbulent model fits well the accretion rates but \BTbis{overpredicts} the disk masses. However, we recall that the model can alternatively fit the disk masses but would then underpredict the accretion rates. The true discrepancy between the turbulent model and the observations is the short apparent disk lifetime $M_D/\dot{M}_*$ found in AGE-PRO. As seen in the $M_D-\dot{M}_*$ plane (Fig. \ref{fig:MD-Macc}) and in the cumulative distribution of $M_D/\dot{M}_*$ (Fig. \ref{fig:CDF_best_fit}, bottom panel), half of the AGE-PRO sample is below 1~Myr with 2 disks below 0.1~Myr. As demonstrated in \citet{2012MNRAS.419..925J} and \citet{2017MNRAS.468.1631R}, turbulent-driven evolution leads to $M_D/\dot{M}_*$ longer or about the disk age, regardless of the details of the turbulence-driven model. This result is fully recovered in our synthetic population in Fig. \ref{fig:MD-Macc} (right panel) and Fig. \ref{fig:CDF_best_fit} (bottom panel) with the difference that the large uncertainty in disk mass makes very short $M_D/\dot{M}_*$ a possible but unlikely outcome of retrieval. In addition, outside-in disk dispersal, discussed by \citet{2017MNRAS.468.1631R} in the context of external photoevaporation, produces low mass disks with short $M_D/\dot{M}_*$ ratio. Still, because these disks are short-lived, they should represent a minor fraction of sources. Moreover, they cluster in the very low disk mass region. Therefore outside-in disk dispersal cannot account for the short disk lifetime found by AGE-PRO. We conclude that the tension with turbulence-driven models arises from the combination of accretion rate and disk mass with a disk mass to accretion rate ratio too short to be matched by population synthesis models.

In contrast, the MHD wind-driven model reproduces very well the distribution of disk gas mass, accretion rate, and disk lifetime (Fig. \ref{fig:MD-Macc}, left panel and Fig. \ref{fig:CDF_best_fit} red curve). The simulated population exhibits a large scatter, even before taking into account the uncertainty on the derived disk mass and the variability in the accretion rate (see black points versus probability distribution in Fig. \ref{fig:MD-Macc}). Overall, the 2D probability distribution encompasses very well the AGE-PRO measurements. The short disk lifetime reproduced by the model is due to the evolutionary tracks of individual disks \citep[see][for a detailed discussion]{2022MNRAS.512.2290T}: before disk dispersal, a disk moves horizontally in the $M_D-\dot{M}_*$ plane, producing a population of low-mass disks with high accretion rates. It is this population of disks about to be dispersed that matches the disks with a very short disk lifetime. Interestingly, disks with a low mass and short lifetime are also compact. This is consistent with MHD wind dispersal process since the disk dispersal time is set by the ratio between disk size and \alphaDW.

%Therefore, the comparison in the disk-mass accretion plane demonstrates that correlations are expected if the sample size is large enough, even with the large uncertainties in the disk mass estimate. The clustering of the AGE-PRO sample is more in line with the prediction of the MHD disk-wind model, but the sample appears to be too small to discriminate models based on the distribution in the $M_D-\dot{M}_*$ plane. Reproducing the median values of the population parameters appears to be a more direct way to constrain and discriminate the evolutionary scenario for small samples.

\subsection{Perspectives on the observational data}

\subsubsection{Representativeness of the AGE-PRO star-forming regions}

One of the fundamental assumptions behind our analysis is that the Ophiuchus, Lupus, and Upper Sco populations are representative of the same population at different ages. Based on the analysis presented in the Oph, Lupus, and Upper Sco AGE-PRO papers \citep{AGEPRO_II_Ophiuchus, AGEPRO_III_Lupus, AGEPRO_IV_UpperSco} the AGE-PRO sample is representative of the whole population in each region, based on the comparison with the available accretion rates (not applicable to Oph) and millimeter luminosities for similar spectral types. Surveys of young star-forming regions of a similar age as Lupus (1-3~Myr) show little variation in their distribution of continuum mm luminosity and accretion rates except for CrA that show weaker mm continuum fluxes \citep{2019A&A...626A..11C}. Our knowledge of old star-forming regions similar to Upper Sco is more limited. It remains unclear whether the median accretion rate declines with cluster age, as the measurements of \citet{2023ApJ...945..112F} used in the present work suggest, or if they remain comparable to young star-forming regions even when corrected for the stellar mass dependency \citep[see e.g.,][Fig. 11]{2022A&A...663A..98T}. Surveys of old star-forming regions in the mm are also lacking, making Upper Sco our unique reference point with a complex star formation history \citep{2022A&A...667A.163M,2023A&A...678A..71R}. Therefore, variations in terms of gas content between different star-forming regions remains to be probed, targeting the well-known middle-aged but also less-known old star-forming regions.

Another important caveat for the Upper Sco region is the importance of external irradiation in the disk evolution. Throughout this work, we neglected external photoevaporation which can affect disk evolution \citep{2022MNRAS.514.2315C}. Only strong external FUV of $G_0>10^3$ are expected to significantly affect the dispersal time and all the properties of the disks \citep{2017AJ....153..240A,2023A&A...679A..82M, garate2024}, but the modest FUV strength experienced by the AGE-PRO Upper Sco sample can slightly affect the disk gas mass but largely decrease gas disk size \citep{AGE-PRO_externalPE}.

\subsubsection{Disk gas mass}

It is commonly suggested that the measurement of the disk size alone constitutes a decisive test for disk evolution models. This comes from the idea that turbulence-driven accretion requires viscous expansion and MHD wind-driven accretion does not. Our work demonstrates that disk sizes constitute an essential constraint for disk evolution models but also shows that the latter idea is somewhat misleading. First, as demonstrated in Fig. \ref{fig:example_turbulence}, turbulence-driven disks can shrink over time due to internal photoevaporation. The requirement is that the disk is compact enough such that the local wind mass-loss rate is large enough out to the disk edge. Since AGE-PRO shows that the disks are on average compact ($R_{CO}\simeq 70-110~$au), internal photoevaporation is expected to affect the evolution of the disk size in the case of turbulence-driven evolution. External photoevaporation would be even more efficient \citep{AGE-PRO_externalPE}. On the other hand, the survivorship bias can greatly affect disk sizes: in MHD wind-driven models, we find that CO disk size slightly increases due to the removal of compact disks. The measurement of CO size alone, \revBT{even in regions of different age,} is not a decisive test for disk evolution scenarios.

However, combining disk gas size, disk gas mass, and stellar accretion rate measurements brings solid constraints on disk evolution models. This is demonstrated in Fig. \ref{fig:fig_stage_2-3_PEW} which shows that turbulence-driven models can not simultaneously reproduce disk mass, accretion rate, and CO size. Overall, we find that disk mass is a crucial parameter along with accretion rates but indubitably the most uncertain. It is the relatively low median disk gas mass found in Lupus and Upper Sco that questions the turbulence-driven paradigm. 

If disk masses in these regions were to be systematically underestimated by a factor of five to ten, turbulence-driven models would be compatible with the data with an apparent lifetime of $M_D/\dot{M}_* > 1~$Myr. MHD wind-driven models would also be able to fit the data with a higher ejection-to-accretion mass ratio $f_M$ and higher initial disk mass $<M_0>$. \revBT{This can be seen in Fig. \ref{fig:stage3_MHD-DW}: if the median disk mass at Lupus age were to be an order of magnitude higher, the best-fit ejection-to-accretion mass ratio would be about $f_M = 1-10$. Such high values of the ejection rate are in fact not irrealistic comprared with theoretical MHD wind models and numerical simulations \citep[e.g.,][]{2019ApJ...874...90W}. In this case, the initial disk mass would need to be increased (see Fig. \ref{fig:stage2_MHD-DW}).}

The estimates of the disk gas mass are subject to considerable challenges \citep{2023ASPC..534..501M}. Within the AGE-PRO large programme, the DALI thermochemical model is used \citep{AGEPRO_V_gasmasses} to retrieve disk gas masses from CO and N$_2$H$^+$ but the retrieval based solely on the CO line fluxes and using \citet{2022ApJ...925...49R} models give similar results in Lupus \citep[see][]{AGEPRO_III_Lupus}. Overall, our estimates depend on our understanding of the chemistry and physics of disks and one can expect considerable progress in the future. Benchmarks between mass estimates based on N$_2$H$^+$ \citep{2019ApJ...881..127A,2022ApJ...926L...2T}, kinematics \citep{2024A&A...686A...9M}, and HD \citep{2017A&A...605A..69T} show relatively consistent results for large and massive disks \citep[see][for a review]{2023ASPC..534..501M}. However, we stress that the lowest disk masses found in AGE-PRO are for very compact disks. Deep observations designed to characterize the chemistry of compact disks ($<R_{CO}> \lesssim 60~$au) should be performed. Our fitting strategy allows one to investigate how the best-fit model will be affected by a revision of the disk gas masses since the constraints from the gas masses are, on purpose, taken into account at the final stage of the fit (see Fig. \ref{fig:scheme_fitting}).

A final caveat behind disk mass estimates is that the molecular lines trace the outer disk where it is assumed that the bulk mass resides. The regions inside of $\simeq 10~$au could hide a large fraction of the mass which would remain poorly traced by ALMA due to optical depth effects. This would require a significant jump in the surface density profile. Since disks accrete, a jump in the surface density profile implies a jump in $\alpha$ values (either radial or vertical torque). In the turbulence-driven scenario, a massive inner disk with very low $\alpha$ values is expected to have a long survival timescale since photoevaporation does not operate in that region \citep{hartmann2006, Morishima2012, garate2021}. One can then speculate that such a setup would lead to a population of extremely compact but accreting disks. Detailed models including radial variation in $\alpha$ values \citep[e.g.,][]{2024MNRAS.533.1211T} are deferred to future work. In parallel, constraints on the absolute value of the column density in the inner disk can be obtained by line pressure broadening of CO \citep{2022ApJ...937L..14Y}.

\subsection{Disk dispersal and the survivorship bias}
\label{subsec:diskdispersal}

\begin{figure}[ht!]
\includegraphics[width=1.0\columnwidth]{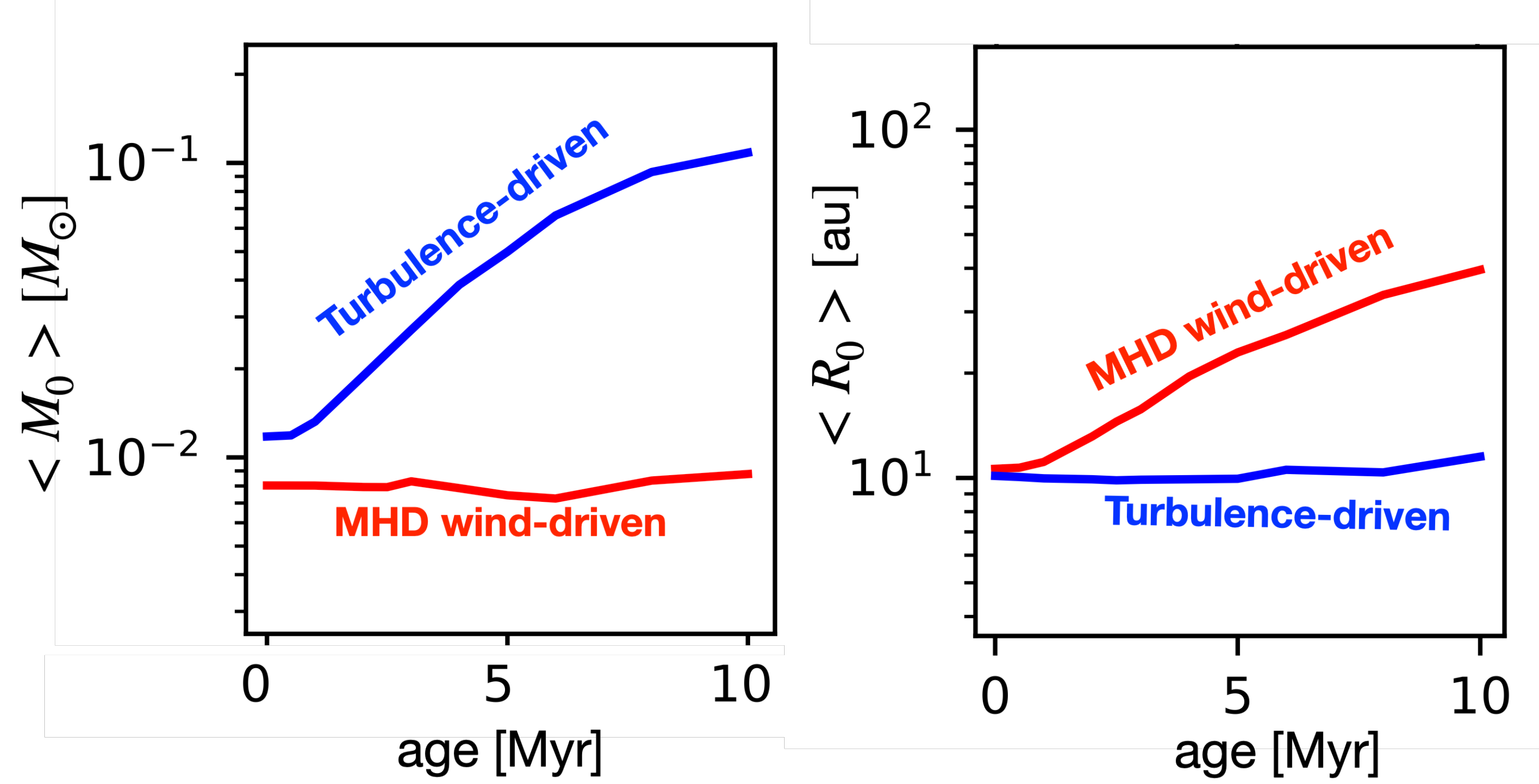}
\caption{The survivorship bias illustrated by the initial median value of the disk mass (left) and size (right) calculated over the surviving disks. The blue and red lines correspond to the best-fit MHD wind-driven and turbulence-driven populations (see Table \ref{table:best-fit}).}
\label{fig:survivors}
\end{figure} 

Because the evolution of a disk cannot be observationally monitored, the use of surveys at a population level is needed. However, in the presence of disk dispersal, the evolution of the properties of a population is not only determined by the evolution of individual disks but also by the properties of the survivors.  
Our population synthesis approach highlights the importance of this bias, a.k.a. \textit{survivorship bias}, when interpreting the results of surveys. The evolution of the median disk mass in the turbulence-driven scenario is undoubtedly the most striking example: whereas the mass of individual disks declines, disk dispersal removes light disks resulting in a shallow evolution of the disk mass. \revBT{This selection process, identified by \citet{2020MNRAS.492.1120S}, is evidenced by the dramatic increase in the initial mass of the surviving disks shown in Fig. \ref{fig:survivors} (see left panel)}. MHD wind-driven evolution also introduces  a survivorship bias: compact disks evolve and disperse faster since they have shorter accretion timescales \tacc. \revBT{This results in an increase in the initial median disk size of the surviving disks highlighted in Fig. \ref{fig:survivors} (right panel)}. This effect drives the small increase in CO gas size seen in Fig. \ref{fig:best-fit-MHD-DW}.

We stress that the survivorship bias is different from the selection bias induced by the definition of a sample. It is a bias that does not depend on the choice of the observer but arises because the defined population of disk loses members due to dispersal. One could instead follow the population of all the stars but in this case, the median disk mass drops to zero as soon as the disk fraction drops below $50\%$. We call it bias because it is a selection process that prevents us from probing directly the evolution of the disk properties. The implication is that the survivorship bias is a fundamental bias that is difficult to correct in the observations. This requires knowing which disk in a young population survives at a later stage. For a given disk evolution model, the effect of survivorship bias can still be qualitatively assessed since the population parameters that are the most affected are those that control disk dispersal. On the other hand, survivorship bias constitutes a promising way to constrain disk evolution mechanisms. As an example, X-ray photoevaporation is expected to disperse not only light disks but also disks around X-ray bright stars. This idea has been explored to test the relevance of X-ray photoevaporation \citep{2005ApJS..160..401P,2021A&A...648A.121F,2022ApJ...935..111L} even though the interpretations of the results are complicated by the possible quenching of X-ray emission by the accretion flow.

\subsection{Spreads and correlations between disk parameters}

Throughout this work, we assumed that $\alpha$, $R_0$, $M_0$, and the parameter controlling the wind mass-loss rate ($\dot{M}_{PEW}$ or $f_M$) are independent and follow lognormal distributions with fixed spreads. We recall that the spreads have been chosen to reproduce the decline of the disk fraction but no further adjustment has been made to fit the spread in the observed disk mass and accretion rates. We however note that our best-fit models reproduce well the spread in the accretion rate (Fig. \ref{fig:CDF_best_fit}, middle panel), especially if we consider the effect of variability. The spread in the distribution of disk mass is also well matched (see Fig. \ref{fig:CDF_best_fit}, top panel) even though the spreads are dominated by the large statistical uncertainties in the retrieved gas mass.

Interestingly, we note that the spreads in the disk parameters impact the evolution of the medians. This is again due to the survivorship bias. The clearest example is the evolution of the median CO disk size in MHD wind-driven evolution. Disks are removed according to their \tacc values which is the ratio between \alphaDW and $R_0$. This drives the increase in $<R_{CO}>$. However, in the absence of any spread in initial disk size, $<R_{CO}>$ would remain constant in time.

The effect of the correlations between the disk parameters is an interesting avenue to explore. Due to survivorship bias, a parameter that is not related to disk dispersal can be greatly affected if it is correlated with a key parameter for dispersal. The most obvious correlation is probably between the initial disk mass and the initial disk size even though observations of Class I disks suggest a rather weak dependency between disk mass and disk size \citep{2020ApJ...890..130T}. For example, the disks could be born with similar values of the surface density and differ in mass only because they differ in size. In the case of our MHD wind-driven model, small disks are dispersed first such that a positive correlation would produce a shallower evolution in disk mass. For our turbulence-driven model, light disks are dispersed first such that a positive correlation would steepen the evolution of the CO disk size.

All in all, our best-fit models constitute a baseline for a full exploration of the effect of the spreads and the correlations between the disk parameters. Such models will add new free parameters and additional observational constraints should be taken into account to limit the degeneracies.

\subsection{Limits and perspective on 1D disk evolution models}

Our work aims to test and quantify disk evolution processes by confronting existing surveys and in particular AGE-PRO to population synthesis models. 
Our goal is not to propose new disk evolution models but to provide reference population models based on existing 1D evolution models. These models are meant to capture the essential features of either turbulence-driven evolution or MHD wind-driven evolution. Here, we outline the simplifications of these models and discuss future disk evolution models that remain to be built.

\subsubsection{Turbulence-driven case}

For the turbulence-driven model, \PP{one of the} uncertainties is the photoevaporative wind model \citep{2023ASPC..534..567P}.  The wind properties are the result of the spectral distribution of high energy photons (from $\simeq10$ eV to $\simeq1$ keV), radiative transfer, and gas thermal balance of the disk upper layers. Over the past decades, several studies have been conducted to establish mass-loss rate profiles along with absolute mass-loss rates with contradictory results \citep{2017ApJ...847...11W,2021ApJ...910...51K,2021MNRAS.508.3611P}. In our work, we kept the total mass-loss rate as a free parameter to avoid biases due to the use of a specific study and we adopted the mass-loss rate profile of \citet{2021MNRAS.508.3611P} for a $0.5~M_{\odot}$ mass star. Our best-fit models have a typical mass-loss rate of $\simeq 3 \times 10^{-9} - 10^{-8}~$M$_{\odot}$/yr, which is smaller than the mass-loss rate of $\simeq 2 \times 10^{-8}~M_{\odot}/yr$ predicted by \citet{2021MNRAS.508.3611P} for a $0.5~M_{\odot}$ stellar mass. Similar results were also found by analyzing the distribution of the accretion rates alone \PP{\citep{alexander2023}}. Our findings are also well in line with the recent work of \citet{2024arXiv240800848S} who showed that the inclusion of collisions of O with neutral H, which was not included in \citet{2021MNRAS.508.3611P}, and subsequent line cooling, decreases the mass-loss rate by a factor of 5-10.

Our study demonstrates that internal photoevaporative winds control not only the disk dispersal time but also the overall evolution of the disk properties, notably the disk size as illustrated in Fig. \ref{fig:example_turbulence}. In this context, the exact mass-loss rate profile matters. If the local mass-loss rate were to be significant at large distances, as in the calculations of \citet{2009ApJ...690.1539G}, disks could be initially more extended.

An important simplification of the model is the absence of attenuation of the stellar radiation by an inner MHD disk-wind. MHD disk-winds launched from the inner disks ($\lesssim 1~$au) are likely ubiquitous \citep{2016ApJ...831..169S,2018ApJ...868...28F,2020ApJ...903...78P} and attenuation could reduce the heating of the outer disk where photoevaporative winds are launched. Our estimate of the median wind mass-loss rate should therefore be considered as time-averaged. Consistent models would require identifying which photon energy drives the wind and the composition of the inner MHD wind (dust content, metallicity). More detailed models including the attenuation by an evolving inner MHD disk-wind should be explored \citep[see also][]{2023A&A...674A.165W}.

The assumption of constant $\alpha_{SS}$ in space and time is also an important simplification of the turbulence-driven model used in this work. The presence of a dead zone with a limited radial extent can create a steep transition in the $\alpha_{SS}$ profile \citep{delage2022}. A systematic study exploring sophisticated profiles of $\alpha_{SS}$ is beyond the scope of the paper. Preliminary tests show that a population of disks with a highly turbulent outer region ($r>20~$au) and low $\alpha_{SS}$ inner disk tends to behave like a population of disks with constant $\alpha_{SS}$ values. In particular, an $\alpha_{SS}$ with step function does not seem to solve the tension with the observations but more work remains to be done in that regard.

\subsubsection{MHD wind-driven case}

In contrast to turbulence-driven models, 1D disk evolution models including MHD disk-winds are relatively recent \citep{2016A&A...596A..74S,2016ApJ...821...80B,2022MNRAS.512.2290T}. Throughout this work, we adopted the simple model of \citet{2022MNRAS.512.2290T} which has the advantage of relying on limited assumptions and being directly comparable to numerical simulations. The wind torque is parameterized by the \alphaDW parameter which is proportional to the mid-plane magnetization, traditionally quantified by the thermal-to-magnetic pressure ratio $\beta$.

The secular evolution of the magnetic field strength, which controls the wind torque, is notably a major uncertainty in secular evolution models. As shown by \citet{2013ApJ...778L..14A} and \citet{2016ApJ...821...80B}, if the disk preserves its magnetic field, the disk can be quickly dispersed. This is the situation explored in this work where the $\omega$ parameter controls the time evolution of \alphaDW. In the framework of this model, we show that a slowly decreasing magnetic field producing an increased magnetization with $\beta \propto \sqrt{M_D}$ can naturally account for the observational data. Numerical simulation running on a secular timescale opened by GPU-accelerated codes \citep[see e.g.][]{2023A&A...677A...9L} are warranted to interpret these constraints. 

The wind mass-loss rate and the impact of the high-energy photons are also a major uncertainty in MHD wind-driven models. In this work, we use the magnetic lever arm parameter $\lambda$ to parameterize the wind mass-loss rate (see Eq.~\ref{eq:sigma_dot_MHD-DW}). In a realistic setup, the value of $\lambda$ is the result of the field morphology, field strength, and heating of the disk surface \citep[see e.g.,][]{2016ApJ...821...80B}. The contrast between photoevaporative winds and MHD disk-winds is commonly made in the literature. We recall that MHD wind winds emerge when a magnetic field with a net flux threatens the disk but thermal effects are fundamental in the wind launching process \citep{2000A&A...361.1178C,2016ApJ...821...80B}.
The transition between photoevaporative winds and MHD disk-winds is often presented as the transition when the MHD wind mass-loss rate is about that predicted in the absence of any net magnetic flux. \citet{2017A&A...600A..75B} found that the wind mass-loss rate decreases with disk magnetization as $\dot{M}_{MHD-DW} \propto \beta^{-1/2}$.  \citet{2020A&A...633A..21R} further confirmed this trend showing that for a given disk setup, the wind mass-loss rate reaches the threshold value of photoevaporation for $\beta \gtrsim 10^7$. Within our modeling framework, it means that $\lambda$ is predicted to decline with $\beta$ or, equivalently increase with \alphaDW. However, in the absence of systematic theoretical constraints on how $\lambda$ varies with disk fundamental properties, we simply assumed $\lambda$ to be constant in space and time. We also note that our best-fit MHD disk-wind model predicts disks with increasing magnetizations but decreasing mass-loss rates. In the literature, 1D evolution models have attempted to couple MHD disk-winds and photoevaporative winds by simply adding the MHD disk-wind mass-loss rate and the photoevaporative mass-loss rate \citep{2020MNRAS.492.3849K,2023A&A...674A.165W}. A promising avenue to consistently describe the transition between photoevaporation and MHD disk-wind would be to determine how $\lambda$ varies with at least \alphaDW, and with the impinging radiation field using a wide grid of numerical simulations including detailed micro-physics.

\section{Conclusion}

In this work, we developed a disk population synthesis approach to challenge existing disk evolution models in the framework of the AGE-PRO large program. The use of population synthesis is essential since the evolution of a population is not only set by the evolution of each individual disk but also by the properties of the survivors. To obtain robust constraints we used not only the results of AGE-PRO, namely the disk gas sizes and masses, but also the disk fractions and the accretion rates. Each free population parameter is constrained sequentially using the disk fractions and the properties of the middle-aged Lupus population. This allows us to considerably reduce the parameter space before confronting the two scenarios with the AGE-PRO results. Our results are the following. 
\begin{itemize}
    \item \revBT{
    %The recent disk surveys probing the gaseous component of the disks, notably AGE-PRO, open a new avenue to probe the disk evolution mechanisms. 
    The analysis of disk surveys requires adopting a disk population synthesis approach to reproduce simultaneously all the available observational constraints. This approach is key to include the fact that the evolution of a population is not only the result of the evolution of individual disks but also of disk dispersal that selects disks with specific properties, a phenomenon that we call here \textit{survivorship bias}.}
    \item MHD wind-driven models mimicking the evolution of the disk magnetization can reproduce disk fraction and the overall properties of the Ophiuchus, Lupus, and Upper Sco populations. It requires initially compact ($R_0 \simeq 10~$au) disks with a moderate mass-loss rate $\dot{M}_W / \dot{M}_* \lesssim 1$, and an accretion timescale of about $t_{acc,0} \simeq 0.4-0.8\,$Myr ($\alpha_{DW,0} = 5 \times 10^{-4} - 10^{-3}$). This corresponds to disks with an initial magnetization of about $\beta \simeq 10^{5}$.
    \item Turbulence-driven models with internal photoevaporation can reproduce disk dispersal and gas size but fail to reproduce simultaneously the accretion rates and the disk masses. The short apparent disk lifetimes of $M_D/\dot{M}_* \simeq 0.1-1~$Myr for half of the middle-aged disks are the major tension with turbulence-driven models.
    Still, detailed studies of compact disks that have the lowest masses \revBT{and the shortest apparent disk lifetimes} should be conducted to confirm their low masses.
\end{itemize}

\revBT{This work tends to favor the emerging MHD wind-driven disk accretion scenario, which has major implications for planet formation and migration. The increasing number of disk wind detections with ALMA and JWST will enable constraining some of the key input parameters of these models such as the wind mass-loss rates and the lever arm.} This work provides the community with observationally constrained synthetic disk populations that are required to study dust evolution and planet formation. We, however, acknowledge that the disk evolution models cannot capture the full complexity of the disk physics. This first study warrants the development of more sophisticated disk evolution models that are timely in the era of disk surveys.

\label{sec:conclusion}

\section*{Acknowledgment}
The authors thank the referee for a rigorous and constructive report. B.T. thanks Min Fang for sharing data on Upper Sco. 
This work was supported by the Programme National PCMI of CNRS/INSU with INC/INP co-funded by CEA and CNES. G.R. acknowledges funding from the Fondazione Cariplo, grant no. 2022-1217, and the European Research Council (ERC) under the European Union’s Horizon Europe Research \& Innovation Programme under grant agreement no. 101039651 (DiscEvol). Views and opinions expressed are however those of the author(s) only, and do not necessarily reflect those of the European Union or the European Research Council Executive Agency. Neither the European Union nor the granting authority can be held responsible for them.
RA acknowledges funding from the Science \& Technology Facilities Council (STFC) through Consolidated Grant ST/W000857/1.
L.T. and K. Z. acknowledge the support of the NSF AAG grant \#2205617. 
P.P. and A.S. acknowledge the support from the UK Research and Innovation (UKRI) under the UK government’s Horizon Europe funding guarantee from ERC (under grant agreement No 101076489).
A.S. also acknowledges support from FONDECYT de Postdoctorado 2022 $\#$3220495.
I.P. and D.D. acknowledge support from Collaborative NSF Astronomy \& Astrophysics Research grant (ID: 2205870).
C.A.G. and L.P. acknowledge support from FONDECYT de Postdoctorado 2021 \#3210520.
L.P. also acknowledges support from ANID BASAL project FB210003.
L.A.C and C.G.R. acknowledge support from the Millennium Nucleus on Young Exoplanets and their Moons (YEMS), ANID - Center Code NCN2021\_080 and 
L.A.C. also acknowledges support from the FONDECYT grant \#1241056.
N.T.K. acknowledges support provided by the Alexander von Humboldt Foundation in the framework of the Sofja Kovalevskaja Award endowed by the Federal Ministry of Education and Research.
K.S. acknowledges support from the European Research Council under the Horizon 2020 Framework Program via the ERC Advanced Grant Origins 83 24 28. GL has received funding from the European Union’s Horizon 2020 research and innovation program under the Marie-Sklodowska Curie grant agreement No. 823823738 (DUSTBUSTERS) and from PRIN-MUR 20228JPA3A. 
All figures were generated with the \texttt{PYTHON}-based package \texttt{MATPLOTLIB} \citep{2007CSE.....9...90H}. This research made use of Astropy,\footnote{http://www.astropy.org} a community-developed core Python package for Astronomy \citep{2018AJ....156..123A}.


\vspace{5mm}
\facilities{ALMA}


\software{astropy \citep{2018AJ....156..123A},  
          \textit{Diskpop} \citep{2024arXiv240721101S}}



\appendix

\section{Constraints on \medMW and \medMO for turbulence-driven populations}

\begin{figure*}[ht!]
\includegraphics[width=\columnwidth]{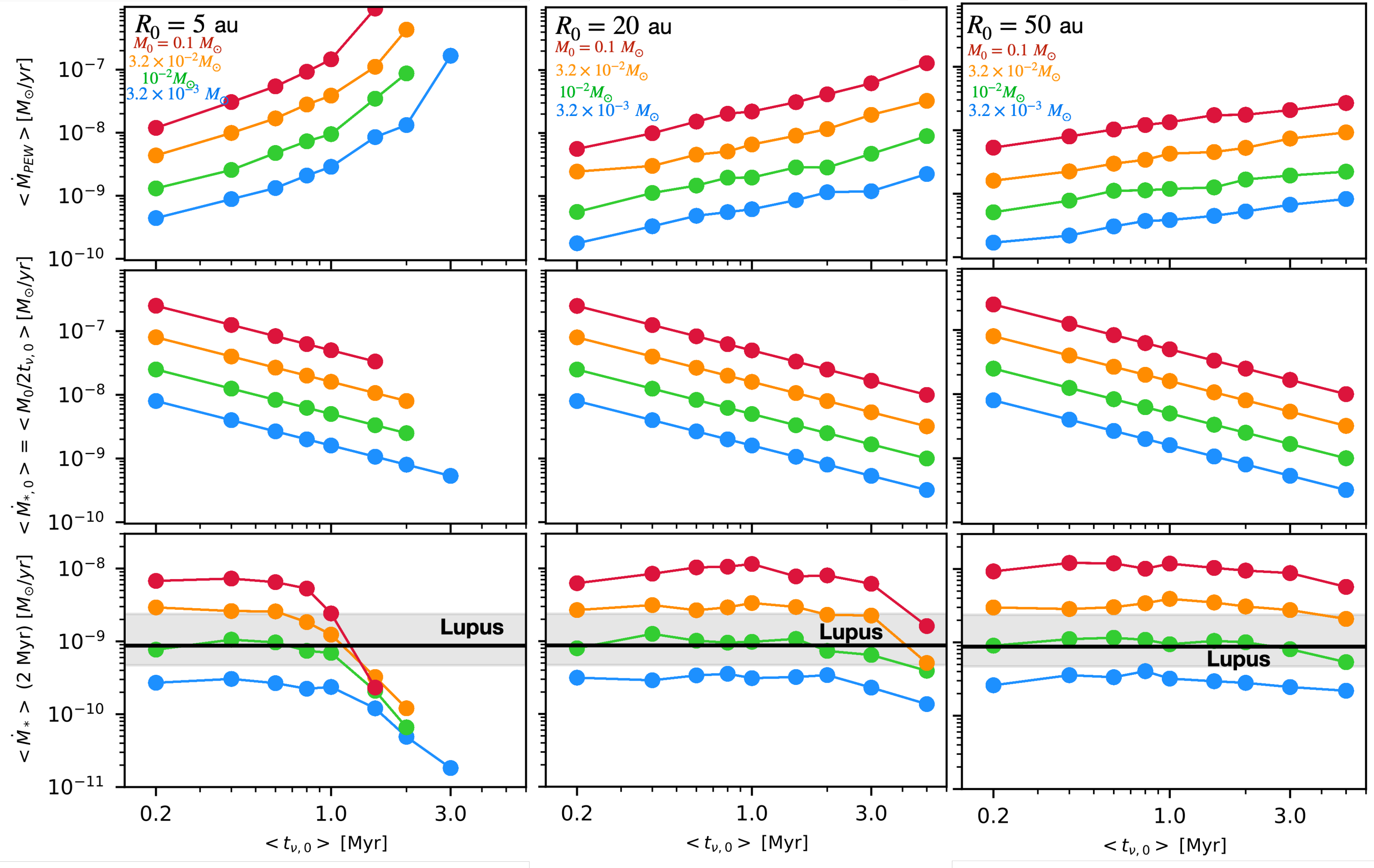}
\centering
\caption{Summary of the constraints on turbulence-driven disk evolution obtained from disk dispersal time. The fitted values of \medMW are plotted as a function of \medtnu for various values of $R_0$ (top row). The median accretion rate at $t=0$ and $t=2$~Myr are shown in the middle and bottom panels.}
\label{fig:appendix_stage1_PEW_overview}
\end{figure*} 

%\bibliography{sample631}{}
\bibliography{aastex631}{}

\begin{thebibliography}{}
\expandafter\ifx\csname natexlab\endcsname\relax\def\natexlab#1{#1}\fi
\providecommand{\url}[1]{\href{#1}{#1}}
\providecommand{\dodoi}[1]{doi:~\href{http://doi.org/#1}{\nolinkurl{#1}}}
\providecommand{\doeprint}[1]{\href{http://ascl.net/#1}{\nolinkurl{http://ascl.net/#1}}}
\providecommand{\doarXiv}[1]{\href{https://arxiv.org/abs/#1}{\nolinkurl{https://arxiv.org/abs/#1}}}

\bibitem[{{Agurto-Gangas} {et~al.}(2025, in press){Agurto-Gangas}, {P{\'e}rez},
  {Sierra}, {Miley}, {Zhang}, {Pascucci}, {Pinilla}, {Deng}, {Carpenter},
  {Trapman}, {Cieza}, {Vioque}, {Kurtovic}, {Rosotti}, {Schwarz},
  {Hogerheijde}, {Anania}, {Tabone}, {Torresvillanueva}, {Ruiz-Rodriguez}, \&
  {Gonz{\'a}lez}}]{AGEPRO_IV_UpperSco}
{Agurto-Gangas}, C., {P{\'e}rez}, L.~M., {Sierra}, A., {et~al.} 2025, in press,
  \apj

\bibitem[{{Alcal{\'a}} {et~al.}(2014){Alcal{\'a}}, {Natta}, {Manara}, {Spezzi},
  {Stelzer}, {Frasca}, {Biazzo}, {Covino}, {Randich}, {Rigliaco}, {Testi},
  {Comer{\'o}n}, {Cupani}, \& {D'Elia}}]{alcala2014}
{Alcal{\'a}}, J.~M., {Natta}, A., {Manara}, C.~F., {et~al.} 2014, \aap, 561,
  A2, \dodoi{10.1051/0004-6361/201322254}

\bibitem[{{Alcal{\'a}} {et~al.}(2017{\natexlab{a}}){Alcal{\'a}}, {Manara},
  {Natta}, {Frasca}, {Testi}, {Nisini}, {Stelzer}, {Williams}, {Antoniucci},
  {Biazzo}, {Covino}, {Esposito}, {Getman}, \&
  {Rigliaco}}]{2017A&A...600A..20A}
{Alcal{\'a}}, J.~M., {Manara}, C.~F., {Natta}, A., {et~al.} 2017{\natexlab{a}},
  \aap, 600, A20, \dodoi{10.1051/0004-6361/201629929}

\bibitem[{{Alcal{\'a}} {et~al.}(2017{\natexlab{b}}){Alcal{\'a}}, {Manara},
  {Natta}, {Frasca}, {Testi}, {Nisini}, {Stelzer}, {Williams}, {Antoniucci},
  {Biazzo}, {Covino}, {Esposito}, {Getman}, \& {Rigliaco}}]{alcala2017}
---. 2017{\natexlab{b}}, \aap, 600, A20, \dodoi{10.1051/0004-6361/201629929}

\bibitem[{{Alexander}(2008)}]{alexander2008}
{Alexander}, R. 2008, \nar, 52, 60, \dodoi{10.1016/j.newar.2008.04.004}

\bibitem[{{Alexander} {et~al.}(2014){Alexander}, {Pascucci}, {Andrews},
  {Armitage}, \& {Cieza}}]{2014prpl.conf..475A}
{Alexander}, R., {Pascucci}, I., {Andrews}, S., {Armitage}, P., \& {Cieza}, L.
  2014, in Protostars and Planets VI, ed. H.~{Beuther}, R.~S. {Klessen}, C.~P.
  {Dullemond}, \& T.~{Henning}, 475,
  \dodoi{10.2458/azu\_uapress\_9780816531240-ch021}

\bibitem[{{Alexander} {et~al.}(2023){Alexander}, {Rosotti}, {Armitage},
  {Herczeg}, {Manara}, \& {Tabone}}]{alexander2023}
{Alexander}, R., {Rosotti}, G., {Armitage}, P.~J., {et~al.} 2023, \mnras, 524,
  3948, \dodoi{10.1093/mnras/stad1983}

\bibitem[{{Alexander} \& {Armitage}(2009)}]{2009ApJ...704..989A}
{Alexander}, R.~D., \& {Armitage}, P.~J. 2009, \apj, 704, 989,
  \dodoi{10.1088/0004-637X/704/2/989}

\bibitem[{{Alexander} {et~al.}(2006){Alexander}, {Clarke}, \&
  {Pringle}}]{2006MNRAS.369..229A}
{Alexander}, R.~D., {Clarke}, C.~J., \& {Pringle}, J.~E. 2006, \mnras, 369,
  229, \dodoi{10.1111/j.1365-2966.2006.10294.x}

\bibitem[{{Anania}(2025)}]{AGE-PRO_externalPE}
{Anania}, R. 2025, \apj.
\newblock \doarXiv{xxxx.xxxx}

\bibitem[{{Anderson} {et~al.}(2019){Anderson}, {Blake}, {Bergin}, {Zhang},
  {Carpenter}, {Schwarz}, {Huang}, \& {{\"O}berg}}]{2019ApJ...881..127A}
{Anderson}, D.~E., {Blake}, G.~A., {Bergin}, E.~A., {et~al.} 2019, \apj, 881,
  127, \dodoi{10.3847/1538-4357/ab2cb5}

\bibitem[{{Andrews} {et~al.}(2013){Andrews}, {Rosenfeld}, {Kraus}, \&
  {Wilner}}]{andrews2013}
{Andrews}, S.~M., {Rosenfeld}, K.~A., {Kraus}, A.~L., \& {Wilner}, D.~J. 2013,
  \apj, 771, 129, \dodoi{10.1088/0004-637X/771/2/129}

\bibitem[{{Andrews} \& {Williams}(2005)}]{andrews2005}
{Andrews}, S.~M., \& {Williams}, J.~P. 2005, \apj, 631, 1134,
  \dodoi{10.1086/432712}

\bibitem[{{Ansdell} {et~al.}(2017){Ansdell}, {Williams}, {Manara}, {Miotello},
  {Facchini}, {van der Marel}, {Testi}, \& {van
  Dishoeck}}]{2017AJ....153..240A}
{Ansdell}, M., {Williams}, J.~P., {Manara}, C.~F., {et~al.} 2017, \aj, 153,
  240, \dodoi{10.3847/1538-3881/aa69c0}

\bibitem[{{Armitage} {et~al.}(2013){Armitage}, {Simon}, \&
  {Martin}}]{2013ApJ...778L..14A}
{Armitage}, P.~J., {Simon}, J.~B., \& {Martin}, R.~G. 2013, \apjl, 778, L14,
  \dodoi{10.1088/2041-8205/778/1/L14}

\bibitem[{{Astropy Collaboration} {et~al.}(2018){Astropy Collaboration},
  {Price-Whelan}, {Sip{\H{o}}cz}, {G{\"u}nther}, {Lim}, {Crawford}, {Conseil},
  {Shupe}, {Craig}, {Dencheva}, {Ginsburg}, {VanderPlas}, {Bradley},
  {P{\'e}rez-Su{\'a}rez}, {de Val-Borro}, {Aldcroft}, {Cruz}, {Robitaille},
  {Tollerud}, {Ardelean}, {Babej}, {Bach}, {Bachetti}, {Bakanov}, {Bamford},
  {Barentsen}, {Barmby}, {Baumbach}, {Berry}, {Biscani}, {Boquien}, {Bostroem},
  {Bouma}, {Brammer}, {Bray}, {Breytenbach}, {Buddelmeijer}, {Burke},
  {Calderone}, {Cano Rodr{\'\i}guez}, {Cara}, {Cardoso}, {Cheedella}, {Copin},
  {Corrales}, {Crichton}, {D'Avella}, {Deil}, {Depagne}, {Dietrich}, {Donath},
  {Droettboom}, {Earl}, {Erben}, {Fabbro}, {Ferreira}, {Finethy}, {Fox},
  {Garrison}, {Gibbons}, {Goldstein}, {Gommers}, {Greco}, {Greenfield},
  {Groener}, {Grollier}, {Hagen}, {Hirst}, {Homeier}, {Horton}, {Hosseinzadeh},
  {Hu}, {Hunkeler}, {Ivezi{\'c}}, {Jain}, {Jenness}, {Kanarek}, {Kendrew},
  {Kern}, {Kerzendorf}, {Khvalko}, {King}, {Kirkby}, {Kulkarni}, {Kumar},
  {Lee}, {Lenz}, {Littlefair}, {Ma}, {Macleod}, {Mastropietro}, {McCully},
  {Montagnac}, {Morris}, {Mueller}, {Mumford}, {Muna}, {Murphy}, {Nelson},
  {Nguyen}, {Ninan}, {N{\"o}the}, {Ogaz}, {Oh}, {Parejko}, {Parley}, {Pascual},
  {Patil}, {Patil}, {Plunkett}, {Prochaska}, {Rastogi}, {Reddy Janga},
  {Sabater}, {Sakurikar}, {Seifert}, {Sherbert}, {Sherwood-Taylor}, {Shih},
  {Sick}, {Silbiger}, {Singanamalla}, {Singer}, {Sladen}, {Sooley},
  {Sornarajah}, {Streicher}, {Teuben}, {Thomas}, {Tremblay}, {Turner},
  {Terr{\'o}n}, {van Kerkwijk}, {de la Vega}, {Watkins}, {Weaver}, {Whitmore},
  {Woillez}, {Zabalza}, \& {Astropy Contributors}}]{2018AJ....156..123A}
{Astropy Collaboration}, {Price-Whelan}, A.~M., {Sip{\H{o}}cz}, B.~M., {et~al.}
  2018, \aj, 156, 123, \dodoi{10.3847/1538-3881/aabc4f}

\bibitem[{{Bai}(2016)}]{2016ApJ...821...80B}
{Bai}, X.-N. 2016, \apj, 821, 80, \dodoi{10.3847/0004-637X/821/2/80}

\bibitem[{{Bai} \& {Stone}(2011)}]{2011ApJ...736..144B}
{Bai}, X.-N., \& {Stone}, J.~M. 2011, \apj, 736, 144,
  \dodoi{10.1088/0004-637X/736/2/144}

\bibitem[{{Bai} \& {Stone}(2013)}]{2013ApJ...769...76B}
---. 2013, \apj, 769, 76, \dodoi{10.1088/0004-637X/769/1/76}

\bibitem[{{Barenfeld} {et~al.}(2016){Barenfeld}, {Carpenter}, {Ricci}, \&
  {Isella}}]{2016ApJ...827..142B}
{Barenfeld}, S.~A., {Carpenter}, J.~M., {Ricci}, L., \& {Isella}, A. 2016,
  \apj, 827, 142, \dodoi{10.3847/0004-637X/827/2/142}

\bibitem[{{B{\'e}thune} {et~al.}(2017){B{\'e}thune}, {Lesur}, \&
  {Ferreira}}]{2017A&A...600A..75B}
{B{\'e}thune}, W., {Lesur}, G., \& {Ferreira}, J. 2017, \aap, 600, A75,
  \dodoi{10.1051/0004-6361/201630056}

\bibitem[{{Blandford} \& {Payne}(1982)}]{1982MNRAS.199..883B}
{Blandford}, R.~D., \& {Payne}, D.~G. 1982, \mnras, 199, 883,
  \dodoi{10.1093/mnras/199.4.883}

\bibitem[{{Booth} {et~al.}(2021){Booth}, {Tabone}, {Ilee}, {Walsh}, {Aikawa},
  {Andrews}, {Bae}, {Bergin}, {Bergner}, {Bosman}, {Calahan}, {Cataldi},
  {Cleeves}, {Czekala}, {Guzm{\'a}n}, {Huang}, {Law}, {Le Gal}, {Long},
  {Loomis}, {M{\'e}nard}, {Nomura}, {{\"O}berg}, {Qi}, {Schwarz}, {Teague},
  {Tsukagoshi}, {Wilner}, {Yamato}, \& {Zhang}}]{2021ApJS..257...16B}
{Booth}, A.~S., {Tabone}, B., {Ilee}, J.~D., {et~al.} 2021, \apjs, 257, 16,
  \dodoi{10.3847/1538-4365/ac1ad4}

\bibitem[{{Casse} \& {Ferreira}(2000)}]{2000A&A...361.1178C}
{Casse}, F., \& {Ferreira}, J. 2000, \aap, 361, 1178.
\newblock \doarXiv{astro-ph/0008244}

\bibitem[{{Cazzoletti} {et~al.}(2019){Cazzoletti}, {Manara}, {Liu}, {van
  Dishoeck}, {Facchini}, {Alcal{\`a}}, {Ansdell}, {Testi}, {Williams},
  {Carrasco-Gonz{\'a}lez}, {Dong}, {Forbrich}, {Fukagawa}, {Galv{\'a}n-Madrid},
  {Hirano}, {Hogerheijde}, {Hasegawa}, {Muto}, {Pinilla}, {Takami}, {Tamura},
  {Tazzari}, \& {Wisniewski}}]{2019A&A...626A..11C}
{Cazzoletti}, P., {Manara}, C.~F., {Liu}, H.~B., {et~al.} 2019, \aap, 626, A11,
  \dodoi{10.1051/0004-6361/201935273}

\bibitem[{{Clarke}(2007)}]{clarke2007}
{Clarke}, C.~J. 2007, \mnras, 376, 1350,
  \dodoi{10.1111/j.1365-2966.2007.11547.x}

\bibitem[{{Clarke} {et~al.}(2001){Clarke}, {Gendrin}, \&
  {Sotomayor}}]{clarke2001}
{Clarke}, C.~J., {Gendrin}, A., \& {Sotomayor}, M. 2001, \mnras, 328, 485,
  \dodoi{10.1046/j.1365-8711.2001.04891.x}

\bibitem[{{Coleman} \& {Haworth}(2022)}]{2022MNRAS.514.2315C}
{Coleman}, G. A.~L., \& {Haworth}, T.~J. 2022, \mnras, 514, 2315,
  \dodoi{10.1093/mnras/stac1513}

\bibitem[{{de Valon} {et~al.}(2020){de Valon}, {Dougados}, {Cabrit}, {Louvet},
  {Zapata}, \& {Mardones}}]{2020A&A...634L..12D}
{de Valon}, A., {Dougados}, C., {Cabrit}, S., {et~al.} 2020, \aap, 634, L12,
  \dodoi{10.1051/0004-6361/201936950}

\bibitem[{{Delage} {et~al.}(2022){Delage}, {Okuzumi}, {Flock}, {Pinilla}, \&
  {Dzyurkevich}}]{delage2022}
{Delage}, T.~N., {Okuzumi}, S., {Flock}, M., {Pinilla}, P., \& {Dzyurkevich},
  N. 2022, \aap, 658, A97, \dodoi{10.1051/0004-6361/202141689}

\bibitem[{{Deng} {et~al.}(2025, in press){Deng}, {Vioque}, {Pascucci},
  {P{\'e}rez}, {Zhang}, {Kurtovic}, {Trapman}, {Torresvillanueva},
  {Agurto-Gangas}, {Carpenter}, {Pinilla}, {Gorti}, {Tabone}, {Sierra},
  {Rosotti}, {Cieza}, {Anania}, {Gonz{\'a}lez}, {Hogerheijde}, {Miley},
  {Ruiz-Rodriguez}, {Ruaud}, \& {Schwarz}}]{AGEPRO_III_Lupus}
{Deng}, D., {Vioque}, M., {Pascucci}, I., {et~al.} 2025, in press, \apj

\bibitem[{{Drazkowska} {et~al.}(2023){Drazkowska}, {Bitsch}, {Lambrechts},
  {Mulders}, {Harsono}, {Vazan}, {Liu}, {Ormel}, {Kretke}, \&
  {Morbidelli}}]{2023ASPC..534..717D}
{Drazkowska}, J., {Bitsch}, B., {Lambrechts}, M., {et~al.} 2023, in
  Astronomical Society of the Pacific Conference Series, Vol. 534, Protostars
  and Planets VII, ed. S.~{Inutsuka}, Y.~{Aikawa}, T.~{Muto}, K.~{Tomida}, \&
  M.~{Tamura}, 717, \dodoi{10.48550/arXiv.2203.09759}

\bibitem[{{Emsenhuber} {et~al.}(2023){Emsenhuber}, {Burn}, {Weder}, {Monsch},
  {Picogna}, {Ercolano}, \& {Preibisch}}]{2023A&A...673A..78E}
{Emsenhuber}, A., {Burn}, R., {Weder}, J., {et~al.} 2023, \aap, 673, A78,
  \dodoi{10.1051/0004-6361/202244767}

\bibitem[{{Fang} {et~al.}(2023){Fang}, {Pascucci}, {Edwards}, {Gorti},
  {Hillenbrand}, \& {Carpenter}}]{2023ApJ...945..112F}
{Fang}, M., {Pascucci}, I., {Edwards}, S., {et~al.} 2023, \apj, 945, 112,
  \dodoi{10.3847/1538-4357/acb2c9}

\bibitem[{{Fang} {et~al.}(2018){Fang}, {Pascucci}, {Edwards}, {Gorti},
  {Banzatti}, {Flock}, {Hartigan}, {Herczeg}, \&
  {Dupree}}]{2018ApJ...868...28F}
---. 2018, \apj, 868, 28, \dodoi{10.3847/1538-4357/aae780}

\bibitem[{{Favre} {et~al.}(2013){Favre}, {Cleeves}, {Bergin}, {Qi}, \&
  {Blake}}]{2013ApJ...776L..38F}
{Favre}, C., {Cleeves}, L.~I., {Bergin}, E.~A., {Qi}, C., \& {Blake}, G.~A.
  2013, \apjl, 776, L38, \dodoi{10.1088/2041-8205/776/2/L38}

\bibitem[{{Fedele} {et~al.}(2010){Fedele}, {van den Ancker}, {Henning},
  {Jayawardhana}, \& {Oliveira}}]{2010A&A...510A..72F}
{Fedele}, D., {van den Ancker}, M.~E., {Henning}, T., {Jayawardhana}, R., \&
  {Oliveira}, J.~M. 2010, \aap, 510, A72, \dodoi{10.1051/0004-6361/200912810}

\bibitem[{{Fernandes} {et~al.}(2019){Fernandes}, {Mulders}, {Pascucci},
  {Mordasini}, \& {Emsenhuber}}]{2019ApJ...874...81F}
{Fernandes}, R.~B., {Mulders}, G.~D., {Pascucci}, I., {Mordasini}, C., \&
  {Emsenhuber}, A. 2019, \apj, 874, 81, \dodoi{10.3847/1538-4357/ab0300}

\bibitem[{{Ferreira}(1997)}]{1997A&A...319..340F}
{Ferreira}, J. 1997, \aap, 319, 340.
\newblock \doarXiv{astro-ph/9607057}

\bibitem[{{Ferreira} {et~al.}(2006){Ferreira}, {Dougados}, \&
  {Cabrit}}]{2006A&A...453..785F}
{Ferreira}, J., {Dougados}, C., \& {Cabrit}, S. 2006, \aap, 453, 785,
  \dodoi{10.1051/0004-6361:20054231}

\bibitem[{{Flaischlen} {et~al.}(2021){Flaischlen}, {Preibisch}, {Manara}, \&
  {Ercolano}}]{2021A&A...648A.121F}
{Flaischlen}, S., {Preibisch}, T., {Manara}, C.~F., \& {Ercolano}, B. 2021,
  \aap, 648, A121, \dodoi{10.1051/0004-6361/202039746}

\bibitem[{{Gammie}(1996)}]{1996ApJ...457..355G}
{Gammie}, C.~F. 1996, \apj, 457, 355, \dodoi{10.1086/176735}

\bibitem[{{G{\'a}rate} {et~al.}(2024){G{\'a}rate}, {Pinilla}, {Haworth}, \&
  {Facchini}}]{garate2024}
{G{\'a}rate}, M., {Pinilla}, P., {Haworth}, T.~J., \& {Facchini}, S. 2024,
  \aap, 681, A84, \dodoi{10.1051/0004-6361/202347850}

\bibitem[{{G{\'a}rate} {et~al.}(2021){G{\'a}rate}, {Delage}, {Stadler},
  {Pinilla}, {Birnstiel}, {Stammler}, {Picogna}, {Ercolano}, {Franz}, \&
  {Lenz}}]{garate2021}
{G{\'a}rate}, M., {Delage}, T.~N., {Stadler}, J., {et~al.} 2021, \aap, 655,
  A18, \dodoi{10.1051/0004-6361/202141444}

\bibitem[{{Gorti} \& {Hollenbach}(2009)}]{2009ApJ...690.1539G}
{Gorti}, U., \& {Hollenbach}, D. 2009, \apj, 690, 1539,
  \dodoi{10.1088/0004-637X/690/2/1539}

\bibitem[{{Hartmann} {et~al.}(2006){Hartmann}, {D'Alessio}, {Calvet}, \&
  {Muzerolle}}]{hartmann2006}
{Hartmann}, L., {D'Alessio}, P., {Calvet}, N., \& {Muzerolle}, J. 2006, \apj,
  648, 484, \dodoi{10.1086/505788}

\bibitem[{{Hartmann} {et~al.}(2016){Hartmann}, {Herczeg}, \&
  {Calvet}}]{2016ARA&A..54..135H}
{Hartmann}, L., {Herczeg}, G., \& {Calvet}, N. 2016, \araa, 54, 135,
  \dodoi{10.1146/annurev-astro-081915-023347}

\bibitem[{{Hollenbach} {et~al.}(1994){Hollenbach}, {Johnstone}, {Lizano}, \&
  {Shu}}]{1994ApJ...428..654H}
{Hollenbach}, D., {Johnstone}, D., {Lizano}, S., \& {Shu}, F. 1994, \apj, 428,
  654, \dodoi{10.1086/174276}

\bibitem[{{Hunter}(2007)}]{2007CSE.....9...90H}
{Hunter}, J.~D. 2007, Computing in Science and Engineering, 9, 90,
  \dodoi{10.1109/MCSE.2007.55}

\bibitem[{{Jones} {et~al.}(2012){Jones}, {Pringle}, \&
  {Alexander}}]{2012MNRAS.419..925J}
{Jones}, M.~G., {Pringle}, J.~E., \& {Alexander}, R.~D. 2012, \mnras, 419, 925,
  \dodoi{10.1111/j.1365-2966.2011.19730.x}

\bibitem[{{Kimmig} {et~al.}(2020){Kimmig}, {Dullemond}, \&
  {Kley}}]{2020A&A...633A...4K}
{Kimmig}, C.~N., {Dullemond}, C.~P., \& {Kley}, W. 2020, \aap, 633, A4,
  \dodoi{10.1051/0004-6361/201936412}

\bibitem[{{Komaki} {et~al.}(2021){Komaki}, {Nakatani}, \&
  {Yoshida}}]{2021ApJ...910...51K}
{Komaki}, A., {Nakatani}, R., \& {Yoshida}, N. 2021, \apj, 910, 51,
  \dodoi{10.3847/1538-4357/abe2af}

\bibitem[{{Kratter} \& {Lodato}(2016)}]{2016ARA&A..54..271K}
{Kratter}, K., \& {Lodato}, G. 2016, \araa, 54, 271,
  \dodoi{10.1146/annurev-astro-081915-023307}

\bibitem[{{Kraus} {et~al.}(2012){Kraus}, {Ireland}, {Hillenbrand}, \&
  {Martinache}}]{kraus2012}
{Kraus}, A.~L., {Ireland}, M.~J., {Hillenbrand}, L.~A., \& {Martinache}, F.
  2012, \apj, 745, 19, \dodoi{10.1088/0004-637X/745/1/19}

\bibitem[{{Kunitomo} {et~al.}(2020){Kunitomo}, {Suzuki}, \&
  {Inutsuka}}]{2020MNRAS.492.3849K}
{Kunitomo}, M., {Suzuki}, T.~K., \& {Inutsuka}, S.-i. 2020, \mnras, 492, 3849,
  \dodoi{10.1093/mnras/staa087}

\bibitem[{{Laos} {et~al.}(2022){Laos}, {Wisniewski}, {Kuchner}, {Silverberg},
  {G{\"u}nther}, {Principe}, {Bonine}, {Kounkel}, \& {The Disk Detective
  Collaboration}}]{2022ApJ...935..111L}
{Laos}, S., {Wisniewski}, J.~P., {Kuchner}, M.~J., {et~al.} 2022, \apj, 935,
  111, \dodoi{10.3847/1538-4357/ac8156}

\bibitem[{{Lega} {et~al.}(2022){Lega}, {Morbidelli}, {Nelson}, {Ramos},
  {Crida}, {B{\'e}thune}, \& {Batygin}}]{2022A&A...658A..32L}
{Lega}, E., {Morbidelli}, A., {Nelson}, R.~P., {et~al.} 2022, \aap, 658, A32,
  \dodoi{10.1051/0004-6361/202141675}

\bibitem[{{Lesur} {et~al.}(2023{\natexlab{a}}){Lesur}, {Flock}, {Ercolano},
  {Lin}, {Yang}, {Barranco}, {Benitez-Llambay}, {Goodman}, {Johansen}, {Klahr},
  {Laibe}, {Lyra}, {Marcus}, {Nelson}, {Squire}, {Simon}, {Turner}, {Umurhan},
  \& {Youdin}}]{2023ASPC..534..465L}
{Lesur}, G., {Flock}, M., {Ercolano}, B., {et~al.} 2023{\natexlab{a}}, in
  Astronomical Society of the Pacific Conference Series, Vol. 534, Protostars
  and Planets VII, ed. S.~{Inutsuka}, Y.~{Aikawa}, T.~{Muto}, K.~{Tomida}, \&
  M.~{Tamura}, 465, \dodoi{10.48550/arXiv.2203.09821}

\bibitem[{{Lesur} {et~al.}(2023{\natexlab{b}}){Lesur}, {Baghdadi},
  {Wafflard-Fernandez}, {Mauxion}, {Robert}, \& {Van den
  Bossche}}]{2023A&A...677A...9L}
{Lesur}, G.~R.~J., {Baghdadi}, S., {Wafflard-Fernandez}, G., {et~al.}
  2023{\natexlab{b}}, \aap, 677, A9, \dodoi{10.1051/0004-6361/202346005}

\bibitem[{{Lodato} {et~al.}(2017){Lodato}, {Scardoni}, {Manara}, \&
  {Testi}}]{2017MNRAS.472.4700L}
{Lodato}, G., {Scardoni}, C.~E., {Manara}, C.~F., \& {Testi}, L. 2017, \mnras,
  472, 4700, \dodoi{10.1093/mnras/stx2273}

\bibitem[{{Louvet} {et~al.}(2018){Louvet}, {Dougados}, {Cabrit}, {Mardones},
  {M{\'e}nard}, {Tabone}, {Pinte}, \& {Dent}}]{2018A&A...618A.120L}
{Louvet}, F., {Dougados}, C., {Cabrit}, S., {et~al.} 2018, \aap, 618, A120,
  \dodoi{10.1051/0004-6361/201731733}

\bibitem[{{Lynden-Bell} \& {Pringle}(1974)}]{1974MNRAS.168..603L}
{Lynden-Bell}, D., \& {Pringle}, J.~E. 1974, \mnras, 168, 603,
  \dodoi{10.1093/mnras/168.3.603}

\bibitem[{{Manara} {et~al.}(2023){Manara}, {Ansdell}, {Rosotti}, {Hughes},
  {Armitage}, {Lodato}, \& {Williams}}]{2023ASPC..534..539M}
{Manara}, C.~F., {Ansdell}, M., {Rosotti}, G.~P., {et~al.} 2023, in
  Astronomical Society of the Pacific Conference Series, Vol. 534, Protostars
  and Planets VII, ed. S.~{Inutsuka}, Y.~{Aikawa}, T.~{Muto}, K.~{Tomida}, \&
  M.~{Tamura}, 539, \dodoi{10.48550/arXiv.2203.09930}

\bibitem[{{Manara} {et~al.}(2016){Manara}, {Rosotti}, {Testi}, {Natta},
  {Alcal{\'a}}, {Williams}, {Ansdell}, {Miotello}, {van der Marel}, {Tazzari},
  {Carpenter}, {Guidi}, {Mathews}, {Oliveira}, {Prusti}, \& {van
  Dishoeck}}]{2016A&A...591L...3M}
{Manara}, C.~F., {Rosotti}, G., {Testi}, L., {et~al.} 2016, \aap, 591, L3,
  \dodoi{10.1051/0004-6361/201628549}

\bibitem[{{Manara} {et~al.}(2017){Manara}, {Testi}, {Herczeg}, {Pascucci},
  {Alcal{\'a}}, {Natta}, {Antoniucci}, {Fedele}, {Mulders}, {Henning},
  {Mohanty}, {Prusti}, \& {Rigliaco}}]{2017A&A...604A.127M}
{Manara}, C.~F., {Testi}, L., {Herczeg}, G.~J., {et~al.} 2017, \aap, 604, A127,
  \dodoi{10.1051/0004-6361/201630147}

\bibitem[{{Manara} {et~al.}(2019){Manara}, {Tazzari}, {Long}, {Herczeg},
  {Lodato}, {Rota}, {Cazzoletti}, {van der Plas}, {Pinilla}, {Dipierro},
  {Edwards}, {Harsono}, {Johnstone}, {Liu}, {Menard}, {Nisini}, {Ragusa},
  {Boehler}, \& {Cabrit}}]{2019A&A...628A..95M}
{Manara}, C.~F., {Tazzari}, M., {Long}, F., {et~al.} 2019, \aap, 628, A95,
  \dodoi{10.1051/0004-6361/201935964}

\bibitem[{{Manara} {et~al.}(2020){Manara}, {Natta}, {Rosotti}, {Alcal{\'a}},
  {Nisini}, {Lodato}, {Testi}, {Pascucci}, {Hillenbrand}, {Carpenter},
  {Scholz}, {Fedele}, {Frasca}, {Mulders}, {Rigliaco}, {Scardoni}, \&
  {Zari}}]{2020A&A...639A..58M}
{Manara}, C.~F., {Natta}, A., {Rosotti}, G.~P., {et~al.} 2020, \aap, 639, A58,
  \dodoi{10.1051/0004-6361/202037949}

\bibitem[{{Martel} \& {Lesur}(2022)}]{2022A&A...667A..17M}
{Martel}, {\'E}., \& {Lesur}, G. 2022, \aap, 667, A17,
  \dodoi{10.1051/0004-6361/202142946}

\bibitem[{{Martire} {et~al.}(2024){Martire}, {Longarini}, {Lodato}, {Rosotti},
  {Winter}, {Facchini}, {Hardiman}, {Benisty}, {Stadler}, {Izquierdo}, \&
  {Testi}}]{2024A&A...686A...9M}
{Martire}, P., {Longarini}, C., {Lodato}, G., {et~al.} 2024, \aap, 686, A9,
  \dodoi{10.1051/0004-6361/202348546}

\bibitem[{{Mauc{\'o}} {et~al.}(2023){Mauc{\'o}}, {Manara}, {Ansdell},
  {Bettoni}, {Claes}, {Alcala}, {Miotello}, {Facchini}, {Haworth}, {Lodato}, \&
  {Williams}}]{2023A&A...679A..82M}
{Mauc{\'o}}, K., {Manara}, C.~F., {Ansdell}, M., {et~al.} 2023, \aap, 679, A82,
  \dodoi{10.1051/0004-6361/202347627}

\bibitem[{{Miotello} {et~al.}(2023){Miotello}, {Kamp}, {Birnstiel}, {Cleeves},
  \& {Kataoka}}]{2023ASPC..534..501M}
{Miotello}, A., {Kamp}, I., {Birnstiel}, T., {Cleeves}, L.~C., \& {Kataoka}, A.
  2023, in Astronomical Society of the Pacific Conference Series, Vol. 534,
  Protostars and Planets VII, ed. S.~{Inutsuka}, Y.~{Aikawa}, T.~{Muto},
  K.~{Tomida}, \& M.~{Tamura}, 501, \dodoi{10.48550/arXiv.2203.09818}

\bibitem[{{Miret-Roig} {et~al.}(2022){Miret-Roig}, {Galli}, {Olivares}, {Bouy},
  {Alves}, \& {Barrado}}]{2022A&A...667A.163M}
{Miret-Roig}, N., {Galli}, P.~A.~B., {Olivares}, J., {et~al.} 2022, \aap, 667,
  A163, \dodoi{10.1051/0004-6361/202244709}

\bibitem[{{Morbidelli} \& {Raymond}(2016)}]{2016JGRE..121.1962M}
{Morbidelli}, A., \& {Raymond}, S.~N. 2016, Journal of Geophysical Research
  (Planets), 121, 1962, \dodoi{10.1002/2016JE005088}

\bibitem[{{Morishima}(2012)}]{Morishima2012}
{Morishima}, R. 2012, \mnras, 420, 2851,
  \dodoi{10.1111/j.1365-2966.2011.19940.x}

\bibitem[{{Mulders} {et~al.}(2017){Mulders}, {Pascucci}, {Manara}, {Testi},
  {Herczeg}, {Henning}, {Mohanty}, \& {Lodato}}]{2017ApJ...847...31M}
{Mulders}, G.~D., {Pascucci}, I., {Manara}, C.~F., {et~al.} 2017, \apj, 847,
  31, \dodoi{10.3847/1538-4357/aa8906}

\bibitem[{{Nakatani} {et~al.}(2018){Nakatani}, {Hosokawa}, {Yoshida}, {Nomura},
  \& {Kuiper}}]{2018ApJ...857...57N}
{Nakatani}, R., {Hosokawa}, T., {Yoshida}, N., {Nomura}, H., \& {Kuiper}, R.
  2018, \apj, 857, 57, \dodoi{10.3847/1538-4357/aab70b}

\bibitem[{{Nazari} {et~al.}(2024){Nazari}, {Tabone}, {Ahmadi}, {Cabrit}, {van
  Dishoeck}, {Codella}, {Ferreira}, {Podio}, {Tychoniec}, \& {van
  Gelder}}]{2024A&A...686A.201N}
{Nazari}, P., {Tabone}, B., {Ahmadi}, A., {et~al.} 2024, \aap, 686, A201,
  \dodoi{10.1051/0004-6361/202348676}

\bibitem[{{Ogihara} {et~al.}(2018){Ogihara}, {Kokubo}, {Suzuki}, \&
  {Morbidelli}}]{2018A&A...615A..63O}
{Ogihara}, M., {Kokubo}, E., {Suzuki}, T.~K., \& {Morbidelli}, A. 2018, \aap,
  615, A63, \dodoi{10.1051/0004-6361/201832720}

\bibitem[{{Ogihara} {et~al.}(2015){Ogihara}, {Morbidelli}, \&
  {Guillot}}]{2015A&A...584L...1O}
{Ogihara}, M., {Morbidelli}, A., \& {Guillot}, T. 2015, \aap, 584, L1,
  \dodoi{10.1051/0004-6361/201527117}

\bibitem[{{Owen} {et~al.}(2012){Owen}, {Clarke}, \&
  {Ercolano}}]{2012MNRAS.422.1880O}
{Owen}, J.~E., {Clarke}, C.~J., \& {Ercolano}, B. 2012, \mnras, 422, 1880,
  \dodoi{10.1111/j.1365-2966.2011.20337.x}

\bibitem[{{Owen} {et~al.}(2010){Owen}, {Ercolano}, {Clarke}, \&
  {Alexander}}]{2010MNRAS.401.1415O}
{Owen}, J.~E., {Ercolano}, B., {Clarke}, C.~J., \& {Alexander}, R.~D. 2010,
  \mnras, 401, 1415, \dodoi{10.1111/j.1365-2966.2009.15771.x}

\bibitem[{{Pascucci} {et~al.}(2023){Pascucci}, {Cabrit}, {Edwards}, {Gorti},
  {Gressel}, \& {Suzuki}}]{2023ASPC..534..567P}
{Pascucci}, I., {Cabrit}, S., {Edwards}, S., {et~al.} 2023, in Astronomical
  Society of the Pacific Conference Series, Vol. 534, Protostars and Planets
  VII, ed. S.~{Inutsuka}, Y.~{Aikawa}, T.~{Muto}, K.~{Tomida}, \& M.~{Tamura},
  567, \dodoi{10.48550/arXiv.2203.10068}

\bibitem[{{Pascucci} {et~al.}(2020){Pascucci}, {Banzatti}, {Gorti}, {Fang},
  {Pontoppidan}, {Alexander}, {Ballabio}, {Edwards}, {Salyk}, {Sacco},
  {Flaccomio}, {Blake}, {Carmona}, {Hall}, {Kamp}, {K{\"a}ufl}, {Meeus},
  {Meyer}, {Pauly}, {Steendam}, \& {Sterzik}}]{2020ApJ...903...78P}
{Pascucci}, I., {Banzatti}, A., {Gorti}, U., {et~al.} 2020, \apj, 903, 78,
  \dodoi{10.3847/1538-4357/abba3c}

\bibitem[{{Pascucci} {et~al.}(2025){Pascucci}, {Beck}, {Cabrit}, {Bajaj},
  {Edwards}, {Louvet}, {Najita}, {Skinner}, {Gorti}, {Salyk}, {Brittain},
  {Krijt}, {Muzerolle Page}, {Ruaud}, {Schwarz}, {Semenov}, {Duch{\^e}ne}, \&
  {Villenave}}]{2025NatAs...9...81P}
{Pascucci}, I., {Beck}, T.~L., {Cabrit}, S., {et~al.} 2025, Nature Astronomy,
  9, 81, \dodoi{10.1038/s41550-024-02385-7}

\bibitem[{{Picogna} {et~al.}(2021){Picogna}, {Ercolano}, \&
  {Espaillat}}]{2021MNRAS.508.3611P}
{Picogna}, G., {Ercolano}, B., \& {Espaillat}, C.~C. 2021, \mnras, 508, 3611,
  \dodoi{10.1093/mnras/stab2883}

\bibitem[{{Preibisch} {et~al.}(2005){Preibisch}, {Kim}, {Favata}, {Feigelson},
  {Flaccomio}, {Getman}, {Micela}, {Sciortino}, {Stassun}, {Stelzer}, \&
  {Zinnecker}}]{2005ApJS..160..401P}
{Preibisch}, T., {Kim}, Y.-C., {Favata}, F., {et~al.} 2005, \apjs, 160, 401,
  \dodoi{10.1086/432891}

\bibitem[{{Ratzenb{\"o}ck} {et~al.}(2023){Ratzenb{\"o}ck}, {Gro{\ss}schedl},
  {Alves}, {Miret-Roig}, {Bomze}, {Forbes}, {Goodman}, {Hacar}, {Lin},
  {Meingast}, {M{\"o}ller}, {Piecka}, {Posch}, {Rottensteiner}, {Swiggum}, \&
  {Zucker}}]{2023A&A...678A..71R}
{Ratzenb{\"o}ck}, S., {Gro{\ss}schedl}, J.~E., {Alves}, J., {et~al.} 2023,
  \aap, 678, A71, \dodoi{10.1051/0004-6361/202346901}

\bibitem[{{Ribas} {et~al.}(2015){Ribas}, {Bouy}, \&
  {Mer{\'\i}n}}]{2015A&A...576A..52R}
{Ribas}, {\'A}., {Bouy}, H., \& {Mer{\'\i}n}, B. 2015, \aap, 576, A52,
  \dodoi{10.1051/0004-6361/201424846}

\bibitem[{{Ribas} {et~al.}(2014){Ribas}, {Mer{\'\i}n}, {Bouy}, \&
  {Maud}}]{2014A&A...561A..54R}
{Ribas}, {\'A}., {Mer{\'\i}n}, B., {Bouy}, H., \& {Maud}, L.~T. 2014, \aap,
  561, A54, \dodoi{10.1051/0004-6361/201322597}

\bibitem[{{Ricci} {et~al.}(2010){Ricci}, {Testi}, {Natta}, {Neri}, {Cabrit}, \&
  {Herczeg}}]{ricci2010}
{Ricci}, L., {Testi}, L., {Natta}, A., {et~al.} 2010, \aap, 512, A15,
  \dodoi{10.1051/0004-6361/200913403}

\bibitem[{{Riols} {et~al.}(2020){Riols}, {Lesur}, \&
  {Menard}}]{2020A&A...639A..95R}
{Riols}, A., {Lesur}, G., \& {Menard}, F. 2020, \aap, 639, A95,
  \dodoi{10.1051/0004-6361/201937418}

\bibitem[{{Rodenkirch} {et~al.}(2020){Rodenkirch}, {Klahr}, {Fendt}, \&
  {Dullemond}}]{2020A&A...633A..21R}
{Rodenkirch}, P.~J., {Klahr}, H., {Fendt}, C., \& {Dullemond}, C.~P. 2020,
  \aap, 633, A21, \dodoi{10.1051/0004-6361/201834945}

\bibitem[{{Rosotti} {et~al.}(2017){Rosotti}, {Clarke}, {Manara}, \&
  {Facchini}}]{2017MNRAS.468.1631R}
{Rosotti}, G.~P., {Clarke}, C.~J., {Manara}, C.~F., \& {Facchini}, S. 2017,
  \mnras, 468, 1631, \dodoi{10.1093/mnras/stx595}

\bibitem[{{Ruaud} {et~al.}(2022){Ruaud}, {Gorti}, \&
  {Hollenbach}}]{2022ApJ...925...49R}
{Ruaud}, M., {Gorti}, U., \& {Hollenbach}, D.~J. 2022, \apj, 925, 49,
  \dodoi{10.3847/1538-4357/ac3826}

\bibitem[{{Ruiz-Rodriguez} {et~al.}(2025, in press){Ruiz-Rodriguez},
  {Gonz\'alez-Ruilova}, {Cieza}, {Zhang}, {Trapman}, {Pinilla}, {Pascucci},
  {P\'erez}, {Deng}, {Agurto-Gangas}, {Carpenter}, {Tabone}, {Rosotti},
  {Sierra}, {Anania}, {Miley}, {Schwarz}, {Vioque}, \&
  {Kurtovic}}]{AGEPRO_II_Ophiuchus}
{Ruiz-Rodriguez}, D.~A., {Gonz\'alez-Ruilova}, C., {Cieza}, L.~A., {et~al.}
  2025, in press, \apj

\bibitem[{{Sellek} {et~al.}(2020){Sellek}, {Booth}, \&
  {Clarke}}]{2020MNRAS.498.2845S}
{Sellek}, A.~D., {Booth}, R.~A., \& {Clarke}, C.~J. 2020, \mnras, 498, 2845,
  \dodoi{10.1093/mnras/staa2519}

\bibitem[{{Sellek} {et~al.}(2024){Sellek}, {Grassi}, {Picogna}, {Rab},
  {Clarke}, \& {Ercolano}}]{2024arXiv240800848S}
{Sellek}, A.~D., {Grassi}, T., {Picogna}, G., {et~al.} 2024, arXiv e-prints,
  arXiv:2408.00848, \dodoi{10.48550/arXiv.2408.00848}

\bibitem[{{Shakura} \& {Sunyaev}(1973)}]{1973A&A....24..337S}
{Shakura}, N.~I., \& {Sunyaev}, R.~A. 1973, \aap, 500, 33

\bibitem[{{Simon} {et~al.}(2016){Simon}, {Pascucci}, {Edwards}, {Feng},
  {Gorti}, {Hollenbach}, {Rigliaco}, \& {Keane}}]{2016ApJ...831..169S}
{Simon}, M.~N., {Pascucci}, I., {Edwards}, S., {et~al.} 2016, \apj, 831, 169,
  \dodoi{10.3847/0004-637X/831/2/169}

\bibitem[{{Somigliana} {et~al.}(2020){Somigliana}, {Toci}, {Lodato}, {Rosotti},
  \& {Manara}}]{2020MNRAS.492.1120S}
{Somigliana}, A., {Toci}, C., {Lodato}, G., {Rosotti}, G., \& {Manara}, C.~F.
  2020, \mnras, 492, 1120, \dodoi{10.1093/mnras/stz3481}

\bibitem[{{Somigliana} {et~al.}(2024){Somigliana}, {Testi}, {Rosotti}, {Toci},
  {Lodato}, {Anania}, {Tabone}, {Tazzari}, {Klessen}, {Lebreuilly},
  {Hennebelle}, \& {Molinari}}]{2024arXiv240721101S}
{Somigliana}, A., {Testi}, L., {Rosotti}, G., {et~al.} 2024, arXiv e-prints,
  arXiv:2407.21101, \dodoi{10.48550/arXiv.2407.21101}

\bibitem[{{Suriano} {et~al.}(2018){Suriano}, {Li}, {Krasnopolsky}, \&
  {Shang}}]{2018MNRAS.477.1239S}
{Suriano}, S.~S., {Li}, Z.-Y., {Krasnopolsky}, R., \& {Shang}, H. 2018, \mnras,
  477, 1239, \dodoi{10.1093/mnras/sty717}

\bibitem[{{Suzuki} {et~al.}(2016){Suzuki}, {Ogihara}, {Morbidelli}, {Crida}, \&
  {Guillot}}]{2016A&A...596A..74S}
{Suzuki}, T.~K., {Ogihara}, M., {Morbidelli}, A., {Crida}, A., \& {Guillot}, T.
  2016, \aap, 596, A74, \dodoi{10.1051/0004-6361/201628955}

\bibitem[{{Tabone} {et~al.}(2022{\natexlab{a}}){Tabone}, {Rosotti}, {Cridland},
  {Armitage}, \& {Lodato}}]{2022MNRAS.512.2290T}
{Tabone}, B., {Rosotti}, G.~P., {Cridland}, A.~J., {Armitage}, P.~J., \&
  {Lodato}, G. 2022{\natexlab{a}}, \mnras, 512, 2290,
  \dodoi{10.1093/mnras/stab3442}

\bibitem[{{Tabone} {et~al.}(2022{\natexlab{b}}){Tabone}, {Rosotti}, {Lodato},
  {Armitage}, {Cridland}, \& {van Dishoeck}}]{2022MNRAS.512L..74T}
{Tabone}, B., {Rosotti}, G.~P., {Lodato}, G., {et~al.} 2022{\natexlab{b}},
  \mnras, 512, L74, \dodoi{10.1093/mnrasl/slab124}

\bibitem[{{Tabone} {et~al.}(2017){Tabone}, {Cabrit}, {Bianchi}, {Ferreira},
  {Pineau des For{\^e}ts}, {Codella}, {Gusdorf}, {Gueth}, {Podio}, \&
  {Chapillon}}]{2017A&A...607L...6T}
{Tabone}, B., {Cabrit}, S., {Bianchi}, E., {et~al.} 2017, \aap, 607, L6,
  \dodoi{10.1051/0004-6361/201731691}

\bibitem[{{Tabone} {et~al.}(2020){Tabone}, {Cabrit}, {Pineau des For{\^e}ts},
  {Ferreira}, {Gusdorf}, {Podio}, {Bianchi}, {Chapillon}, {Codella}, \&
  {Gueth}}]{2020A&A...640A..82T}
{Tabone}, B., {Cabrit}, S., {Pineau des For{\^e}ts}, G., {et~al.} 2020, \aap,
  640, A82, \dodoi{10.1051/0004-6361/201834377}

\bibitem[{{Testi} {et~al.}(2014){Testi}, {Birnstiel}, {Ricci}, {Andrews},
  {Blum}, {Carpenter}, {Dominik}, {Isella}, {Natta}, {Williams}, \&
  {Wilner}}]{2014prpl.conf..339T}
{Testi}, L., {Birnstiel}, T., {Ricci}, L., {et~al.} 2014, in Protostars and
  Planets VI, ed. H.~{Beuther}, R.~S. {Klessen}, C.~P. {Dullemond}, \&
  T.~{Henning}, 339--361, \dodoi{10.2458/azu_uapress_9780816531240-ch015}

\bibitem[{{Testi} {et~al.}(2022){Testi}, {Natta}, {Manara}, {de Gregorio
  Monsalvo}, {Lodato}, {Lopez}, {Muzic}, {Pascucci}, {Sanchis}, {Miranda},
  {Scholz}, {De Simone}, \& {Williams}}]{2022A&A...663A..98T}
{Testi}, L., {Natta}, A., {Manara}, C.~F., {et~al.} 2022, \aap, 663, A98,
  \dodoi{10.1051/0004-6361/202141380}

\bibitem[{{Tobin} {et~al.}(2020){Tobin}, {Sheehan}, {Megeath},
  {D{\'\i}az-Rodr{\'\i}guez}, {Offner}, {Murillo}, {van 't Hoff}, {van
  Dishoeck}, {Osorio}, {Anglada}, {Furlan}, {Stutz}, {Reynolds}, {Karnath},
  {Fischer}, {Persson}, {Looney}, {Li}, {Stephens}, {Chandler}, {Cox},
  {Dunham}, {Tychoniec}, {Kama}, {Kratter}, {Kounkel}, {Mazur}, {Maud},
  {Patel}, {Perez}, {Sadavoy}, {Segura-Cox}, {Sharma}, {Stephenson}, {Watson},
  \& {Wyrowski}}]{2020ApJ...890..130T}
{Tobin}, J.~J., {Sheehan}, P.~D., {Megeath}, S.~T., {et~al.} 2020, \apj, 890,
  130, \dodoi{10.3847/1538-4357/ab6f64}

\bibitem[{{Toci} {et~al.}(2023){Toci}, {Lodato}, {Livio}, {Rosotti}, \&
  {Trapman}}]{2023MNRAS.518L..69T}
{Toci}, C., {Lodato}, G., {Livio}, F.~G., {Rosotti}, G., \& {Trapman}, L. 2023,
  \mnras, 518, L69, \dodoi{10.1093/mnrasl/slac137}

\bibitem[{{Toci} {et~al.}(2021){Toci}, {Rosotti}, {Lodato}, {Testi}, \&
  {Trapman}}]{2021MNRAS.507..818T}
{Toci}, C., {Rosotti}, G., {Lodato}, G., {Testi}, L., \& {Trapman}, L. 2021,
  \mnras, 507, 818, \dodoi{10.1093/mnras/stab2112}

\bibitem[{{Tong} {et~al.}(2024){Tong}, {Alexander}, \&
  {Rosotti}}]{2024MNRAS.533.1211T}
{Tong}, S., {Alexander}, R., \& {Rosotti}, G. 2024, \mnras, 533, 1211,
  \dodoi{10.1093/mnras/stae1748}

\bibitem[{{Trapman} {et~al.}(2017){Trapman}, {Miotello}, {Kama}, {van
  Dishoeck}, \& {Bruderer}}]{2017A&A...605A..69T}
{Trapman}, L., {Miotello}, A., {Kama}, M., {van Dishoeck}, E.~F., \&
  {Bruderer}, S. 2017, \aap, 605, A69, \dodoi{10.1051/0004-6361/201630308}

\bibitem[{{Trapman} {et~al.}(2020){Trapman}, {Rosotti}, {Bosman},
  {Hogerheijde}, \& {van Dishoeck}}]{2020A&A...640A...5T}
{Trapman}, L., {Rosotti}, G., {Bosman}, A.~D., {Hogerheijde}, M.~R., \& {van
  Dishoeck}, E.~F. 2020, \aap, 640, A5, \dodoi{10.1051/0004-6361/202037673}

\bibitem[{{Trapman} {et~al.}(2023){Trapman}, {Rosotti}, {Zhang}, \&
  {Tabone}}]{2023ApJ...954...41T}
{Trapman}, L., {Rosotti}, G., {Zhang}, K., \& {Tabone}, B. 2023, \apj, 954, 41,
  \dodoi{10.3847/1538-4357/ace7d1}

\bibitem[{{Trapman} {et~al.}(2022){Trapman}, {Zhang}, {van't Hoff},
  {Hogerheijde}, \& {Bergin}}]{2022ApJ...926L...2T}
{Trapman}, L., {Zhang}, K., {van't Hoff}, M. L.~R., {Hogerheijde}, M.~R., \&
  {Bergin}, E.~A. 2022, \apjl, 926, L2, \dodoi{10.3847/2041-8213/ac4f47}

\bibitem[{{Trapman} {et~al.}(in press, 2025{\natexlab{a}}){Trapman}, {Zhang},
  {Rosotti}, {Pinilla}, {Tabone}, {Pascucci}, {Agurto-Gangas}, {Anania},
  {Carpenter}, {Cieza}, {Deng}, {Garate}, {González}, {Gorti}, {Hogerheijde},
  {Kurtovic}, {Kuznetsova}, {Miley}, {Pérez}, {Ruiz-Rodriguez}, {Schwarz},
  {Sierra}, {Torresvillanueva}, \& {Vioque}}]{AGEPRO_V_gasmasses}
{Trapman}, L., {Zhang}, K., {Rosotti}, G., {et~al.} in press,
  2025{\natexlab{a}}, \apj

\bibitem[{{Trapman} {et~al.}(in press, 2025{\natexlab{b}}){Trapman}, {Vioque},
  {Kurtovic}, {Zhang}, {Rosotti}, {Pinilla}, {Carpenter}, {Cieza}, {Pascucci},
  {Anania}, {Agurto-Gangas}, {Deng}, {Miley}, {P{\'e}rez}, {Sierra}, {Tabone},
  {Ruiz-Rodriguez}, {Gonz{\'a}lez-Ruilova}, \&
  {TorresVillanueva}}]{AGEPRO_XI_gas_disk_sizes}
{Trapman}, L., {Vioque}, M., {Kurtovic}, N., {et~al.} in press,
  2025{\natexlab{b}}, \apj

\bibitem[{{van Terwisga} {et~al.}(2022){van Terwisga}, {Hacar}, {van Dishoeck},
  {Oonk}, \& {Portegies Zwart}}]{2022A&A...661A..53V}
{van Terwisga}, S.~E., {Hacar}, A., {van Dishoeck}, E.~F., {Oonk}, R., \&
  {Portegies Zwart}, S. 2022, \aap, 661, A53,
  \dodoi{10.1051/0004-6361/202141913}

\bibitem[{{Venuti} {et~al.}(2014){Venuti}, {Bouvier}, {Flaccomio}, {Alencar},
  {Irwin}, {Stauffer}, {Cody}, {Teixeira}, {Sousa}, {Micela}, {Cuillandre}, \&
  {Peres}}]{2014A&A...570A..82V}
{Venuti}, L., {Bouvier}, J., {Flaccomio}, E., {et~al.} 2014, \aap, 570, A82,
  \dodoi{10.1051/0004-6361/201423776}

\bibitem[{{Wafflard-Fernandez} \& {Lesur}(2023)}]{2023A&A...677A..70W}
{Wafflard-Fernandez}, G., \& {Lesur}, G. 2023, \aap, 677, A70,
  \dodoi{10.1051/0004-6361/202245305}

\bibitem[{{Wang} {et~al.}(2019){Wang}, {Bai}, \&
  {Goodman}}]{2019ApJ...874...90W}
{Wang}, L., {Bai}, X.-N., \& {Goodman}, J. 2019, \apj, 874, 90,
  \dodoi{10.3847/1538-4357/ab06fd}

\bibitem[{{Wang} \& {Goodman}(2017)}]{2017ApJ...847...11W}
{Wang}, L., \& {Goodman}, J. 2017, \apj, 847, 11,
  \dodoi{10.3847/1538-4357/aa8726}

\bibitem[{{Weder} {et~al.}(2023){Weder}, {Mordasini}, \&
  {Emsenhuber}}]{2023A&A...674A.165W}
{Weder}, J., {Mordasini}, C., \& {Emsenhuber}, A. 2023, \aap, 674, A165,
  \dodoi{10.1051/0004-6361/202243453}

\bibitem[{{Winter} \& {Haworth}(2022)}]{2022EPJP..137.1132W}
{Winter}, A.~J., \& {Haworth}, T.~J. 2022, European Physical Journal Plus, 137,
  1132, \dodoi{10.1140/epjp/s13360-022-03314-1}

\bibitem[{{Yap} \& {Batygin}(2024)}]{2024Icar..41716085Y}
{Yap}, T.~E., \& {Batygin}, K. 2024, \icarus, 417, 116085,
  \dodoi{10.1016/j.icarus.2024.116085}

\bibitem[{{Yoshida} {et~al.}(2022){Yoshida}, {Nomura}, {Tsukagoshi}, {Furuya},
  \& {Ueda}}]{2022ApJ...937L..14Y}
{Yoshida}, T.~C., {Nomura}, H., {Tsukagoshi}, T., {Furuya}, K., \& {Ueda}, T.
  2022, \apjl, 937, L14, \dodoi{10.3847/2041-8213/ac903a}

\bibitem[{{Zagaria} {et~al.}(2022){Zagaria}, {Rosotti}, {Clarke}, \&
  {Tabone}}]{2022MNRAS.514.1088Z}
{Zagaria}, F., {Rosotti}, G.~P., {Clarke}, C.~J., \& {Tabone}, B. 2022, \mnras,
  514, 1088, \dodoi{10.1093/mnras/stac1461}

\bibitem[{{Zhang}(2025)}]{AGE-PRO_I-overview}
{Zhang}, K. 2025, \apj.
\newblock \doarXiv{xxxx.xxxx}

\bibitem[{{Zhang} {et~al.}(2025, in press){Zhang}, {P{\'e}rez}, {Pascucci},
  {Pinilla}, {Cieza}, {Carpenter}, {Trapman}, {Deng}, {Vioque},
  {Agurto-Gangas}, {Sierra}, {Miley}, {Ruiz-Rodriguez}, {Kurtovic},
  {Gonz{\'a}lez}, {Tabone}, {Garate}, {Anania}, {Rosotti}, {Kuznetsova},
  {Schwarz}, {Gorti}, {Hogerheijde}, \& {Torresvillanueva}}]{AGEPRO_I_overview}
{Zhang}, K., {P{\'e}rez}, L., {Pascucci}, I., {et~al.} 2025, in press, \apj

\end{thebibliography}
\bibliographystyle{aasjournal}



\end{document}